\documentclass[aps, prd, reprint, nofootinbib, longbibliography,floatfix, superscriptaddress,twocolumn,english]{revtex4-2}
\usepackage[utf8]{inputenc}
\usepackage{float}
\usepackage{soul}
\usepackage{subfigure}
\usepackage{amssymb,amsmath,amsfonts,amsthm,epsfig,epstopdf,array,mathrsfs}
\usepackage{braket}
\usepackage{graphicx}
\usepackage{dcolumn}

\usepackage[normalem]{ulem}
\usepackage{placeins}
\usepackage{afterpage}
\usepackage{bm}
\usepackage{hyperref}

\usepackage{tikz}
\usepackage{comment}

\theoremstyle{definition}

\newtheorem{defn}{Definition}

\newsavebox\mybox

\def\bea{\begin{eqnarray}}
\def\eea{\end{eqnarray}}
\def\be{\begin{equation}}
\def\ee{\end{equation}}

\begin{document}

\preprint{}

\title{How ubiquitous is entanglement in quantum field theory?}

\author{Ivan Agullo}
\email{agullo@lsu.edu}
\affiliation{Department of Physics and Astronomy, Louisiana State University, Baton Rouge, LA 70803, USA}
\author{B\'eatrice Bonga}
\email{bbonga@science.ru.nl}
\affiliation{Institute for Mathematics, Astrophysics and Particle Physics,
Radboud University, 6525 AJ Nijmegen, The Netherlands}
\author{Patricia Ribes-Metidieri}
\email{patricia.ribesmetidieri@ru.nl}
\affiliation{Institute for Mathematics, Astrophysics and Particle Physics,
Radboud University, 6525 AJ Nijmegen, The Netherlands}\affiliation{Department of Physics and Astronomy, Louisiana State University, Baton Rouge, LA 70803, USA}
\author{Dimitrios Kranas}
\email{dkrana1@lsu.edu}
\affiliation{Department of Physics and Astronomy, Louisiana State University, Baton Rouge, LA 70803, USA}
\author{Sergi Nadal-Gisbert}
\email{sergi.nadal@uv.es}
\affiliation{Departamento de Fisica Teorica and IFIC, Centro Mixto Universidad de Valencia-CSIC. Facultad de Fisica, Universidad de Valencia, Burjassot-46100, Valencia, Spain}\affiliation{Department of Physics and Astronomy, Louisiana State University, Baton Rouge, LA 70803, USA}

\date{\today}

\begin{abstract}
It is well known that entanglement is widespread  in quantum field theory, in the following sense: every Reeh-Schlieder state contains entanglement between any two  spatially separated regions. This applies, in particular, to the vacuum of a noninteracting scalar theory in Minkowski spacetime. Discussions on entanglement in field theory have focused  mainly on subsystems containing infinitely many degrees of freedom---typically, the field modes that are supported within a compact region of space. 
In this article, we study  entanglement in subsystems made of finitely many field degrees of freedom, in a free scalar theory in $D+1$-dimensional Minkowski spacetime. The focus on finitely many modes of the field is   motivated by the finite capabilities of real experiments. We find that entanglement between  finite-dimensional subsystems is {\em not common at all} and that one needs to carefully select the support of modes for entanglement to show up. We also find that  entanglement is increasingly sparser in higher dimensions. We conclude that entanglement in Minkowski spacetime is significantly less ubiquitous than normally thought.

\end{abstract}

\maketitle

\section{\label{sec:intro}Introduction}

Quantum field theory has revealed unexpected and nonintuitive lessons about the way nature works. Arguably, one of the most notorious results of this paradigm is the Reeh-Schlieder theorem \cite{reehschlider}. It applies to free and interacting theories alike. To discuss its consequences in the simplest possible context, we will restrict to  free real scalar field theories in $D+1$-dimensional Minkowski spacetimes. This restriction ensures that the concepts discussed here cannot be attributed to the interactions of the field theory under consideration; they are intrinsic properties of any quantum field theory.

Consider operators of the form $\hat \Phi_F:= \int dV\, F(x)\, \hat \Phi(x)$, where $F(x)$ is a smooth function and $dV$ the spacetime volume element. These are called smeared field operators, and $F(x)$ are smearing functions (the smearing ensures that $\hat \Phi_F$ is a well-defined operator in the Hilbert space\footnote{In the sense that it maps states to other states. This is not the case without smearing; for instance,  $\hat \Phi(x)$ acting on the vacuum produces a state with infinite norm, $\langle 0|\hat \Phi(x)\hat \Phi(x)|0\rangle\to \infty$, which is clearly not part of the Hilbert space. Smeared field operators do not have this problem and are suitable candidates for the elementary observables of the theory, from which one can generate the full algebra of observables.}). It is well-known that the Hilbert space of the theory can be generated from states  of the form 
 \be \label{state}  |\Psi\rangle =\hat \Phi_{F_1}\hat \Phi_{F_2}\cdots \hat \Phi_{F_N}|0\rangle\, ,\ee
 in the sense that any state can be approximated arbitrarily well by such states, for appropriate choices of smearing functions $F_1(x),\cdots, F_N(x)$. This is not surprising, and  simply tells us that we can create any excitation of the field by acting with an appropriate combination of operators. Intuitively, one can imagine creating an excitation with support in a small laboratory by acting with a suitable set of smeared operators supported within the laboratory.   
 
What is rather surprising---and this is the content of the Reeh-Schlieder theorem---is that 
one can generate the entire Hilbert space from states of the form \eqref{state}  even if we  restrict the smearing functions to be supported within an arbitrarily small open set of Minkowski spacetime. In simple words, one can excite the field in an arbitrary  corner of the Universe by acting on the vacuum with  operators  supported exclusively within our small lab. (One cannot use this fact, however, to produce faster-than-light communication~\cite{Witten:2018,Fewster:2018qbm,Bostelmann:2020unl}.)

Although puzzling at first, this is reminiscent of the properties of maximally entangled states in quantum mechanics~\cite{Witten:2018}. Consider two quantum mechanical systems with Hilbert spaces $\mathcal{H}_A$ and $\mathcal{H}_B$ of the same dimension $n$, and let  $|\Psi\rangle$ be a pure maximally entangled state. It is well known that {\em every}  state in $\mathcal{H}_A\otimes \mathcal{H}_B$ can be obtained by acting on $|\Psi\rangle$ with an operator {\em restricted to subsystem $A$}:
\bea 
&\forall& |\alpha\rangle \in \mathcal{H}_A\otimes \mathcal{H}_B\ {\rm there \ exist} \ \hat O_A \ {\rm such \ that } \nonumber \\ \nonumber
& & |\alpha\rangle=\hat O_A \otimes \hat{\mathbb{I}}_B |\Psi\rangle  \, ,
\eea
where $\hat{\mathbb{I}}_B$ is the identity operator in $\mathcal{H}_B$ (see, for instance, \cite{nielsen2002quantum}). \footnote{This is true not only for maximally entangled states, but also for any state  whose Schmidt form
\be 
 |\Psi\rangle=\sum_i^n c_i\, |i\rangle_A| i\rangle_B\,,  \ee
 has {\em all} coefficients $c_i$ different from zero. These are sometimes called totally entangled states (maximally entangled states correspond to $c_i=1/\sqrt{n}$ for all $i$). 

The proof goes as follows. One basis state $|i\rangle_A|j\rangle_B \in \mathcal{H}_A\otimes \mathcal{H}_B$ can be obtained from  $|\Psi\rangle$ by acting on it with the operator $\frac{1}{c_j}|i 
 \rangle\langle j|_A\otimes \hat{\mathbb{I}}_B$. Similarly, we can create any other basis element in $\mathcal{H}_A\otimes \mathcal{H}_B$. Hence, 
for each and every state in  $\mathcal{H}_A\otimes \mathcal{H}_B$, there is a  linear combination of such operators whose action on $|\Psi\rangle$ produces the desired state, and such linear combination can be written in the form $\hat O_A\otimes \hat{\mathbb{I}}_B$. Note that this argument fails if any of the coefficients $c_j$ are equal to zero.}
 
This is similar to the content of the Reeh-Schlieder theorem for quantum field theory, if we identify the field degrees of freedom inside our small lab with subsystem A and all the rest  with subsystem B. The Reeh-Schlieder theorem reveals that the vacuum state is an extraordinarily rich state regarding its entanglement structure \cite{Verch:2004vj}. In particular, it has been shown that the Reeh-Schlieder theorem implies that, if $A$ and $B$ are  subsystems made of all the field degrees of freedom contained within two regions of spacetime $V_A$ and $V_B$, respectively, and the two regions are spacelike separated, subsystems $A$ and $B$ are  always entangled when the field is prepared in the vacuum \cite{Verch:2004vj,Hollands:2017dov}. The entanglement content of quantum field theory has been reinforced by calculations of the geometric entanglement entropy associated with an open region of space $V$ (see \cite{Sorkin:1985bu,Bombelli:1986rw,Srednicki:1993im,Solodukhin:2011gn,Bianchi:2012ev,Bianchi:2019pvv}, and references therein).

These results have taught us a profound lesson about quantum field theory: entanglement is ``ubiquitous" in the vacuum; and since the short-distance behavior is the same for all states, entanglement is equally ubiquitous in any other state (in Minkowski spacetime, every state with bounded energy satisfies the Reeh-Schlieder property \cite{haag_local_1996}), reflecting the fact that  entanglement between spatially separated regions is an intrinsic property of quantum field theory. 

The results summarized so far involve subsystems containing {\em infinitely many} degrees of freedom (typically, all the field modes supported within a region $V$). Although this is of interest to understand the conceptual and mathematical content of quantum field theory, it would be desirable to extend the discussion to {\em finite dimensional} subsystems of this theory. This is the goal of this paper. This extension is of direct practical interest since experimentalists have access only to a finite set of such field modes. 
\begin{figure*}
\centering
\begin{tikzpicture}
\node at (0,0) { \includegraphics[width=\textwidth]{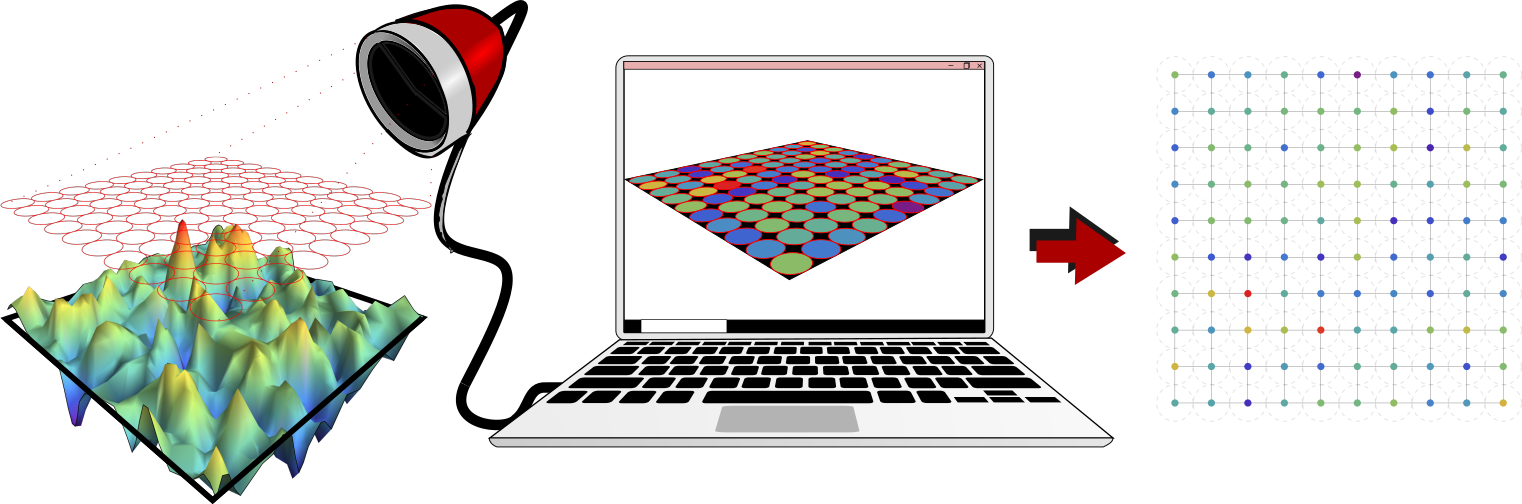}}; 
\node[rotate=-40] at (-7.25,-2.75) {$\Sigma_t$};
\node at (-7,3) {Input}; 
\node at (0.5,3) {Output}; 
\node at (7,3) {Lattice}; 
\end{tikzpicture}
\caption{Representation of a quantum field in a portion of a Cauchy hypersurface, $\Sigma_t$, that a detector might access at an instant of time. The detector is made of a finite number of pixels (represented by the red circles on top of $\Sigma_t$) and captures a simplified (smeared) version of the field in each of its pixels, as represented by the output of the detector on the laptop. The smearing process provides a way to define a lattice  theory out of the continuum field theory, by assigning the smeared field in each of the pixels of the detector with a lattice node, as depicted in the rightmost part of the figure.}
\label{fig:intro_quantum_field_and_detector}
\end{figure*}
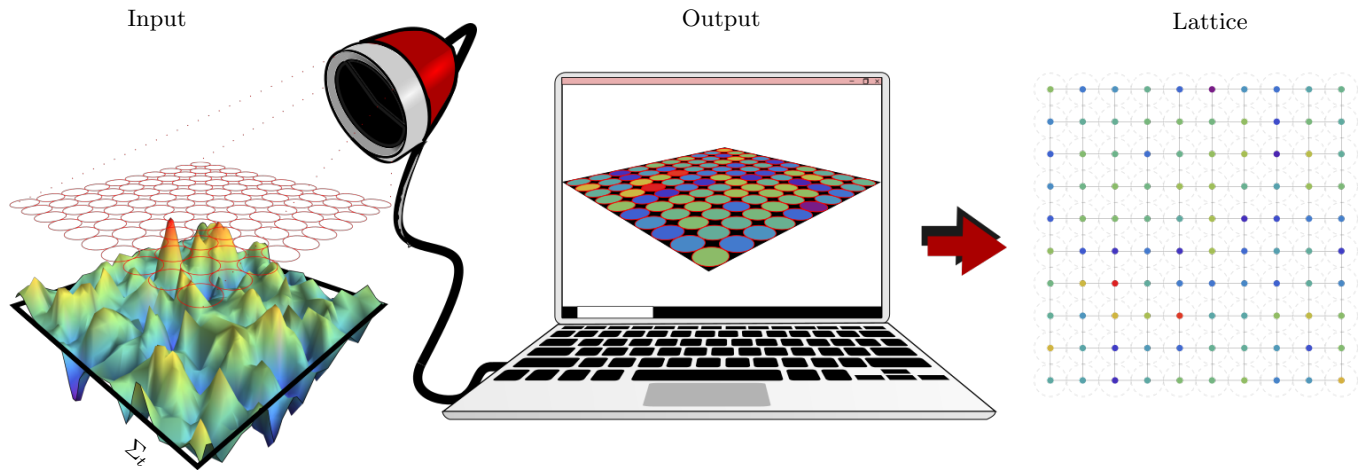

There is a common belief that  entanglement is  ubiquitous in quantum field theory, even if we restrict to finite dimensional subsystems. In particular, it is usually taken for granted that {\em any pair} of field degrees of freedom  are entangled in the vacuum state. This intuition is supported by the following fact. Given a fixed field mode compactly supported in a region $B$ of spacetime, if we choose an arbitrary compact region $A$ separated from the first, the Reeh-Schlieder theorem guarantees that there is at least one mode within region $A$ that is entangled to the fixed mode in region $B$, when the field is in the vacuum state \cite{Hollands:2017dov}. However, the theorem does not tell us how many modes in  $A$ are entangled with the fixed mode in $B$, or how complicated such modes are. Since region A hosts infinitely many modes, the belief that  any  pair of modes, one in  $A$ and one in $B$, are entangled in the vacuum is an (unjustified) extrapolation of the actual content of the Reeh-Schlieder theorem. The primary goal of this paper is to check whether this extrapolation is actually true. We find that it is not. 

We proceed as follows. We construct a family of locally defined individual modes of a scalar field theory in $D+1$-dimensional Minkowski spacetime by smearing the field and its conjugate momentum in space (see Sec.~\ref{sec:individual-modes} for the relation between smearing in space and in spacetime). The smearing function can be intuitively thought of as defining a ``pixel'' of the field theory: the support of the smearing determines the size of the pixel, corresponding to the maximum resolution of a detector, while the shape of the smearing function determines the resolution of the detector within the pixel. In this way, one can divide the space into disjoint pixels, each describing a single degree of freedom of the field theory (this is illustrated in Fig.~\ref{fig:intro_quantum_field_and_detector}). Any finite region contains a finite number of such pixels.

This strategy has the advantage that, given any two regions, each containing $N_A$ and $N_B$ degrees of freedom, one can use standard techniques in quantum mechanics of finite dimensional systems to quantify correlations and entanglement. All the difficulties and subtleties intrinsic to quantum field theory are removed. In particular, the calculations are free of the divergences that plague the calculation of the geometric entanglement entropy associated with a region in quantum field theory.  A similar strategy has been used before in \cite{Martin:2015qta,martin_real-space_2021,Martin:2021qkg,Espinosa-Portales:2022yok,bianchi_entropy_2019} to evaluate mutual information, entropy, quantum discord, and to search for violations of a type of  Bell inequalities, with interesting applications in cosmology (see also \cite{Morgan:2022cdm}).

One can think of our ``pixelation'' of space as a way of defining a lattice field theory out of the continuum theory. With the crucial difference that one is not restricting the degrees of freedom before quantization---something we want to avoid since the entanglement content of lattice field theory can be very different from the theory in the continuum---in particular, it is far from obvious if some analog of the Reeh-Schlieder theorem exists for lattice theory. Our strategy contains the benefits and kindness of lattice field theories, while keeping the richness of the continuum. The lattice constructed in this way is defined by the capabilities of experimentalists, rather than by a drastic truncation of the degrees of freedom prior to quantization. (We extend this strategy in different directions, for instance, by allowing different pixels to overlap.)

The main lesson of this article is that entanglement is significantly less ubiquitous than one would have thought. In particular, for $D\geq 2$ we do not find entanglement between pairs of modes supported in nonoverlapping regions, unless we fine-tune the family of field modes to maximize the contact between the subsystems (this fact can be explained by the analysis in \cite{deSLTorres:2023aws}, which shows that entanglement between regions is sharply concentrated close to the boundary). We also observe that entanglement is weaker in higher dimensions.

In the rest of this article, we proceed  as follows. In Sec.~\ref{sec:2}, we describe the way we isolate individual field degrees of freedom that are localized in a region of space in  a free  scalar theory. We describe how to compute the reduced state describing a finite number of such modes and how to obtain  properties of interest such as von Neumann entropy, correlations, mutual information, and entanglement. In Sec.~\ref{sec:3},  we apply this formalism to two modes belonging to a simple, yet physically interesting, family of modes. We increase the number of modes in each subsystem in Sec.~\ref{sec:4} and evaluate whether entanglement shows up between these ``richer" subsystems. In Sec.~\ref{sec:5}, we extend our analysis to a larger family of smearing functions.  In Sec.~\ref{sec:6}, we discuss choices of pairs of field modes for which we do find entanglement. Finally, Sec.~\ref{sec:discussion} collects the main results of this article, discusses their relevance, and puts them in a larger perspective.  Some details of the calculations in Sec.~\ref{sec:3} have been relegated to Appendix~\ref{app:details}. In Appendix~\ref{Ap:B}, we argue that the smearing functions introduced in Secs.~\ref{sec:3},~\ref{sec:5}, and~\ref{sec:6} give rise to well defined observables, despite the fact that some of them are not smooth. 

Throughout this paper, we use units in which ${\hbar=c=1}$.

\section{Subsystems, reduced states and entanglement}\label{sec:2}

Field theories describe physical systems with infinitely many degrees of freedom. In  experiments, however, we only have access to a finite subset of them. We describe in this section the way we isolated individual field degrees of freedom localized in a region of space in  a free scalar field theory (generalization to other types of free fields is straightforward). We then describe how to compute the reduced quantum state restricted to a finite set of such degrees of freedom when the field is prepared in the vacuum and how to compute properties of interest from it, such as the von Neumann entropy, correlations, mutual information, and entanglement.

\subsection{Defining individual  modes of the field}
\label{sec:individual-modes}

In order to fix the basic concepts, let us consider first an analog situation in standard quantum mechanics. Let us consider a set of $N$ harmonic oscillators, and let $\hat {\vec r}:= (\hat x_1,\hat p_1,\cdots, \hat x_N,\hat p_N)$ be the vector of canonical operators. The canonical commutation relations can be succinctly written as $[\hat r^i,\hat r^j]=i\, \Omega^{ij}$, where $\Omega$ is the (inverse of the) symplectic structure of the classical phase space
\begin{equation*}
   \Omega_N = \bigoplus_{i = 1}^N \Omega_2\,, \qquad  \Omega_{2}= \left(\begin{matrix} 
    0 & 1 \\ -1 & 0\end{matrix}\right) \, .
\end{equation*}
A general observable that is linear in the canonical variables can be written as $\hat O_{\vec v}:= v_i \hat r^i$, (sum over repeated indices is understood) with $\vec v \in \mathbb{R}^{2N}$.  Vectors $\vec v$ can be identified with  elements of $\Gamma^*$, the dual of the classical phase space, establishing a correspondence between linear observables in the classical and quantum theories. Written in this way, all $\hat O_{\vec v}$'s have dimensions of action, and their commutation relations are given by the symplectic product of the corresponding $\vec v$'s,
\be [\hat O_{\vec v},\hat O_{\vec v'}]=i\, v_iv'_j\Omega^{ij}\, . \ee

Any {\em noncommuting} pair of linear observables  $(\hat O_{\vec v},\hat O_{\vec v'})$ {\em defines a subsystem with a single degree of freedom} (we will refer to subsystems like this as ``modes of the system'')---more precisely, the subsystem is defined by the algebra generated by the pair $(\hat O_{\vec v},\hat O_{\vec v'})$ \cite{haag_local_1996}. For instance, subsystems corresponding to each individual oscillator are defined by the pairs $(\hat x_I,\hat p_I)$, $I=1,\cdots, N$. However, the definition is more general and includes modes which are  combinations of several oscillators (when the oscillators are coupled to each other by springs, the normal modes of the Hamiltonian are familiar examples of such combinations). This procedure provides a simple recipe to extract individual modes of our systems. This  idea can be extended to field theory as follows.

A field theory hosts infinitely many degrees of freedom. This is true even if we restrict to an arbitrarily small open region of space. Intuitively,  at each point in space $\vec x$, we have an independent pair of canonically conjugated operators $(\hat \Phi (\vec x),\hat \Pi (\vec x))$, and each such pair defines a single mode of the system; since any region contains infinitely many points, the region hosts as many independent modes. This is only heuristic because, as mentioned in the Introduction, neither of the objects $\hat \Phi (\vec x)$ nor $\hat \Pi (\vec x)$ are well-defined operators. We need to smear them out. The standard procedure is to smear the covariant operator $\hat \Phi (x)$ against a function in spacetime\footnote{Convenient choices for smearing functions in Minkowski spacetime are functions in Schwartz space \cite{Schwartz} or functions of compact support sufficiently differentiable (see Appendix \ref{Ap:B}). We will restrict to the latter since this will allow us to localize field modes in compact regions.} $ \hat \Phi_F:= \int dV F(x) \hat \Phi(x)$, with $F(x)$ a smooth function compactly supported in a region $V$. This is the set of linear observables in the theory---in this covariant formulation, the conjugate momentum $\hat \Pi=\frac{d}{dt}\hat {\Phi}$ is not needed. 
Given two operators defined in this way, their commutation relations are
\be \label{covcom} [\hat \Phi_{F_1},\hat \Phi_{F_2}]=i\, \Delta(F_1,F_2) \, ,\ee 
where 
\be \Delta(F_1,F_2):=\int dV dV'\, F_1(x) F_2(x)\, \Delta(x,x')\, ,\ee 
and $\Delta(x,x'):=G_{\rm Ad}(x,x')-G_{\rm Ret}(x,x')$ is the difference between the advanced and retarded Green's functions of the Klein-Gordon equation. Equation.~\eqref{covcom} is  simply the smeared   version of the  familiar covariant commutation relations $[\hat \Phi(x),\hat \Phi(x')]=i\, \Delta(x,x')$.  

With these definitions, given any two smearing functions $F_1$ and $F_2$ compactly supported in a region $V$, and such that the associated field operators do not commute, the pair $(\hat \Phi_{F_1},\hat \Phi_{F_2})$ defines---again, via the algebra it generates---an individual mode of the system localized in region $V$. This strategy provides  a simple way of extracting from the field theory individual degrees of freedom  localized in a given region.

It is essential to keep in mind that there are infinitely many independent modes within any open region $V$ and that a noncommuting pair $(\hat \Phi_{F_1},\hat \Phi_{F_2})$ defines just one of them. Put plainly, one should not identify a region $V$ with a single mode.

To finish this subsection, we summarize how, in free field theories,  the discussion above can be translated to a canonical picture, where instead of  smearing fields in spacetime, one smears the field  and its conjugate momentum only in space. This reformulation looks closer to the example of $N$ harmonic oscillators given above, and we will  use it in the rest of this article. 

In the canonical picture, linear observables are operators of the form
\be 
\hat O_{f,g}:=\int_{\Sigma_t} d^Dx \, \big(g(\vec x)\, \hat \Phi(\vec x,t)-f(\vec x)\, \hat \Pi(\vec x,t)\big)\, , 
\ee
where the integral is restricted to a Cauchy hypersurface $\Sigma_t$ of a $D+1$-dimensional Minkowski spacetime, which for simplicity in this paper will be chosen to be a hypersurface defined  by a constant value of the time coordinate $t$ of  any arbitrary inertial frame---although nothing will change in the discussion if we use a more complicated choice.  The functions $f(\vec x)$ and $g(\vec x)$ are  compactly supported in a region $R$ of such Cauchy $t=$constant hypersurface, and $\hat \Pi(\vec x,t):=\frac{d}{dt} \hat \Phi(\vec x,t)$. As for the example of harmonic oscillators, we can identify  pairs of functions $(f(\vec x),g(\vec x))$ with elements of the dual phase space $\Gamma^*$; then, all operators $ \hat O_{f,g}$ have dimensions of action. Operators defined from pairs of the form $(0,g(\vec x))$ are called pure field operators; similarly, pairs of the form $(f(\vec x),0)$ are called pure momentum operators. 

The commutation relations in the canonical formulation are given by the symplectic product of the smearing functions 
\be [\hat O_{f,g},\hat O_{f',g'}]=i\, \Omega\big( (f,g), (f',g')\big):=i\, \int_t d^Dx\ (f g'-gf')\, . \nonumber \ee
In this context, individual modes of the system localized in a region of space $R$ are selected by choosing two pairs of functions $(f,g)$ and $(f',g')$ supported within $R$ and such that the commutator in the previous equation is different from zero. This is the way we will define localized field modes in this article.  
The simplest choice is a subsystem defined from a pure field and a pure momentum operator, $\hat O_1=\int_td^Dx \, g \, \hat \Phi$,  $\hat O_2=\int_td^Dx\,  f\, \hat  \Pi$ such that $\int_td^Dx f g=1$, and consequently $[\hat O_1,\hat O_2]=i$. However, more general combinations will also be considered in this article.

The relation between the covariant and canonical pictures is given by the following map between functions $F$ of compact support in spacetime and pairs of functions $(f,g)$ compactly supported in space (see, for instance, \cite{Ashtekar:2021dab,Ashtekar:1975zn,Wald:1995yp} for further details). Recall that the commutator bidistribution $\Delta(x,x')$ satisfies the field equations in both its variables. Hence, by smearing the $x'$ dependence of $\Delta$ with $F$, we are left with a solution of the field equations, $s(x):=\int dV \Delta(x,x') F(x')$. By reading Cauchy data from $s(x)$ corresponding to a $t$=constant hypersurface, and $f(\vec x)=s(\vec x, t),g(\vec x)=\frac{d}{dt}s(\vec x,t)$, we obtain a map $F(x)\to (f(\vec x),g(\vec x))$, which in turn defines a map $\hat \Phi[F]\to \hat O_{f,g}$ between operators smeared in spacetime and operators smeared in space.\footnote{This map is onto but not invertible. The reason is that the operator-valued-distribution $\hat \Phi[x]$ has a Kernel, given by functions of the form $(\Box-m^2)G$, with $G$ a function of compact support in spacetime; i.e.,  $\hat \Phi(x)$ smeared with $(\Box-m^2)G$ vanishes for all $G$. In passing  to the canonical formulation, this kernel is conveniently eliminated by the map $F(x)\to (f(\vec x),g(\vec x))$ defined by $\Delta(x,x')$ because $\Delta$ has the same Kernel as $\hat \Phi(x)$. Hence, while the map $F\to \hat \Phi_F$ has a Kernel, implying that different smearing functions do not necessarily define different operators, 
the map $(f,g)\to \hat O_{f,g}$ has the advantage that is faithful.} This map preserves the commutation relations because the properties of $\Delta$ guarantee that $\Delta(F,F')$ is mapped to $\Omega((f,g),(f',g'))$ (see \cite{Ashtekar:2021dab} for a simple proof).

\subsection{Finite-dimensional subsystems and reduced states} 

Consider now a finite set of independent modes of the field. They are defined by a set of $2N$ operators $\hat O^i$, which can be straightforwardly normalized to satisfy  
\be [\hat O^i,\hat O^j]=i\, \Omega^{ij}\, .\ee
The algebra generated by these observables is isomorphic to the algebra of a quantum mechanical system with $N$ bosonic degrees of freedom (mathematically, the associated Weyl algebra is  a type I von Neumann algebra). Hence, the difficulties intrinsic to field theories are removed by restricting to such a finite set of modes; one is in the realm of standard quantum mechanics to define reduced states, entropies, and entanglement. Physically, we think of this $N$-dimensional subsystem as encoding the degrees of freedom that a particular experimentalist may be able to measure. 

We now describe how to compute the {\em reduced quantum state}  for a  finite  set of modes when the field is prepared in the vacuum state (see \cite{Martin:2021qkg,bianchi_entropy_2019} for previous similar calculations). It is well known that such state is {\em always} mixed  \cite{Verch:2004vj,Hollands:2017dov,Ruep:2021fjh}, something that we will confirm with several examples. This is an important message to keep in mind: in quantum field theory {\em all physically allowed} reduced states describing subsystems localized within a compact region of space are mixed; this is a drastic departure from standard quantum mechanics. 

Given an arbitrary state $\hat \rho$ in the full theory, the task of finding the reduced state for a subsystem of $N$ modes is complicated. However, there is a significant  simplification when $\hat \rho$ is a  Gaussian state. This is the case for the standard vacuum in Minkowski spacetime; we will restrict in the following to such states, although the generalization to other Gaussian states is straightforward. 

Recall that a Gaussian state in field theory is completely and uniquely characterized by its one- and two-point distributions, $\langle \hat \Phi(x) \rangle$ and $\langle \hat \Phi(x) \hat\Phi(x') \rangle$, respectively.  From these distributions, one obtains the first and second moments for any mode of the field by smearing them out. For the vacuum, the one-point distribution is zero. Higher-order correlators can all be obtained from $\langle \hat \Phi(x) \hat\Phi(x') \rangle$. Recall also that the reduced state of a Gaussian state is also Gaussian. Therefore, the reduced state for our $N$-mode subsystem is completely characterized by
\be \langle \hat O^i\rangle\, \quad \text{and} \quad \langle \hat O^i \hat O^j\rangle \, . \ee
For the vacuum state, $\langle \hat O^i\rangle=0$ for all $i$. Hence, the characterization of the reduced quantum state for our subsystem reduces to merely compute the second moments $\langle \hat O^i \hat O^j\rangle$, a task that we will do repeatedly in this article. 

Furthermore, we can decompose the second moments in their symmetric and antisymmetric parts,
\be \langle 0|\hat O^i\hat O^j|0 \rangle=\frac{1}{2} \, \braket{0|\{\hat O^i,\hat O^j\}|0} +\frac{1}{2} \, \braket{0| [\hat O^i ,\hat O^j]|0}\, , \ee
where curly brackets represent the standard anticommutator. Notice that the antisymmetric part, the commutator, is equal to $i\,\Omega^{ij}$ and is state independent. Therefore, all information of the reduced state is actually encoded in the anticommutator part. It is common to call that part the {\em covariance matrix} of the reduced state,
\be \sigma^{ij}:=\langle 0 | \{\hat O^i, \hat O^j\}|0\rangle \, . \ee
Therefore, {\em the calculation of the reduced state of an $N$-mode subsystem when the field is in the vacuum reduces to computing the covariance matrix $\sigma$}. It completely and uniquely characterizes the reduced state (keeping in mind that the first moments are all zero).

Several interesting aspects of the reduced state can be readily obtained from $\sigma$ in an elegant manner. For instance, Heisenberg's uncertainty principle is encoded in the statement that the matrix $\sigma + i \Omega$ is positive semidefinite\footnote{In the sense that $(\sigma^{ij} + i \, \Omega^{ij})\bar v_iv_j\geq 0$ for all $v_i\in \mathbb{C}^{2N}$.}, and the reduced state is pure if and only if the eigenvalues of $\sigma^{ik}\Omega_{kj}$ are all $\pm i$ (in which case the pair ($\sigma^{ij}, \Omega_{ij}$) defines a K\"{a}hler structure in the classical phase space \cite{Ashtekar:1975zn,Hackl:2020ken}). We will use this criterion to confirm that all reduced states we will obtain from the Minkowski vacuum are  mixed.

The information contained in the covariance matrix $\sigma$ can be invariantly characterized by  its {\em symplectic eigenvalues}, denoted by $\nu_I$, with $I=1,\cdots, N$, and defined as the modulus of the eigenvalues of the matrix $\sigma^{ik}\Omega_{kj}$, understood as a linear map in $\mathbb{C}^{2N}$. Many  quantities of interest for us can be readily obtained from the symplectic eigenvalues. For instance, the von Neumann  entropy of the reduced state is given by \cite{serafini2017quantum}
\bea \label{S} S[\bm\sigma]=\sum_I^N \Big[\left( \frac{\nu_I+1}{2}\right) \log_2\left( \frac{\nu_I+1}{2}\right)\nonumber \\ -\left( \frac{\nu_I-1}{2}\right) \log_2\left( \frac{\nu_I-1}{2}\right) \Big ]. \eea

\subsection{Correlations and entanglement} \label{subsec:2}

Given two subsystems, $A$ and $B$, made of $N_A$ and $N_B$ modes, respectively, we will be  interested in computing the correlations and entanglement between them when the field is prepared in the vacuum. 

The correlations between concrete pairs of observables $\hat O^i$ and $\hat O^j$ can be computed straightforwardly since they correspond to the elements of the covariance matrix.  

On the other hand, the total amount of correlations between the two subsystems can be quantified by means of the {\em mutual information} $\mathcal{I}(A,B)$, given by 
\begin{equation} \label{MI}
    \mathcal{I}(A,B) =   S_{A} +   S_{B} -   S_{AB}\,,
\end{equation}
where $S_A$, $S_B$, and $S_{AB}$  are the von Neumann entropies of subsystems $A$, $B$ and the joined system $AB$, respectively. We will check with concrete examples that the mutual information of general subsystems is different from zero in field theory, as expected, since correlations are ubiquitous. However, it is important to keep in mind that a nonzero mutual information $\mathcal{I}(A,B)$ {\em does not} imply that the two subsystems are entangled since $\mathcal{I}(A,B)$ quantifies all correlations,  both of classical and quantum origins. To evaluate whether the subsystems are entangled, we need to go beyond mutual information.

The evaluation of the entanglement between arbitrary subsystems is a subtle issue in quantum field theory; a rigorous strategy to quantify such entanglement has been proposed only recently in \cite{Hollands:2017dov}. The difficulty originates, of course, from the possibly infinite number of degrees of freedom each subsystem may have. Although subsystems made of infinitely many modes are important to understand the conceptual and mathematical  structure of quantum field theory, in practical situations one  has access to a finite number of modes. For finite-dimensional subsystems, one can apply the techniques developed in quantum mechanics to define and quantify entanglement. This is the strategy we follow in this paper.

Given  a finite set of modes of the field,  we are interested in dividing them into  two subsystems, $A$ and $B$, and evaluating whether---and how much---they are entangled when the field is prepared in the Minkowski vacuum. For this task, we need to find an appropriate entanglement quantifier. Entanglement entropy, commonly used in many applications in quantum mechanics, is unfortunately useless  for our task: entanglement entropy  is a quantifier of entanglement only when the total state describing the two subsystems is pure. As discussed above, this is never the case for a finite set of modes in quantum field theory. 

Quantifying entanglement for mixed states is a subtle question, and there is generally not a simple necessary and sufficient criterion for entanglement. However, such a necessary and sufficient criterion {\em does exist} in restricted situations, such as the setup we investigate in this paper, as we  explain now.

One easily-computable measure of entanglement for pure and mixed states alike is the {\em logarithmic negativity} (LN) \cite{peres96, plenio05}, which we will denote by  $E_{\mathcal{N}}$. A nonzero value of the LN implies a violation of the positivity of partial transpose (PPT) criterion \cite{plenio05}. This in turn implies that a nonzero value of the  LN is a sufficient condition for entanglement; but it is not necessary for general quantum states. 
However, when restricting to Gaussian states and when, additionally, one of the subsystems is made of a single mode, regardless of the size of the other subsystem, the LN is different from zero {\em  if and only if} the state is entangled. Furthermore, under these circumstances the LN is a faithful quantifier of entanglement, in the sense that higher LN means more entanglement \cite{serafini2017quantum}. 

The LN is a lower bound for the entanglement that can be distilled from the system via local operations and classical communications. 
For Gaussian quantum states, the value of the LN has an operational meaning as the exact cost (where the cost is measured in Bell pairs or entangled bits) that is required to prepare or simulate the quantum state under consideration \cite{wilde2020ent_cost, wilde2020alpha_ln}. 

The LN for a Gaussian quantum state can be directly computed from its covariance matrix. Consider an $N_A+N_B$-mode Gaussian state $\hat{\rho}$ of a bipartite system, with covariance matrix $\sigma_{AB}$, where $N_A$ and $N_B$ are the number of modes in each subsystem.  The LN for the bipartition can be computed as
 \begin{equation}\label{LN}
      E_{\mathcal{N}} = \sum_{J=1}^{N_A+N_B}\max\{0,-\log_{2} \tilde{\nu}_J\}\,,
\end{equation} where $\tilde{\nu}_J$ are the symplectic eigenvalues of $\tilde{\sigma}$, defined as
\begin{equation} \label{PT}
  \widetilde{\sigma}_{AB} = \bm T\sigma_{AB}\bm T,
\end{equation}
where $\bm T= \mathbb{I}_{2N_A}\oplus\bm\Sigma_{N_B}$ and $\bm\Sigma_{N_B}=\oplus_{N_B} \sigma_z$ is a direct sum of $N_B$ $2\times2$ Pauli-$z$ matrices.  The relation between the LN and the PPT criterion can be understood by noticing that $\tilde{\sigma}$ is actually the covariance matrix of  the {\em partially transposed} density matrix $\hat{\rho}^{\top_B}$, where the transpose is taken only in the $B$ subsystem; therefore, the nonpositivity of $\hat{\rho}^{\top_B}$
implies that some of  the symplectic eigenvalues $\tilde \nu_I$ are smaller than 1, producing $E_{\mathcal{N}} >1$ (see, e.g., Ref.~\cite{serafini2017quantum} for further details).

Observe also that a sufficient condition for quantum entanglement is $\min\{\tilde{\nu}_J\}<1$.

We will mostly use the LN in situations in which it is faithful (that is, when $N_A=1$). Nevertheless, we will also analyze the LN of bipartitions of ``many versus many'' modes in  Sec.~\ref{sec:5}, which will be of interest since $E_{\mathcal{N}} >0$ is always a sufficient condition for entanglement; although in this  case it is not necessary, so $E_{\mathcal{N}} = 0$ does not imply the absence of entanglement. In any case, since the LN is a lower bound for distillable entanglement, $E_{\mathcal{N}} = 0$ indicates that  whatever entanglement may be contained in the system cannot be distilled.

\section{Correlations and entanglement between two degrees of freedom} \label{sec:3}

In this section, we apply the formalism presented above to a simple, yet  illustrative, situation: two modes supported in disjoint regions of space and each defined by a pure field and a pure momentum operator. We evaluate correlations, entropy, mutual information, and entanglement between the two modes. 
We use a family of smearing functions which, for massless fields, permits to derive  analytical expressions in any number of space dimensions $D>1$. (The $D=1$ case requires attention since one needs to introduce a mass to avoid infrared divergences; we solve the massive case numerically.)

We find no entanglement between the two modes for all values of $D>1$, i.e., we find that the reduced state is separable for the family of modes used in this section.  

We generalize this calculation to include a larger number of modes 
in Sec.~\ref{sec:4},  to other smearing functions, and more general definitions of modes in Sec.~\ref{sec:5}.

\subsection{Smearing functions, correlations and covariance matrix}

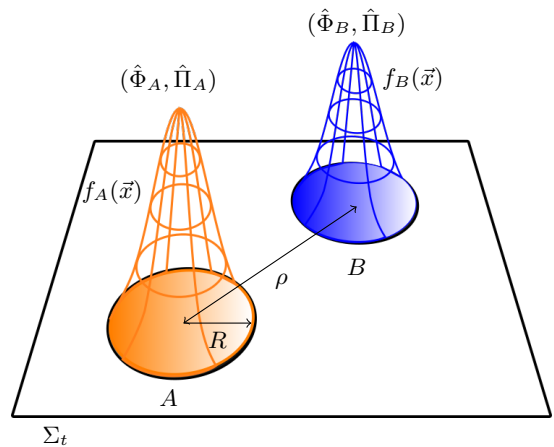
\begin{figure}
    \centering
    \begin{tikzpicture}
       \node at (0,0) {\includegraphics[width=0.4\textwidth]{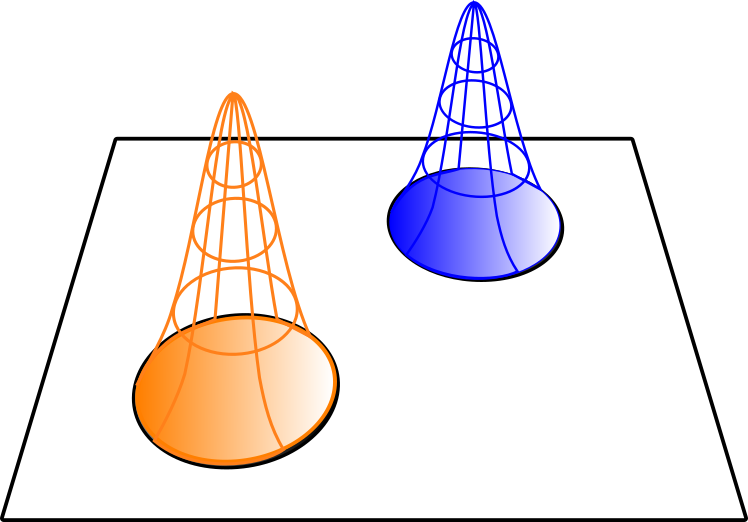}}; 
     \node at (-3,-2.75) {$\Sigma_t$}; 
     \node at (-1.5,-2.25) {$A$};
     \node at (-1.5,2) {$(\hat{\Phi}_A,\hat{\Pi}_A)$}; 
     \node at (-2.25,0.5) {$f_A(\vec{x})$};
     \node at (1.75,2.) {$f_B(\vec{x})$};
    \node at (1.,-0.5) {$B$}; 
    \node at (1,2.75) {$(\hat{\Phi}_B,\hat{\Pi}_B)$};
    \draw[<->] (-1.3,-1.25)--(1,0.3);
    \draw[->] (-1.3,-1.25)--(-0.4,-1.25);
    \node[below] at (0,- 0.475) {$\rho$};
    \node[below] at (-0.85,-1.25) {$R$};
    \end{tikzpicture}
    \caption{Illustration of two spacelike separated balls of radius $R$ in a $t$=constant Cauchy hypersurface in $D+1$-dimensional Minkowski spacetime. A function $f_{A(B)}(\vec{x})$  compactly supported in region A (B) defines a single field-mode $(\hat{\Phi}_{A(B)}, \hat{\Pi}_{A(B)})$, as shown in Eq.~\eqref{eq:phiA2dof_def}.  }
    \label{fig:setup_2dof}
\end{figure}

Consider two $D$ dimensional balls, $A$ and $B$, with radius $R$ in a $D+1$-dimensional Minkowski spacetime, and let $\rho$ be the distance between their centers in units of $R$. The balls are assumed to be disjoint so that $\rho>2$ (see Fig.\ \ref{fig:setup_2dof}). 

We consider in this section two modes, each supported within  region $A$ and $B$, respectively, and defined as follows. The mode in $A$ is defined by a pair of noncommuting operators of the form
\bea\label{eq:phiA2dof_def}
    \hat{\Phi}_A &:=&  \int \mathrm{d}^D x \,  f_A(\vec x) \,\hat{\Phi}(\vec{x})\, , \nonumber \\
     \hat{\Pi}_A &:=&  c \int \mathrm{d}^D x \,  f_A(\vec x) \,\hat{\Pi}(\vec{x})\, ,  
\eea
where $f_A(\vec x)$ is a function compactly supported in region $A$, and 
$c$ is  an arbitrary constant with dimensions of inverse energy (mutual information and entanglement between two subsystems will not depend on the value of $c$ since changing $c$ amounts to performing a symplectic transformation restricted to one subsystem, and these quantities are invariant under such ``local'' transformations). In this section, we denote the pair of noncommuting operators defining the modes of interest as $(\hat{\Phi}_A,\hat{\Pi}_A)$, rather than $(\hat{O}_A^{1},\hat{O}_A^{2})$, as we did in the last section, in order to emphasize that we choose them to be a pure field and pure momentum operators, respectively. The mode $B$ is similarly defined  by using a function $f_B(\vec x)$ compactly supported in region $B$. 

\begin{figure}
 
\includegraphics[width=0.5\textwidth]{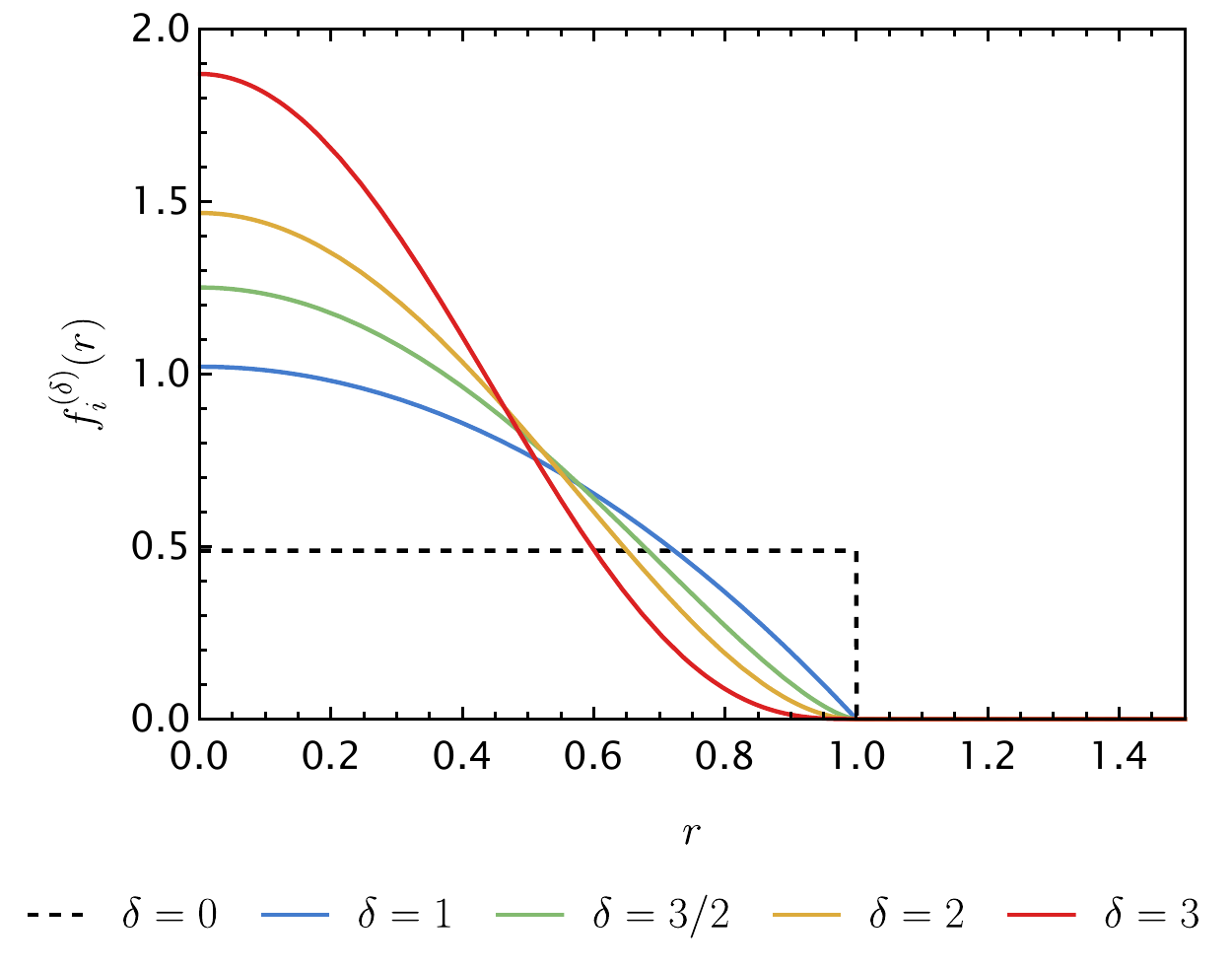}
    \caption{\label{fig:W_delta_vs_r}
    Shape of the smearing functions $f^{(\delta)}_i(\vec x)$ for a few values of $\delta$. Since these functions are spherically symmetric around their center $\vec x_i$, we plot them versus the dimensionless radial coordinates $r:=|\vec x-\vec x_i|/R$. Note that the larger $\delta$ is, the less support $f^{(\delta)}_i$ has near the boundary.}
\end{figure}

For the smearing functions $f_i(\vec x)$, $i=A,B$, in this section we will use the following one-parameter family of {\em non-negative} functions:
\begin{equation}\label{eq:family_test_funcs1}
    f^{(\delta)}_i(\vec x)= A_{\delta}\  \left(1 -\frac{|\vec{x} - \vec{x}_i|^2}{R^2}\right)^{\delta}\,  \Theta\left( 1 - \frac{|\vec{x} - \vec{x}_i|}{R}\right)\,,
\end{equation} where $\vec x_i$ is the center of the ball $i$, and $\Theta(x)$ is the Heaviside step function---which ensures that $f^{(\delta)}_i(\vec x)$ is compactly supported within a ball of radius $R$ centered at $\vec x_i$; $A_{\delta}$ is a normalization constant determined below and $\delta$ a positive real number. Figure~\ref{fig:W_delta_vs_r} shows the shape of $f^{(\delta)}_i(\vec x)$ for some values of $\delta$.

The parameter $\delta$ determines the differentiability class of $f^{(\delta)}_i(\vec x)$. For example, for $\delta=0$, $f^{(\delta)}_i(\vec x)$ reduces to the Heaviside function, which is discontinuous.  For $\delta=1$,  the function is continuous, but its first derivative is not. The differentiability class of $f^{(\delta)}_i(\vec x)$ is  $C^{\delta-1}$ for integer $\delta$. In order for the smeared operators $\hat{\Phi}_i$ and $\hat{\Pi}_i$ to be well defined, it suffices to choose $\delta\geq 1$, as we will see below by explicitly computing their quantum moments in the vacuum. Furthermore, although these and other smearing functions we use in this article are not infinitely differentiable, we argue in Appendix \ref{Ap:B} that this is actually not a restriction. This is because there always exists smooth functions of compact support defining modes whose physical properties are the same as for the modes we actually use, up to arbitrarily high accuracy.

In the following, we will explore modes defined from $f^{(\delta)}_i(\vec x)$ for different finite $\delta\geq 1$, even considering noninteger values. 

An advantage of this family of smearing functions is that their Fourier transform has a simple expression in terms of Bessel functions  (see Appendix~\ref{app:details}). This is true in any spatial dimension $D$ and makes it possible to obtain analytical expressions for the quantities of interest for all $D$. Later in this article, we consider other families of smearing functions, including infinitely differentiable ones. The results are qualitatively similar,  although in those cases we perform calculations numerically.

The commutator between the four operators $\Phi_i$, $\Pi_i$, with $i=A,B$ are
   \bea
       &[\hat{\Phi}_A,\hat{\Phi}_B]&=[\hat{\Pi}_A,\hat{\Pi}_B]= 0 \\ 
 &[\hat{\Phi}_A,\hat{\Pi}_B] &=[\hat{\Phi}_B,\hat{\Pi}_A]=i\, c \int d^Dx\, f^{(\delta)}_A\, f^{(\delta)}_B=0\,. \nonumber
   \eea

The last integral vanishes because $f^{(\delta)}_A(\vec x)$ and $f^{(\delta)}_B(\vec x)$ are supported in disjoint regions. On the other hand,  
\be [\hat{\Phi}_i,\hat{\Pi}_i]=i\, c \int d^Dx\, (f^{(\delta)}_i)^2 \neq 0\, , \ \ i=A,B.\ee
We fix the (dimensionful) constant $A_{\delta}$ in the definition of $f^{(\delta)}_i$ by demanding that $[\hat{\Phi}_i,\hat{\Pi}_i]=i$.  This implies that
\begin{equation}
    A_{\delta} = c^{-1/2}\, R^{-D/2}\pi^{-D/4} \sqrt{\frac{\Gamma(1 + D/2 +2 \delta)}{\Gamma(1+2\delta)}}\,. 
\end{equation}
The covariance matrix of the reduced state for the two modes of interest can be readily obtained by computing vacuum expectation values of symmetrized products of $\hat{\Phi}_i$ and $\hat{\Pi}_i$. It is easy to check that 
$\braket{\{\hat{\Phi}_i,\hat{\Pi}_j\}} =0$ in the Minkowski vacuum and that $\braket{\hat{\Phi}_A} =\braket{\hat{\Phi}_B}=: \braket{\hat{\Phi}}$ and $\braket{\hat{\Pi}_A} =\braket{\hat{\Pi}_B}=: \braket{\hat{\Pi}}$ since we are using the same smearing function in both regions. With this, the covariance matrix of the total system takes the form
 \begin{equation}\label{sAB}
     \sigma_{AB} = \left(\begin{matrix}
  \sigma_{A}^{\mathrm{red}} & C \\ 
  C^T &   \sigma_{B}^{\mathrm{red}}
     \end{matrix} \right) \,,
 \end{equation} 
where
\begin{equation}\label{sA}     \sigma_{A}^{\mathrm{red}}=\sigma_{B}^{\mathrm{red}} = 2\left(\begin{matrix}
         \braket{\hat{\Phi}^2} & 0\\ 
        0 &     \braket{\hat{\Pi}^2}
    \end{matrix}
    \right) \,
\end{equation}
is the covariance matrix of each mode, and  $C = \mathrm{diag}(\braket{\{\hat{\Phi}_A, \hat{\Phi}_B\}},\braket{\{\hat{\Pi}_A, \hat{\Pi}_B\}})$ describes their correlations.  

When the field is massless and for $D>1$, we obtain (see Appendix \ref{app:details} for details of the calculation)
\begin{equation}\label{phicorr}
     \braket{\{\hat{\Phi}_i, \hat{\Phi}_j\}} = 2\, N_{\delta}^2\,\frac{R}{c}\, \left\{ \begin{matrix}
     \mathcal{J}^{D}(-1,\delta) & i=j\\
     \mathcal{L}^{D}(-1,\delta, \rho)& i\neq j
     \end{matrix}\right.\,,
\end{equation}
\begin{equation}\label{picorr}
     \braket{\{\hat{\Pi}_i, \hat{\Pi}_j\}} = 2N_{\delta}^2\,\frac{c}{R}\, \left\{ \begin{matrix}
     \mathcal{J}^{D}(1,\delta) & i=j\\
     \mathcal{L}^{D}(1,\delta, \rho)& i\neq j
     \end{matrix}\right.\,,
\end{equation} 
where
\begin{equation}\begin{split}
      \mathcal{J}^{D}(\lambda, \delta)&=2^{-1-2\delta +\lambda}\frac{\Gamma\left(\frac{D+\lambda}{2}\right)\Gamma\left(1+2\delta - \lambda\right)}{\Gamma\left(1+\delta-\frac{\lambda}{2} \right)^2 \Gamma\left(\frac{D-\lambda}{2}+2\delta+1 \right)}\,,
\end{split}
\end{equation}
\begin{equation}\begin{split}
    \mathcal{L}^{D}(\lambda, \delta, \rho) =&\rho^{-(D+\lambda)} \frac{ \Gamma\left(\frac{D+\lambda}{2}\right)\Gamma\left(D/2\right)}{2^{1+2\delta-\lambda}\Gamma\left(\frac{D}{2}+1 + \delta\right)^2\Gamma\left(-\frac{\lambda}{2}\right)}\times \\
    &{}_3F_2\left[\begin{matrix}
    1+\frac{\lambda}{2},\frac{D+\lambda}{2},\frac{D+1}{2}+\delta\\
    \frac{D}{2} + 1+\delta, D+1+2\delta
    \end{matrix}\,; \frac{4}{\rho^2}\right]\,,
\end{split}
\end{equation} and 
\begin{equation}
    N_{\delta}^2 = \frac{2^{2\delta} \Gamma\left(1 + \frac{D}{2} + 2\delta\right)\Gamma\left(1 + \delta\right)^2}{\Gamma\left(1+2\delta\right)\Gamma\left(D/2\right)}\,.
\end{equation}

A few comments are in order. First of all, from these expressions one can check that, as expected, the field and momentum self-correlations are positive and bounded functions of $\delta$, for $\delta \geq 1$ (for $\delta=0$,  the momentum self-correlations diverge). 
 On the other hand, these expressions show that the correlations between both modes behave,  for large separations $\rho\gg 1$, as
 \begin{equation}
    \braket{\{\hat{\Phi}_A, \hat{\Phi}_B\}} =\rho ^{-(D-1)}\frac{R}{c}\,\Big( u(\delta, D) + \mathcal{O}(\rho^{-2})\Big)\,,
\end{equation}  and 
\begin{equation}
    \braket{\{\hat{\Pi}_A, \hat{\Pi}_B\}}= -\rho^{-(D+1)}\frac{c}{R}\,\Big(v(\delta, D)+\mathcal{O}(\rho^{-2}) \Big)\,
\end{equation}

for $D>1$,  where $$u(\delta, D) = \frac{2^{ -2\delta -1} \Gamma \left(\frac{D-1}{2}\right) \Gamma (\delta +1) \Gamma \left(\frac{1}{2} (D+4 \delta +2)\right)  }{\Gamma \left(\delta +\frac{1}{2}\right) \Gamma \left(\frac{D}{2}+\delta +1\right)^2}$$ and $$v(\delta, D) = \frac{2^{-2 \delta} \delta  \Gamma \left(\frac{D+1}{2}\right) \Gamma (\delta ) \Gamma \left(\frac{1}{2} (D+4 \delta +2)\right)}{\Gamma \left(\delta +\frac{1}{2}\right) \Gamma \left(\frac{D}{2}+\delta +1\right)^2}$$ are positive functions\footnote{The correlator $\braket{\{\hat{\Pi}_A, \hat{\Pi}_B\}}$ is, therefore, negative. This, in turn, implies that the submatrix $C$ of $\sigma_{AB}$ is negative, even when we are using non-negative smearing functions. Consequently, the covariance matrix $\sigma_{AB}$ in \eqref{sAB} describes a Gaussian state which is not manifestly separable, according to Simon's separability  criterion \cite{SimonSeparability2000}. Additional calculations are needed to show that this state is indeed separable.} that depend on $\delta$ and the spacetime dimension.

The dependence $\braket{\{\hat{\Phi}_A, \hat{\Phi}_B\}} \sim\rho ^{-(D-1)}$ and $\braket{\{\hat{\Pi}_A, \hat{\Pi}_B\}}\sim -\rho^{-(D+1)}$ is precisely what is expected, providing a good check to our expressions. 
 
\begin{figure*}[t]
     \hspace*{-.75cm}
     \includegraphics[width=1.05\textwidth]{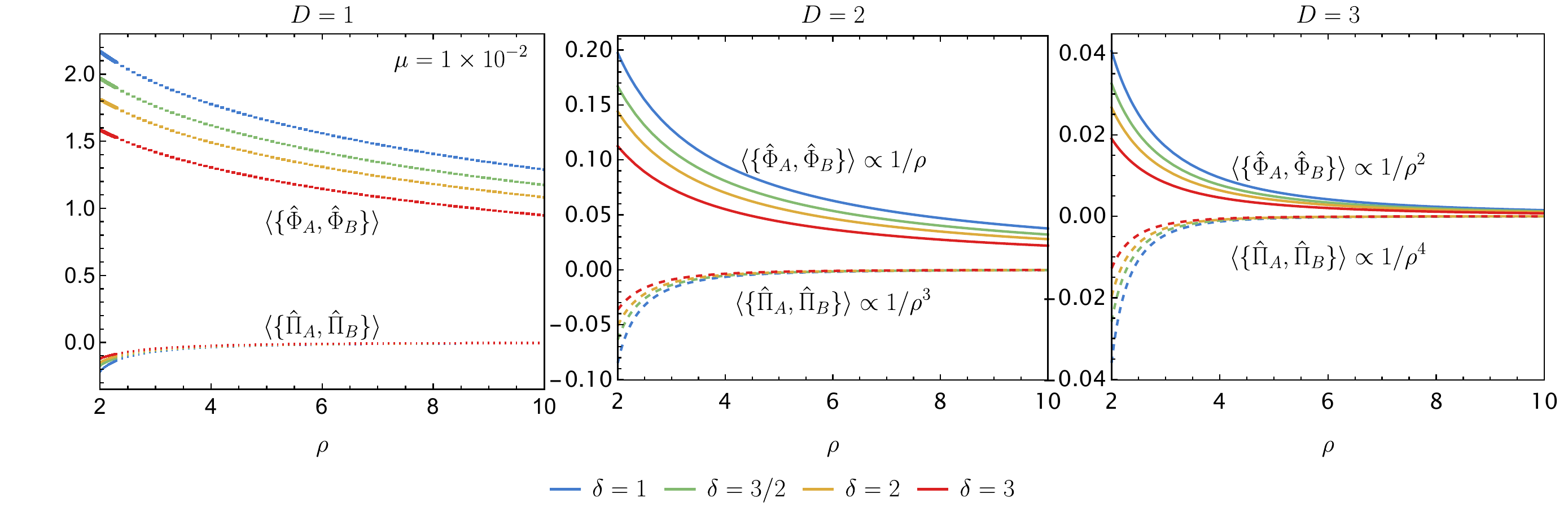}
     \caption{Correlations between two field modes versus the dimensionless distance between the centers of the spherical regions where each mode is supported. $\rho=2$ corresponds to the two regions touching each other. The plots for $D=2$ and $D=3$ describe the correlations of a massless field and are obtained analytically, while for $D=1$ we introduce a small mass $\mu =mR= 10^{-2}$  to avoid infrared divergences, and compute the correlations numerically.  All plots show that  correlations depend on the details of the smearing functions, particularly on the value of the parameter $\delta$. On the other hand, at large separations the fall-off behavior of all correlations is as expected.}
     \label{fig:Correlations_D1D2D3}
 \end{figure*}

 Figure~\ref{fig:Correlations_D1D2D3} shows these correlations for $D=2$ and $D=3$ dimensions of space and for a massless field. One can see that, although at small separations $\rho\approx 2$,  the cross-correlations depend on the details of the smearing functions, in particular, on the value of $\delta$; at large separations they behave as expected for all $\delta\geq 1$. 
 
 We have also included in Fig.~\ref{fig:Correlations_D1D2D3} the $D=1$ case. As mentioned before, in this case one needs to introduce a mass $m$ to the field to avoid infrared divergences. When $m\neq0$, we do not find closed analytical expressions for the  correlation functions, and the results presented in Fig.~\ref{fig:Correlations_D1D2D3} have been obtained numerically.

\subsection{Mutual information and entropy}

The results in the previous subsection confirm that, within the family of modes we have considered, any pair of them with one mode supported in region $A$ and the other in region $B$ are correlated. 

As discussed in Sec.~\ref{sec:2}, the total amount of correlations between both modes can be quantified using  their mutual information, defined in expression \eqref{MI}. To compute this quantity, we first need the von Neumann entropies of each system separately, $S_A$ and $S_B$, and the entropy of the joined system, $S_{AB}$. 

\begin{figure}
    \centering
    \includegraphics[width=0.5\textwidth]{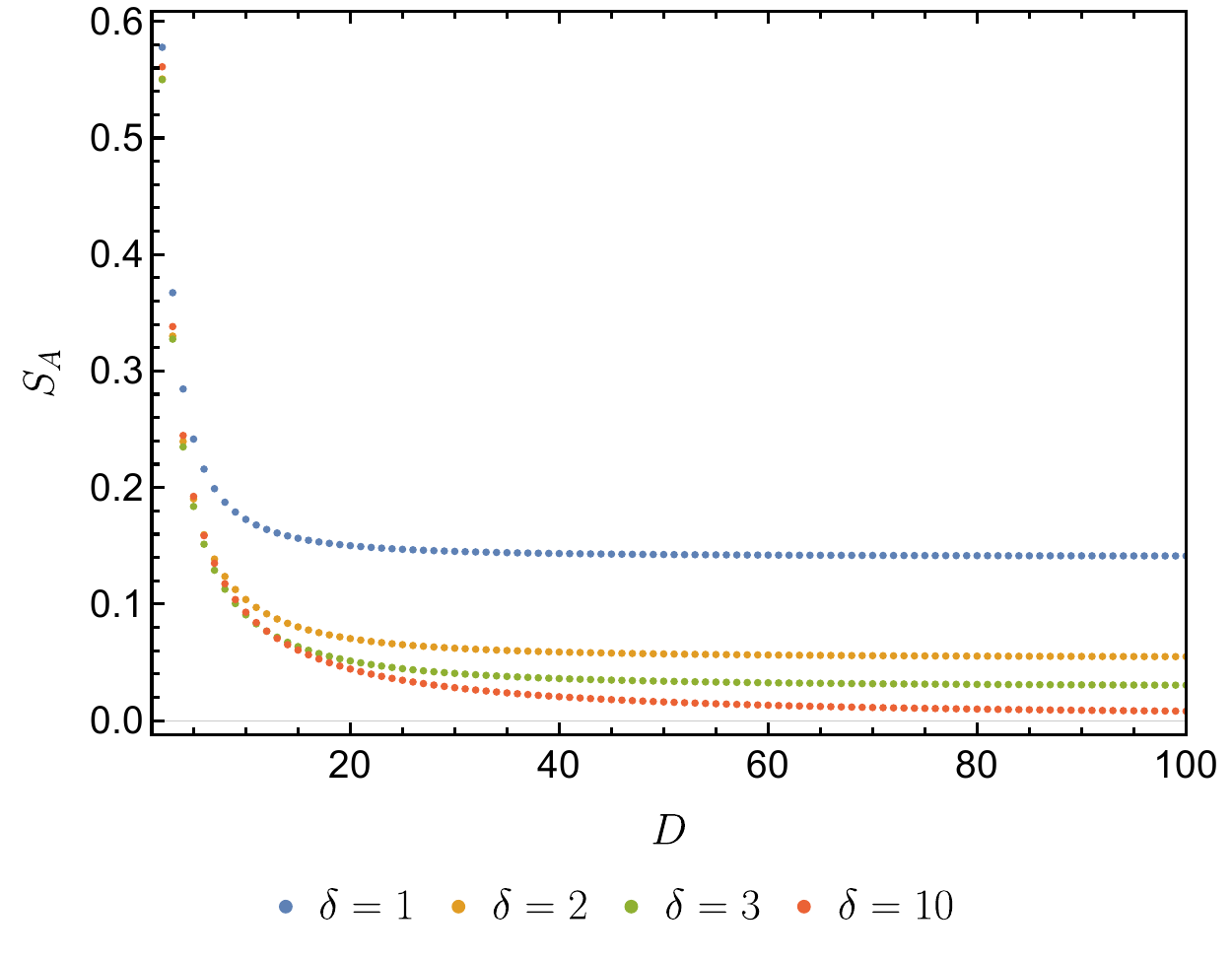}
    \caption{The von Neumann entropy $S_A$ of a single mode as a function of the dimension of space $D$ and different values of $\delta$.}
    \label{fig:vNEntropy_vs_D}
\end{figure}

Since the reduced covariance matrices for each subsystem are identical when the field is in the Minkowski vacuum, $\sigma^{\rm red}_A=\sigma^{\rm red}_B$, so are their entropies. Expression \eqref{S} shows that all we need to compute this entropy is the symplectic eigenvalue of $\sigma^{\rm red}_A$. Using the form of $\sigma^{\rm red}_A$ given in \eqref{sA}, this symplectic eigenvalue is, for $D>1$, 
\bea
   & &  \nu_{A} = 2 \sqrt{\braket{\hat{\Phi}^2}\braket{\hat{\Pi}^2}}=\\ \nonumber 
   & &  \frac{\Gamma (\delta +1)^2 \Gamma \left(\frac{D}{2}+2 \delta +1\right) \sqrt{\frac{\Gamma \left(\frac{D-1}{2}\right) \Gamma \left(\frac{D+1}{2}\right) \Gamma (2 \delta ) \Gamma (2 \delta +2)}{\Gamma \left(\frac{1}{2} (D+4 \delta +1)\right) \Gamma \left(\frac{1}{2} (D+4 \delta +3)\right)}}}{\Gamma \left(\frac{D}{2}\right) \Gamma \left(\delta +\frac{1}{2}\right) \Gamma \left(\delta +\frac{3}{2}\right) \Gamma (2 \delta +1)} \,.
\eea
This quantity is larger than 1, confirming that the reduced state corresponding to a single mode is a mixed quantum state \cite{bianchi_entropy_2019,Hollands:2017dov}. 
From this, we obtain an analytical expression for $S_A$, which we plot in Fig.~\ref{fig:vNEntropy_vs_D}. 

We see that the entropy depends on $\delta$; hence it depends on the details of the smearing function. This is expected since the smearing function actually defines the concrete mode, whose entropy we are evaluating. We observe that the larger $\delta$ is, the smaller the entropy is. Larger $\delta$ corresponds to smearing functions with more weight around the center of the region and less support close to the boundary. In other words, we find that modes supported closer to the boundary have larger entropy. 

We also observe, interestingly, that $S_A$  decreases monotonically with $D$. In the limit $D\to \infty$, for a fixed $\delta$, we have 
\begin{equation}
 \lim_{D\to \infty} \nu^2_{\mathrm{1dof}}=  \frac{\Gamma (2 \delta ) \Gamma (\delta +1)^4 \Gamma (2 \delta +2)}{\Gamma \left(\delta +\frac{1}{2}\right)^2 \Gamma \left(\delta +\frac{3}{2}\right)^2 \Gamma (2 \delta +1)^2} \,.
\end{equation}
 from which we obtain a finite value of $S_A$.\footnote{It is intriguing to note that the double limit $\delta \to \infty$ and $D\to \infty$ produces $S_A\to 0$ (pure state). However, since the $\delta \to \infty$ limit of our smearing functions \eqref{fig:W_delta_vs_r} produces a Dirac-delta distribution, we do not find a clear physical interpretation for this mathematical result.}

It is tempting to interpret $S_A$ as a quantifier of the entanglement between a single mode in region A and the rest of the degrees of freedom of the field theory (infinitely many, some supported within A and some outside).

However, as emphasized in \cite{Hollands:2017dov}, such an interpretation is an unjustified extrapolation of results in standard quantum mechanics because there the Hilbert space of the total system is always a product of the Hilbert spaces of the two subsystems; this is not true in quantum field theory, if the subsystems $A$ and $B$ are complementary. 

The second ingredient entering the expression for the mutual information is the entropy of the joined system of the two modes, $S_{AB}$. This quantity can also be obtained analytically for $D>1$, by plugging in Eq.~\eqref{S} the form of the two symplectic eigenvalues of $\sigma_{AB}$:
\begin{equation} \label{symeigs}
    \nu_{\pm} = \sqrt{(2\braket{\hat{\Phi}^2} \pm \braket{\{\hat{\Phi}_A, \hat{\Phi}_B\}} )(2\braket{\hat{\Pi}^2} \pm \braket{\{\hat{\Pi}_A, \hat{\Pi}_B\}} )} \, .
\end{equation}
Notice that although both field and momentum correlations depend on the radius $R$ of the regions where the modes are supported, this dependence cancels out in the combination of correlation functions appearing in \eqref{symeigs}. Consequently, $\nu_{\pm}$ and quantities derived from it, like mutual information, entropies, and entanglement, remain invariant under rescalings.

\begin{figure*}[t]
\hspace{-1cm}
    \includegraphics[width=1.05\textwidth]{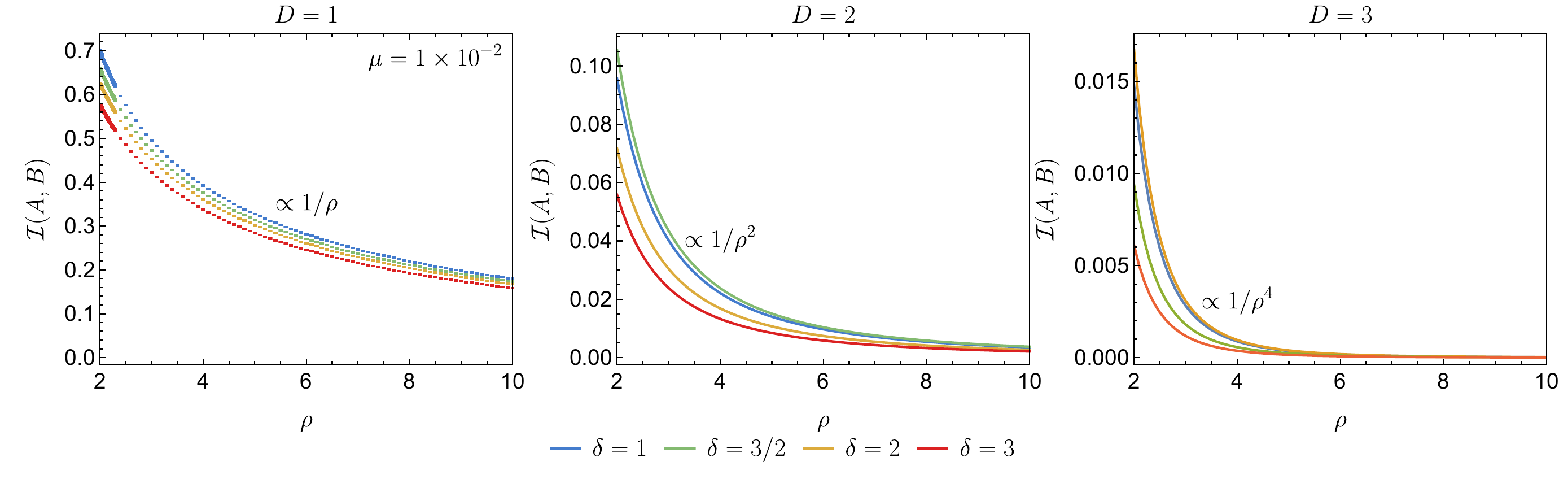}
    \caption{Mutual information for two field modes versus the dimensionless distance between the centers of the spherical regions where each mode is supported. $\rho=2$ corresponds to the two regions touching each other. The plots for $D=2$ and $D=3$ correspond to a massless field and are obtained analytically, while for $D=1$ we introduce a small mass $\mu\equiv m R= 10^{-2}$ to avoid infrared divergences, and compute the mutual information numerically. All plots show that the short-distance behavior of the mutual information depends on the details of the smearing functions through the value of the parameter $\delta$. However, at large distances the dependence on $\delta$ decreases and the fall-off behavior is as expected. }
    \label{fig:MI_vsrho_D1D2D3_v2}
\end{figure*}

We plot the mutual information $\mathcal{I}_{AB}$  versus the distance between the two regions in Fig.~\ref{fig:MI_vsrho_D1D2D3_v2}, for $D=1$ (left panel), $D=2$ (middle figure), and when $D=3$ (right panel), for different values of $\delta$. (As in the previous subsection, the $D=1$ case is computed numerically.) 

 We observe that  the mutual information is finite, and its short-distance behavior ($\rho \gtrsim 2$) depends on the details of the smearing functions. However, its long-distance behavior ($\rho \gg 2$) is given by $\mathcal{I}(A,B) \sim \rho^{-2(D-1)}$ for $D>1$ (we have explicitly checked this up to $D=10$ and expect this fall-off behavior to be true for all $D$). This is the expected result, and it is compatible with results obtained previously in \cite{Martin:2015qta,Martin:2021qkg,Espinosa-Portales:2022yok} for $D=3$. 
 
 An important lesson we extract from this analysis is that, for a fixed distance between the regions supporting two modes, the total correlations (classical and quantum) between them are weaker the larger the dimension $D$ of space is. 

\subsection{Entanglement}

We use the LN to evaluate whether the correlations between two single modes discussed in the previous subsection contain any entanglement. The LN, defined in \eqref{LN}, is obtained from the symplectic eigenvalues of $\tilde \sigma_{AB}$, the partial-transposition of the covariance matrix $\sigma_{AB}$, which are given by
\begin{equation}\label{tnu}
    \tilde{\nu}_{\pm} = \sqrt{(2\braket{\hat{\Phi}^2} \pm \braket{\{\hat{\Phi}_A, \hat{\Phi}_B\}} )(2\braket{\hat{\Pi}^2} \mp \braket{\{\hat{\Pi}_A, \hat{\Pi}_B\}} )}\,. 
\end{equation}
Recall that LN is different from zero only if at least one of these symplectic eigenvalues is smaller than 1.  By comparing this expression with \eqref{symeigs} and by taking into account that $\braket{\{\hat{\Pi}_A, \hat{\Pi}_B\}}$ is negative while $\braket{\{\hat{\Phi}_A, \hat{\Phi}_B\}}$ is positive for our smearing functions, one can see that $\tilde \nu_+ \geq \nu_+$, while 
$\tilde \nu_- \leq \nu_+$ (recall, $\nu_{\pm}$ are the symplectic eigenvalues of $\sigma_{AB}$). Since both $\nu_{\pm}$ are larger than 1, only $\tilde \nu_-$ can possibly contribute to the LN. 

Furthermore, we also see from the expression above that  $\tilde{\nu}_-$ would be smaller than 1 only if the  momentum cross-correlation $\braket{\{\hat{\Pi}_A, \hat{\Pi}_B\}}$ is ``negative enough'' and  that the LN (if different from zero) must fall off with the distance between the regions at a rate dictated by the dimension $D$ of space.  For this reason, we organize the discussion in the rest of this section in terms of the number of spacetime dimensions.

\subsubsection{$D=1$}

\begin{figure}
    \centering
    \includegraphics[width=0.5\textwidth]{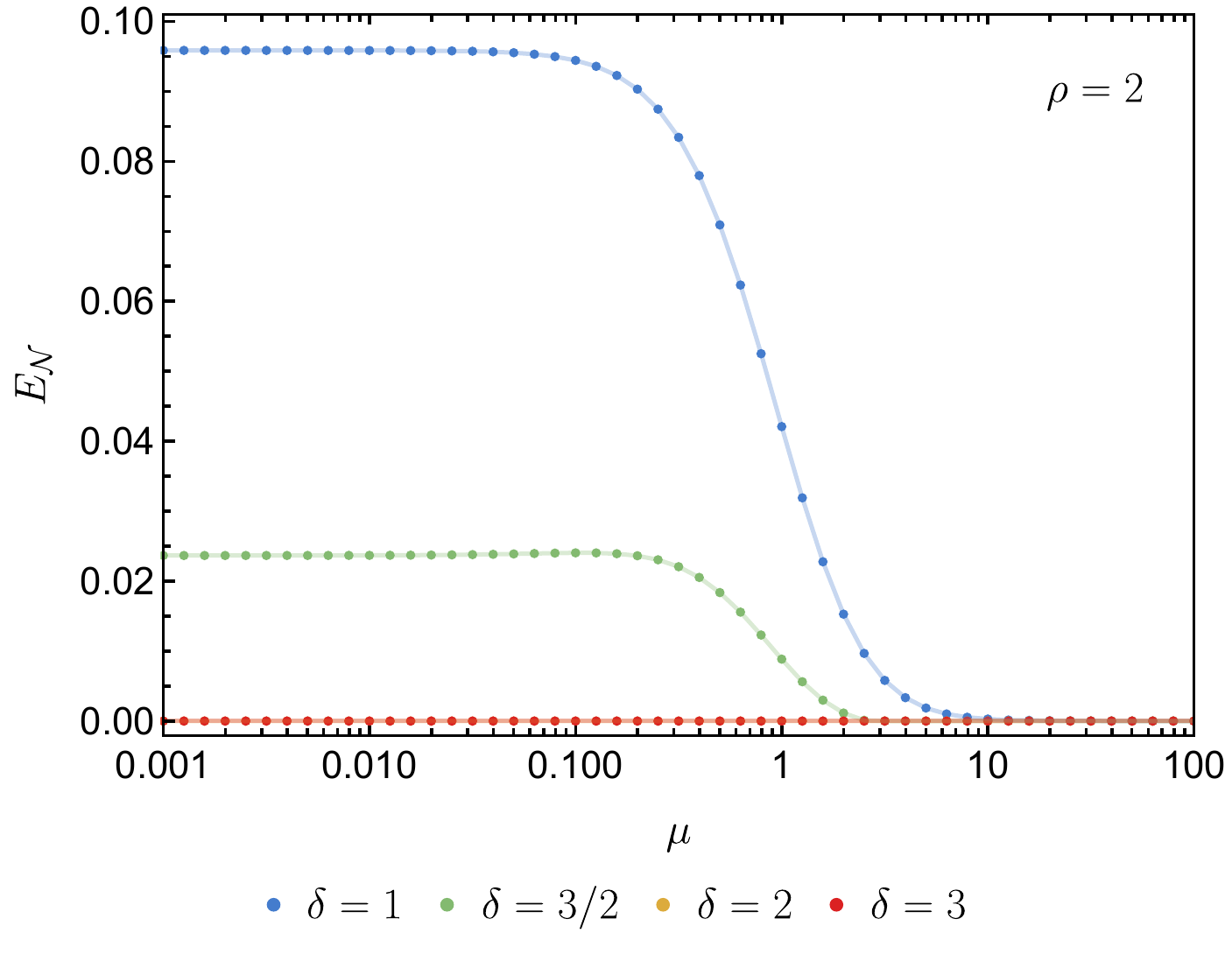}
    \caption{LN for $D=1$ as a function of the dimensionless mass, $\mu=m R$ ($R$ is the radius of the regions of support of A and B), when the regions $A$ and $B$ are kept at a fixed dimensionless distance  $\rho=2$. The LN reaches a maximum when $\mu\ll 1$,  decreases monotonically, and vanishes when the mass of the field reaches a threshold that depends on the details of the smearing function. The LN is zero for all $\mu$ if $\delta\gtrsim 1.7$.}
    \label{fig:LND1mu}
\end{figure}

\begin{figure}
    \centering
    \includegraphics[width=0.5\textwidth]{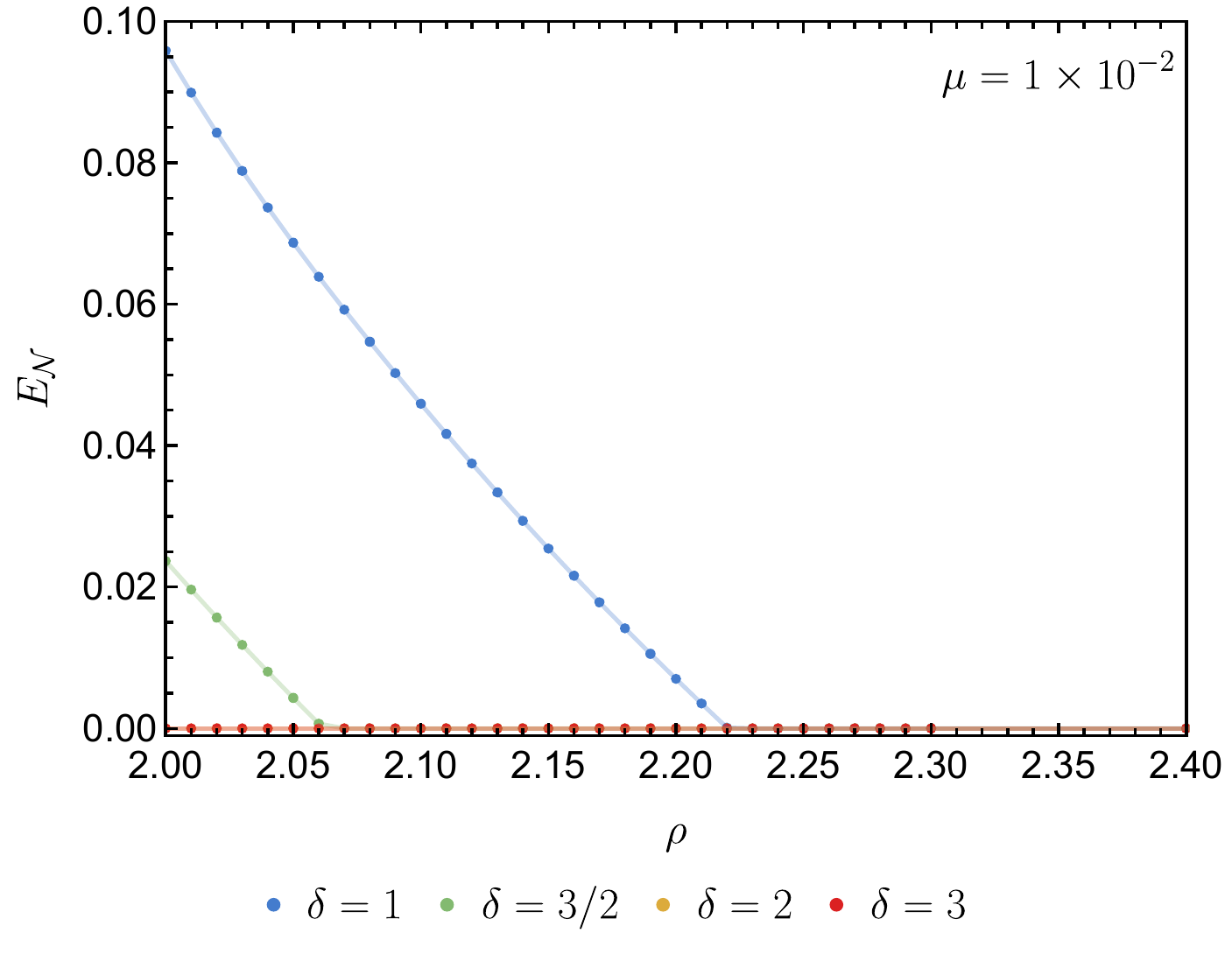}
    \caption{LN for $D=1$ between the two field modes as a function of the  dimensionless distance between the centers of the spherical regions where each mode is supported, for a fixed dimensionless mass of the scalar field ($\mu=10^{-2}$) and for different smearing functions. }
    \label{LND1rho}
\end{figure}

As discussed above, in $D=1$ we compute the correlation functions numerically, and from this we evaluate $\tilde \nu_-$, from which we compute the LN using Eq.~\eqref{LN}. We are interested in understanding: (i) whether the LN is different from zero; and if the answer is in the affirmative, (ii) how the LN depends on the mass $m$ of the field, on the distance between the regions $A$ and $B$, and on the details of the smearing function. The answers to these questions are contained in Figs.~\ref{fig:LND1mu} and \ref{LND1rho}. 

On the one hand, Fig.~\ref{fig:LND1mu} shows the LN versus the dimensionless mass $\mu$,
for a fixed distance $\rho$ between the two regions and for different values of $\delta$. Since the LN decreases with $\rho$, we choose in this plot the minimum possible value of $\rho$, namely $\rho=2$. The main messages we extract from this figure are the following: (1) For $\delta=1$, we find that the LN is different from zero and, consequently, the two modes {\em are entangled}. (2) For $\delta\geq 1.7$ the LN vanishes for any value of the mass $\mu$. This shows that the LN is very sensitive to the shape of the smearing function. The smearing functions $f^{(\delta)}$ that we use in this section have more support close to the boundary of the region for smaller $\delta$. Since correlations fall off with distance, it is therefore expected that modes with larger support close to the boundary ($\delta$ close to $1$) are more correlated (and entangled) than modes defined with $\delta\geq1$. Figure.~\ref{fig:LND1mu} confirms that this intuition is correct and, furthermore, shows that pairs of modes supported in disjoint regions {\em are not entangled at all} if we use $\delta\gtrapprox 1.7$. (3) The LN between these two modes is very sensitive to the mass of the field. For $\delta=1$, the LN reaches a maximum when $\mu\ll1$, decreases very fast when $\mu \approx 1$, and completely vanishes when $\mu\gtrsim 10$. 

On the other hand, Fig.~\ref{LND1rho} shows the way the LN changes with the distance $\rho$ between the regions A and B. This plot is computed for a small value of the dimensionless mass, $\mu=10^{-2}$, for which we know from the previous plot that LN is close to its maximum. Again, we find that the LN is different from zero only for $\delta$ close to 1. We observe that  the LN falls off rapidly with the distance $\rho$ and completely vanishes beyond $\rho\approx 2.2$ (i.e., when the separation between the boundaries of the two regions is about 20\% of their radius). Note that this falloff is much faster than the one we obtained for the mutual information $\mathcal{I}_{AB}\sim \rho^{-1}$.

\subsubsection{$D> 1$}
For the number of spatial dimensions larger than 1, we can compute the LN analytically for massless fields (this is the most interesting case since entanglement is expected to be larger when $\mu\to 0$, as we saw above). Substituting the value of the correlation functions reported in \eqref{phicorr} and \eqref{picorr} in expression \eqref{tnu}, we see that $\tilde \nu_-$ is larger than 1 for all distances $\rho$ and all values of $\delta$, including $\delta=1$. Consequently, we obtain 
\begin{equation}
    E_{\mathcal{N}} = 0\, \quad \forall \,\delta\geq 1 \text{ and }\,  \forall D>1\,. 
\end{equation}
The fact that we find less entanglement in $D\geq 2$ than in $D=1$ is compatible with the fact that correlations are stronger in lower dimensions, as shown in the last subsection. 

In summary, we conclude that the correlations captured by the mutual information computed in the previous section are  mainly classical correlations and do not originate from entanglement. For the family of modes we have explored in this section, only in the special case $D=1$, small mass $\mu$, $\delta$ close to 1, and small separation between the two regions, we find that the two modes are entangled. {\em In all other cases, the reduced state is separable.} 

In Sec.~\ref{sec:5}, we extend the family of smearing functions and show that these conclusions are not peculiar to the special smearings used in this section. In Sec.~\ref{sec:6}, we will show how we can find entanglement between two modes by carefully selecting their region of support.

\begin{figure*}
    \centering
   \begin{tikzpicture}

 \node at (0,6) {\includegraphics[width=\textwidth]{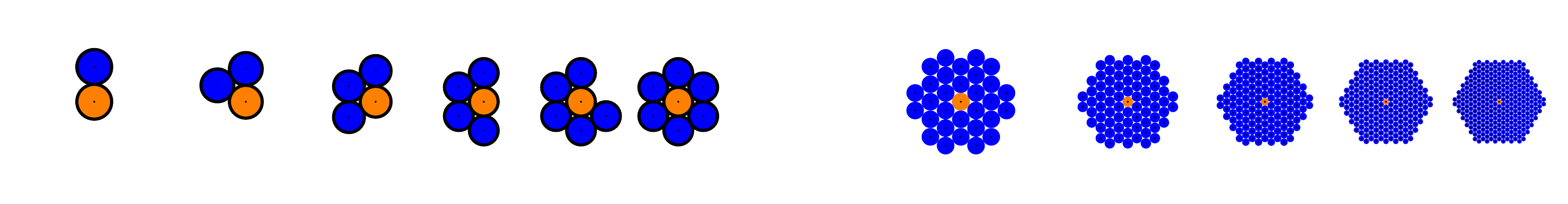}};
 \draw[ultra thin,dash pattern=on 2pt  off 2.5pt] (-7.25,4.8)--(-7.85,5.5); 
 \draw[ultra thin,dash pattern=on 2pt  off 2.5pt] (-5.46,4.8)--(-6.15,5.5); 
 \draw[ultra thin,dash pattern=on 2pt  off 2.5pt] (-4.3,4.8)--(-4.65,5.5);
 \draw[ultra thin,dash pattern=on 2pt  off 2.5pt] (-3.51,4.8)--(-3.45,5.5);
 \draw[ultra thin,dash pattern=on 2pt  off 2.5pt] (-2.9, 4.8) -- (-2.3,5.5);
 \draw[ultra thin,dash pattern=on 2pt  off 2.5pt] (-2.4,4.8) --(-1.24,5.5);
    \node at (0.1,0.2) {\includegraphics[width=.99\textwidth]{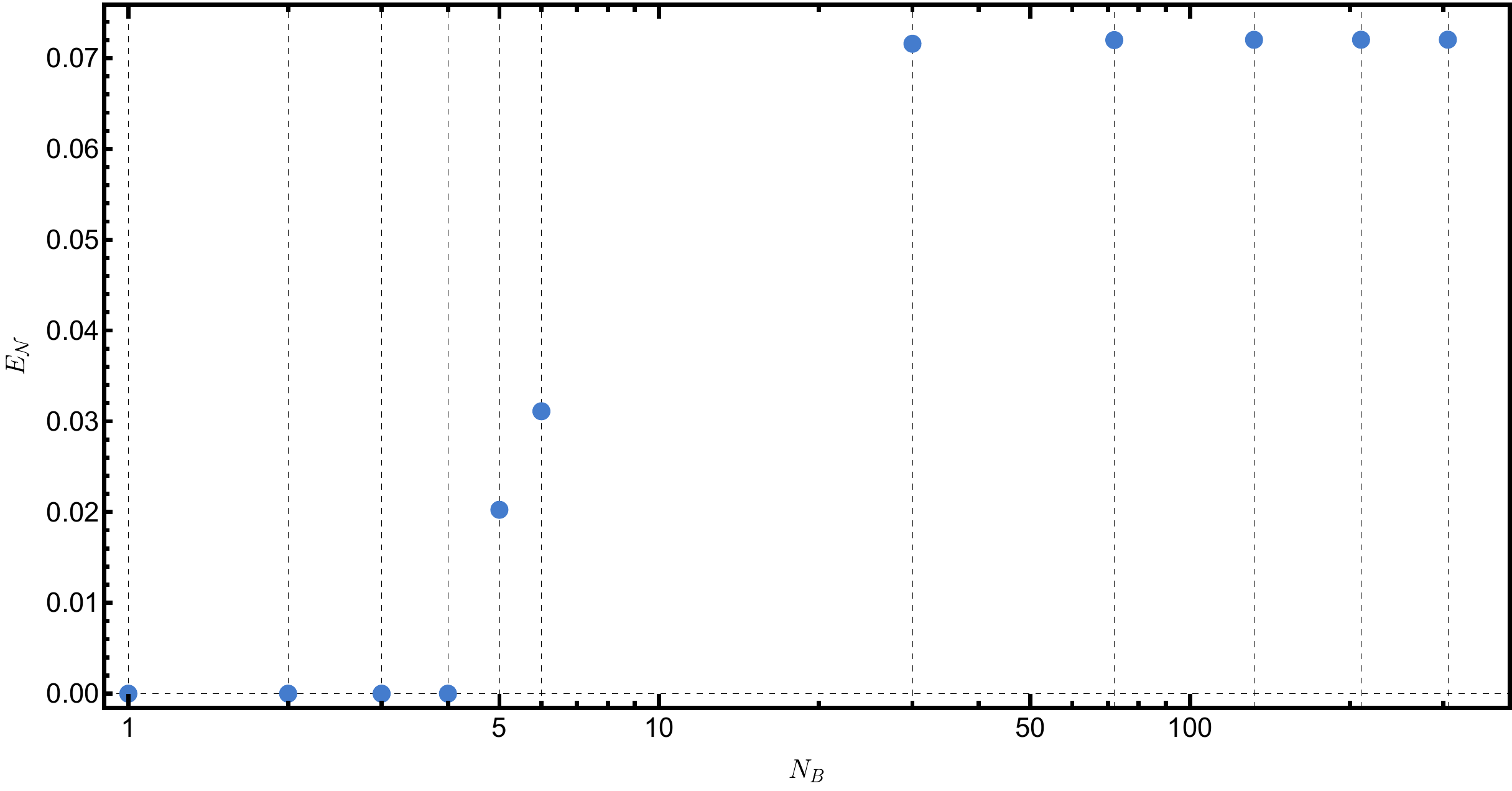}};
   \end{tikzpicture}
     \caption{LN between subsystems $A$ and $B$ for different values of the number of modes $N_B$ in subsystem $B$ and $N_A=1$, in two spatial dimensions, $D=2$. The $N_B$ modes are distributed in space as illustrated in the top part of the figure, where the orange central disk represents the region of support of the single mode in $A$, and the blue disks are the regions where each of the $N_B$ modes are supported. We use smearing functions $f^{(\delta)}$ with $\delta=1$ in this plot. 
     The plot shows that $A$ and $B$ are entangled as long as $N_B\geq 5$. }
    \label{fig:LN_vs_Ndof_subsystem2_hexagon_2dims}
\end{figure*}
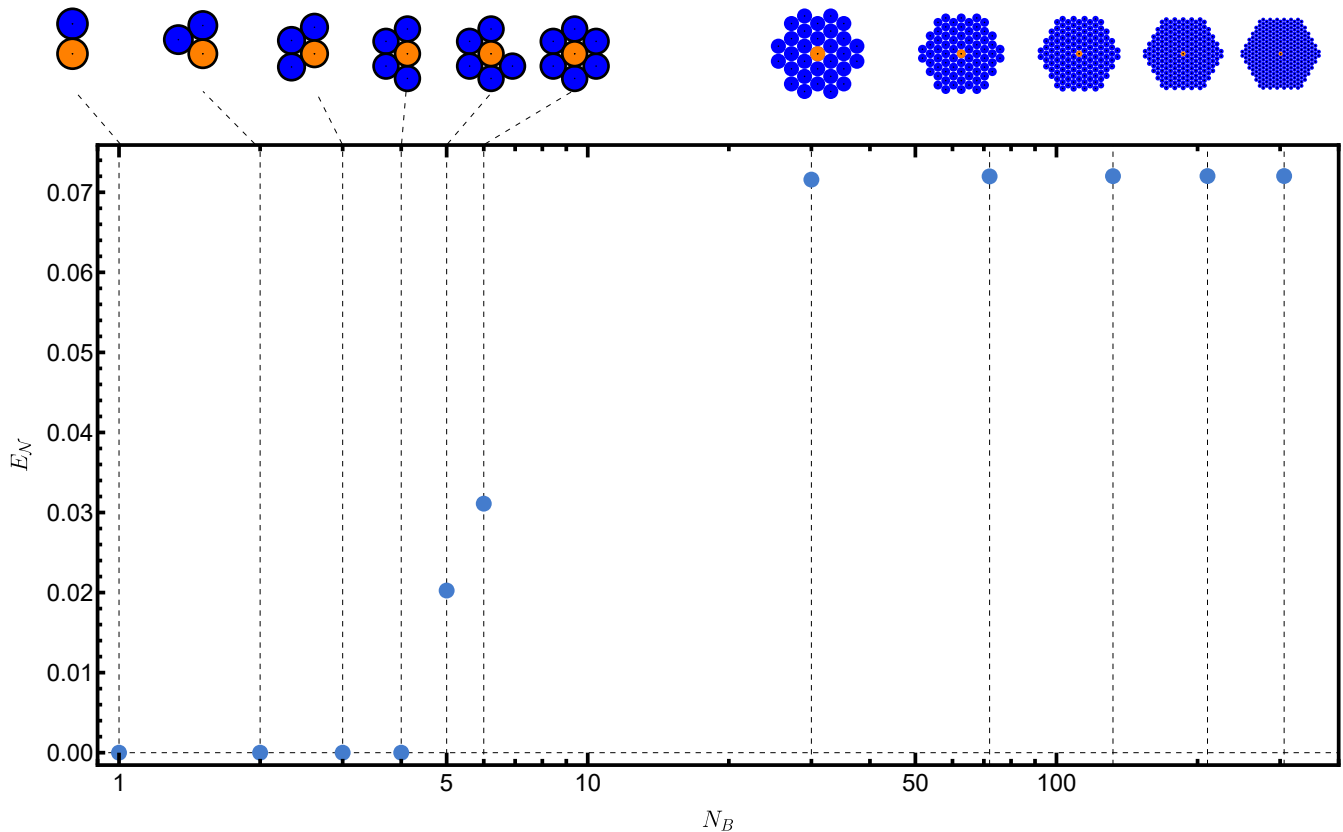

\section{Enlarged subsystems}\label{sec:4}

In this section, we extend the previous calculations by enlarging the number of modes in our two subsystems, with the  goal of studying whether such an enlargement leads to the emergence of entanglement. We will show that it does in $D=2$; in contrast, for larger dimensions we have not been able to find entanglement in any of the multimode configurations described below.  We work in this section in $D\geq 2$ spatial dimensions and massless scalar fields and use the same smearing functions as in the previous section, for which all results can be obtained analytically. We will use smearing functions with $\delta=1$ since this is the case for which entanglement is more likely to appear. The extension to other values of $\delta$ is straightforward. 

Our setup in this section is made of $N_A+N_B$ modes, defined  in disjoint regions in a similar manner as done in the last section. Having more than two modes raises the question of how to distribute them in space. We will restrict attention to configurations that are more likely to show  bipartite entanglement. These are configurations for which the two subsystems are as close as possible. Since we are considering spherical regions, this problem is tantamount to packing spheres as densely as possible in a $D$-dimensional space.

\subsection{$D=2$}
\savebox{\mybox}{\includegraphics[width=0.5\textwidth]{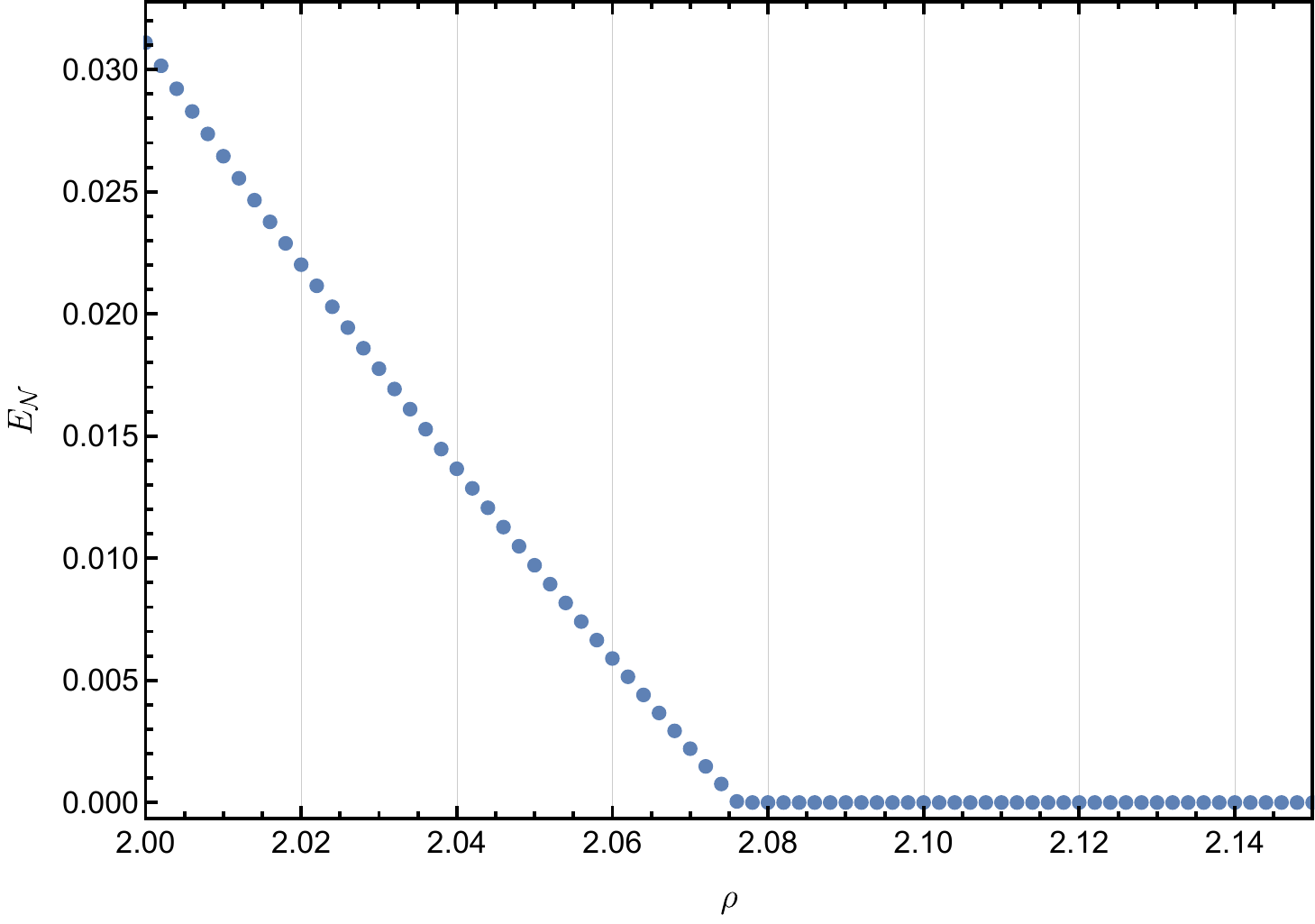}}
\begin{figure*}
    \centering
    
     \begin{minipage}{0.45\textwidth}
     \subfigure[]{
        \centering
        \vbox to \ht\mybox{%
            \vfill
            \hspace{-1cm}
            \begin{tikzpicture}
            \node at (0,0) {    \includegraphics[width=1.05\textwidth]{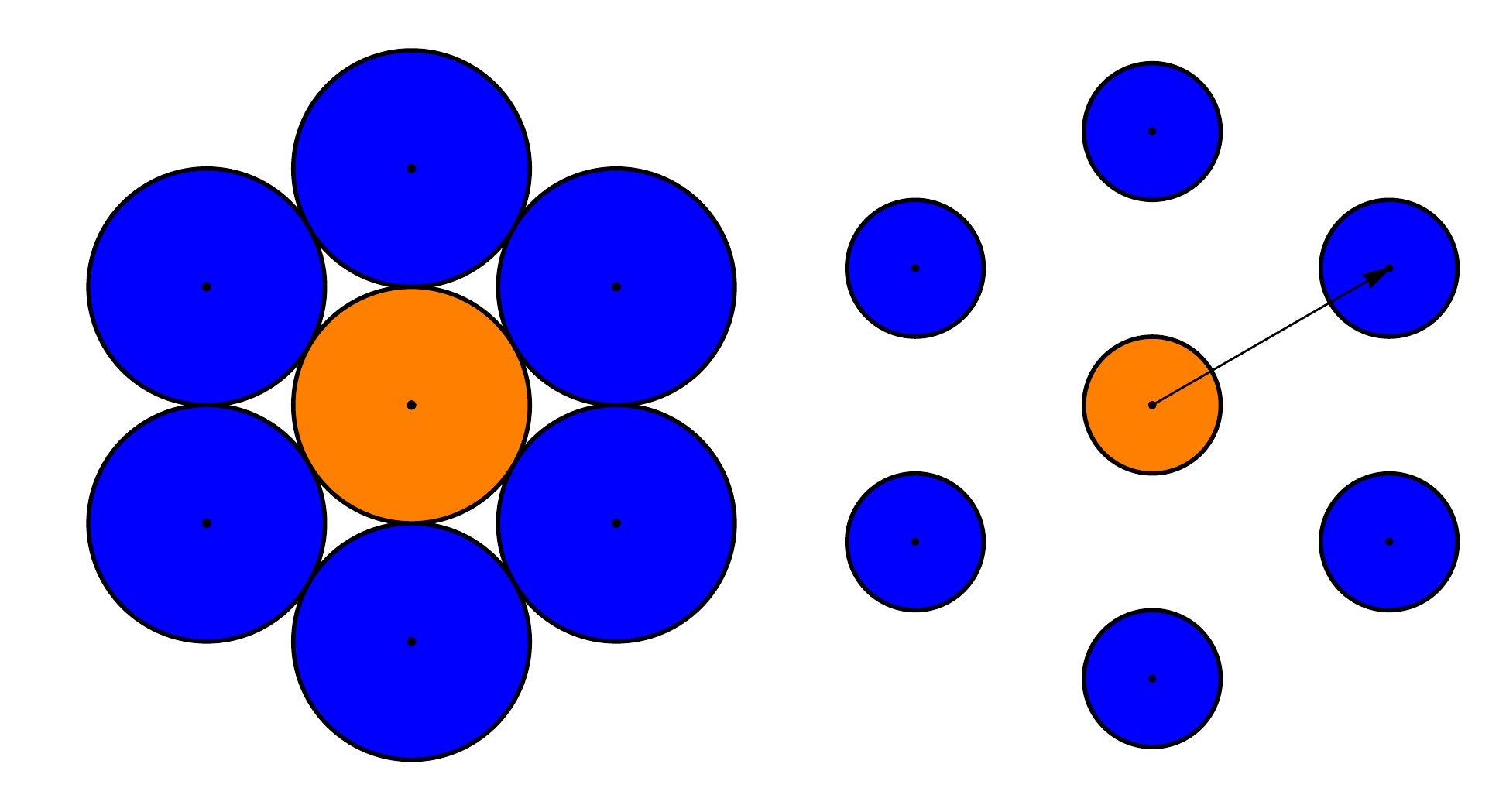}};
            \node at (2.825,0.45) {$\rho$};
            \draw[->] (-1.92,0)--(-1.92,-0.65);
            \node at (-2.1,-0.35) {$R$};
            \draw[->] (2.25,0)--(2.25,-0.39);
            \node at (2.08,-0.15) {$R$};
            \end{tikzpicture}
            \vfill
        }
        }
    \end{minipage}\begin{minipage}{0.45\textwidth}
    \subfigure[]{
        \centering
        \usebox{\mybox}
     }
    \end{minipage}
    
    \caption{(a) Configuration we use to test the way the LN varies with distance in $D=2$. The blue disks represent the regions of support of the six modes forming subsystem $B$, while the orange disk is where the single mode in $A$ is supported. (b) LN between subsystems $A$ and $B$, corresponding to the configurations showed in Fig.~\ref{fig:LN_vs_rho_hexagon_scheme}(a), versus $\rho$, defined as the distance between centers in units of the radius (hence, $\rho\geq 2$). We obtain the covariance matrix for this system analytically. However, since for $N_B=6$ it is a relatively big matrix, we compute its symplectic eigenvalues numerically; this is why we only show $E_{\mathcal{N}}$ for a discrete set of $\rho$'s (blue dots).}
    \label{fig:LN_vs_rho_hexagon_scheme}
\end{figure*}
A natural way of generalizing the results in the last section is by adding new modes to subsystem $B$, while minimizing the distance to subsystem $A$---the latter will be kept composed of a single mode for the moment; i.e., $N_A=1$. This is achieved in $D=2$ in the way shown in Fig.~\ref{fig:LN_vs_Ndof_subsystem2_hexagon_2dims}, namely by locating the modes in subsystem $B$ forming a hexagonal configuration around the mode $A$~\cite{circle_packing}.  Figure~\ref{fig:LN_vs_Ndof_subsystem2_hexagon_2dims} shows the results for the LN between $A$ and $B$ as a function of $N_B$, when the distance between modes is as small as possible (the expression for $E_{\mathcal{N}}$ is lengthy and not particularly illuminating, and we do not write it explicitly). The main feature of Fig.~\ref{fig:LN_vs_Ndof_subsystem2_hexagon_2dims} is that the LN becomes {\em different from zero if}  $N_B>4$. It is interesting to see that, in contrast to $D=1$,  in two spatial dimensions we need to enlarge our subsystems to be able to capture any entanglement. 

 Figure~\ref{fig:LN_vs_Ndof_subsystem2_hexagon_2dims} also shows that the LN saturates as the number of ``layers'' in subsystem $B$ increases, in the sense that adding more layers does not change its value. The interpretation here is that the outer layers are too far away from subsystem A to contribute to the entanglement. 
\savebox{\mybox}{\includegraphics[width=0.5\textwidth]{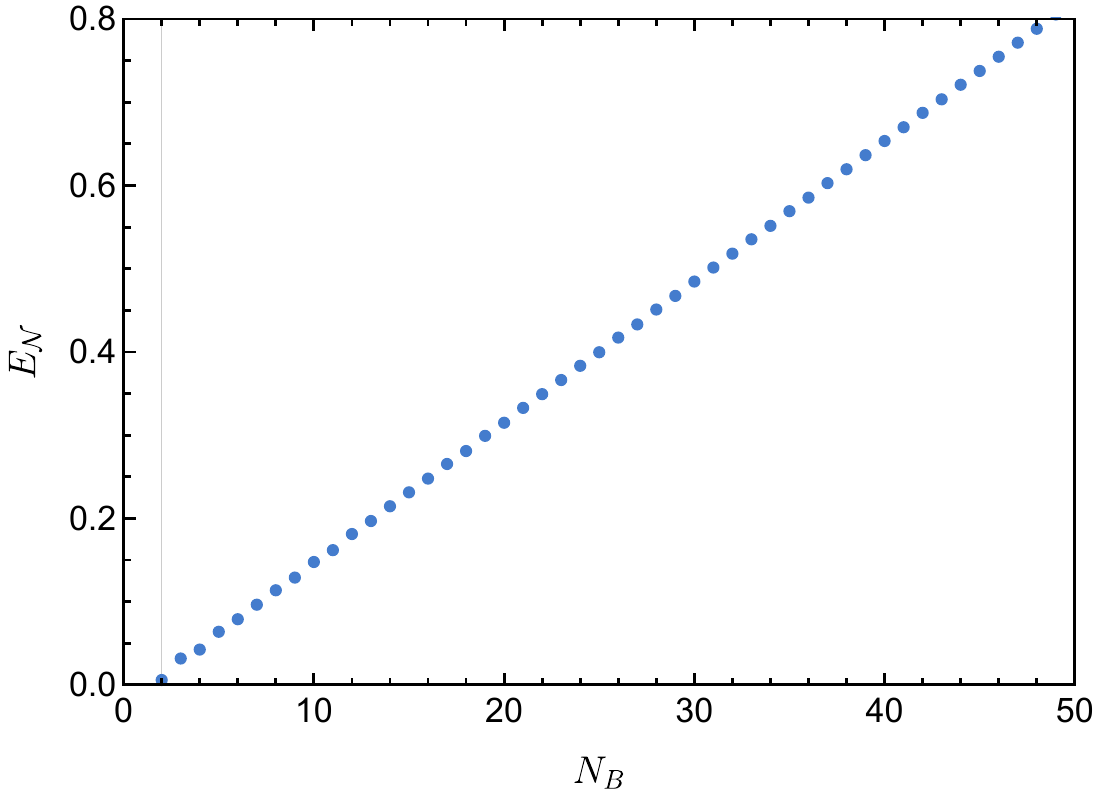}}
\begin{figure*}
    \centering
    \subfigure[]{
     \begin{minipage}{0.45\textwidth}
        \centering
        \vbox to \ht\mybox{%
            \vfill
            \hspace{-1cm}
            \includegraphics[width=1.05\textwidth]{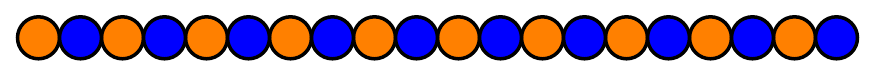}

            \vfill
        }
       
    \end{minipage}}~\subfigure[]{\begin{minipage}{0.45\textwidth}
        \centering
        \usebox{\mybox}
      
    \end{minipage}}
    
    \caption{(a) Configuration in $D=2$ consisting of $N_A=N_B=10$ modes placed alongside a straight line. The orange discs represent the regions of support of the modes forming subsystem $A$, while the blue discs represent the regions of support of the modes that constitute subsystem $B$. (b) LN between subsystems $A$ and $B$, corresponding to the  configuration shown in Fig.~\ref{fig:Alternating_line_NB10} (a), as a function of the number of modes $N_B$ ($N_A$) within subsystem $B$ ($A$). }
    \label{fig:Alternating_line_NB10}
    \end{figure*}
 Now that we have found a configuration containing entanglement, we study the way this entanglement changes with the ``distance'' between the two subsystems. Such distance  can be varied in many different manners. As an illustrative example, we consider the configuration depicted in Fig.~\ref{fig:LN_vs_rho_hexagon_scheme}(a). In Fig.~\ref{fig:LN_vs_rho_hexagon_scheme}(b) we plot  the LN versus the distance $\rho$ for this system, containing  $N_B=6$ modes in subsystem B. We observe that the LN falls off very fast,   completely vanishing when the distance between the surfaces of the regions is less than 10\% of their radius.  

Next, we consider configurations where both subsystems, $A$ and $B$, are made of multiple modes (recall that when both $N_A$ and $N_B$ are larger than 1, a nonzero value of LN is a sufficient but not a necessary  condition for entanglement). In this case, there are plenty of geometric configurations that one can consider. In what follows, we mention two that we find particularly interesting. The first one is depicted in Fig.~\ref{fig:Alternating_line_NB10} and consists of  $2N$ disjoint regions placed alongside a straight line, with  alternating regions belonging to each subsystem. This configuration is interesting because, for $N=N_A=N_B\geq 2$, the LN is different from zero and grows linearly with $N$, as shown in Fig.~\ref{fig:Alternating_line_NB10}. 

\savebox{\mybox}{\includegraphics[width=0.5\textwidth]{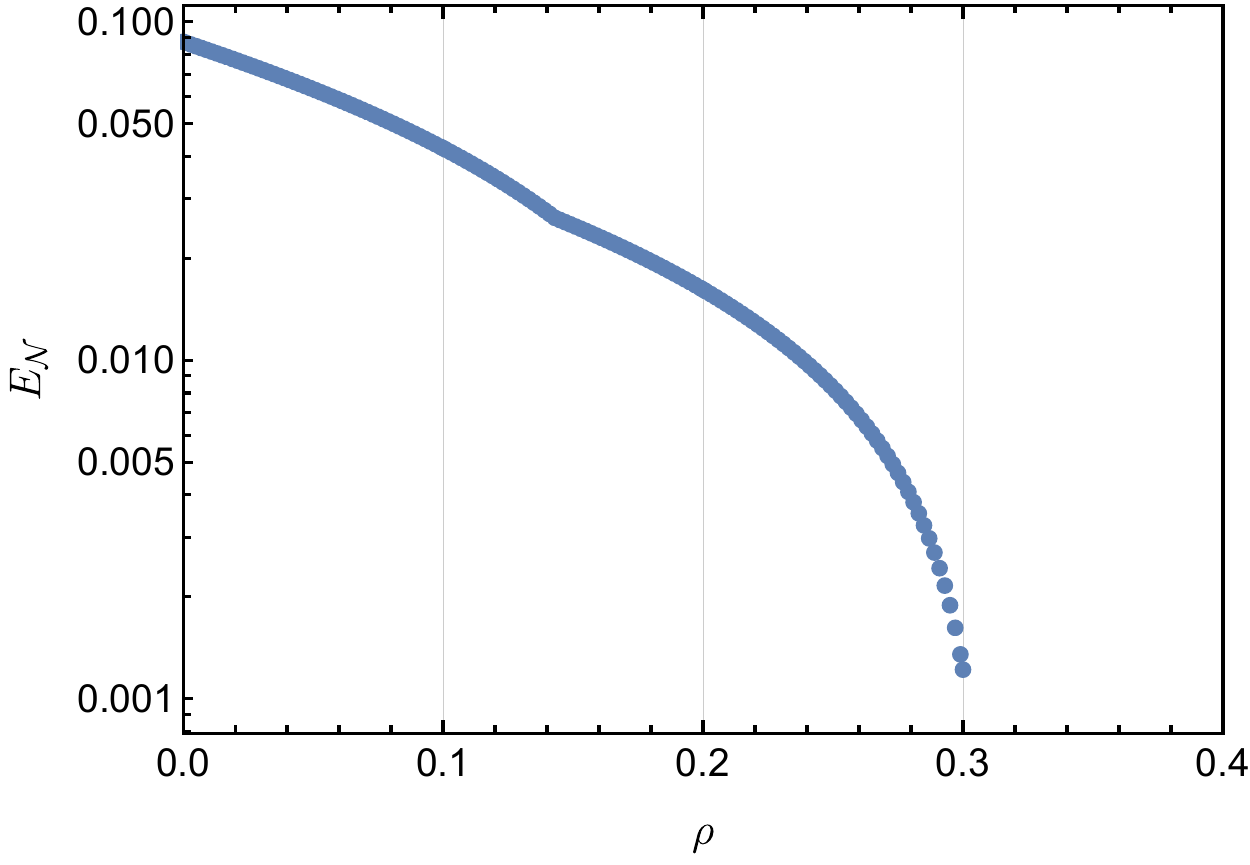}}
\begin{figure*}
    \centering
    \subfigure[]{
     \begin{minipage}{0.45\textwidth}
        \centering
        \vbox to \ht\mybox{%
            \vfill
            \hspace{-1cm}
           \begin{tikzpicture}
    \node at (0,0) { \includegraphics[width=1.05\textwidth]{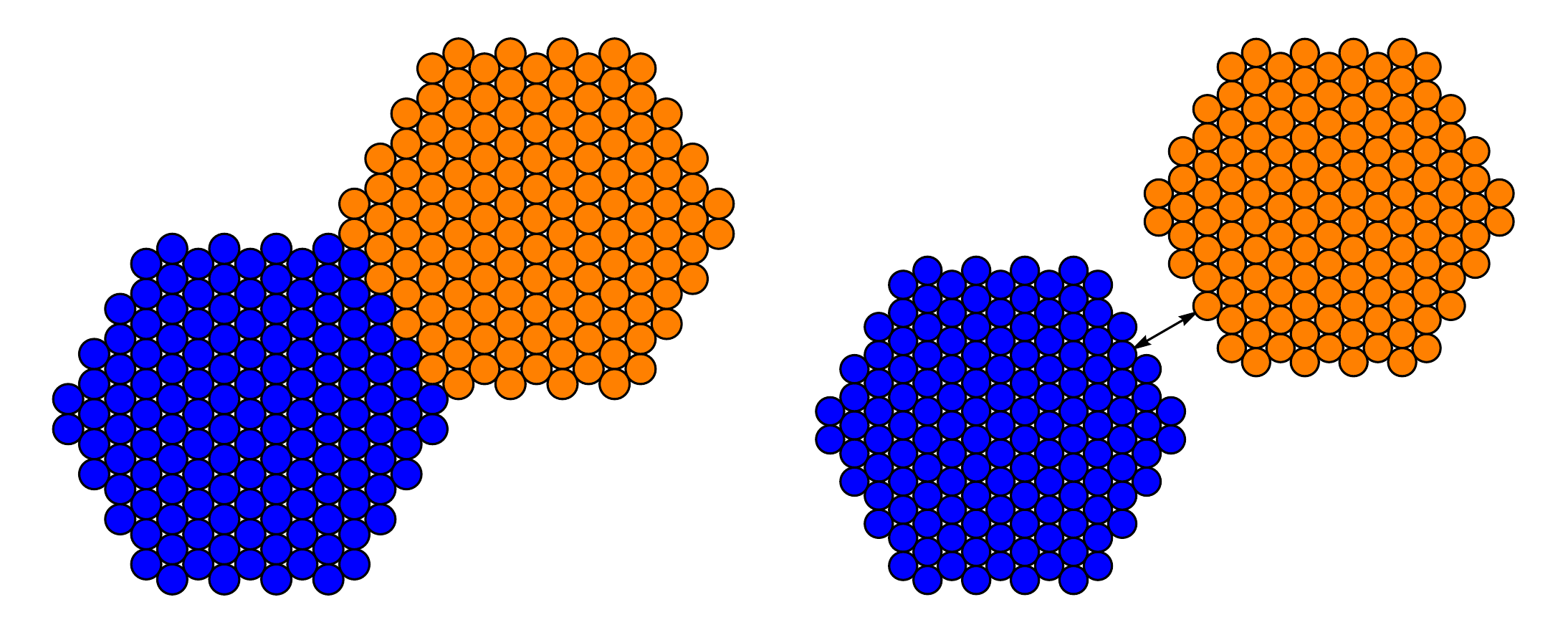}};
    \node at (2.,0.15) {$\rho$};
    \end{tikzpicture}

            \vfill
        }
       
    \end{minipage}}~\subfigure[]{\begin{minipage}{0.45\textwidth}
        \centering
        \usebox{\mybox}
      
    \end{minipage}}
    
    \caption{(a) $D=2$ configuration of two subsystems made of $N_A=N_B$ modes. (b) LN versus the distance $\rho$ (in units of the radius of the individual small regions) corresponding to the $D=2$ configuration of Fig.~\ref{fig:Two_Hexagons_diff_distances} (a).The small kink around $\rho \sim 0.14$ is a boundary effect caused by the concrete geometric configuration we use in this example.} 
    \label{fig:Two_Hexagons_diff_distances}
    \end{figure*}

A second configuration we explore consists of $A$ and $B$ each made of a hexagonal cell with $N$ disjoint modes and separated from each other  as shown in Fig.~\ref{fig:Two_Hexagons_diff_distances}.  This figure shows that for this configuration the LN decreases rapidly  with the separation and completely vanishes when the distance $\rho$ (defined as depicted in Fig.~\ref{fig:Two_Hexagons_diff_distances} and measured in units of the radius of the individual disks) is larger than $0.3$. 

This configuration is inspired by the type of systems considered in lattice field theory. The behavior of the LN in $1+2$ dimensional Minkowski spacetime has been  investigated in such context \cite{Klco:2021biu,Klco:2020rga}. In lattice field theory, each field degree of freedom ``lives'' at the nodes of the lattice, and subsystems $A$ and $B$ are each made of $N_A$ and $N_B$ modes, respectively. One can then consider two regions, each containing $N$ nodes, and evaluate the entanglement between the two regions when the field is prepared in the vacuum. 

Numerical computations \cite{Klco:2021biu,Klco:2020rga} have revealed that entanglement between two regions in lattice field theory has many similarities with our findings. In particular, lattice calculations also show that, for a finite number of modes, the LN becomes zero abruptly beyond some threshold separation distance. The fall off of the LN with distance is found to be exponential~\cite{Klco:2021biu}. Although one cannot compare our calculations with the result of lattice field theory in a detailed manner, mainly because we have included only a finite number of modes in our calculations, our results indicate that  in the continuum theory the LN for a finite number of modes falls off significantly faster with the distance between the two regions than its counterpart in lattice field theory. It would be interesting to have a more detailed comparison, but this is beyond the scope of this work.

\subsection{$D=3$} 
 
\begin{figure*}
     \centering
     \subfigure[\label{fig:1p5HCP_Cell}]{
            \includegraphics[width=0.4\textwidth]{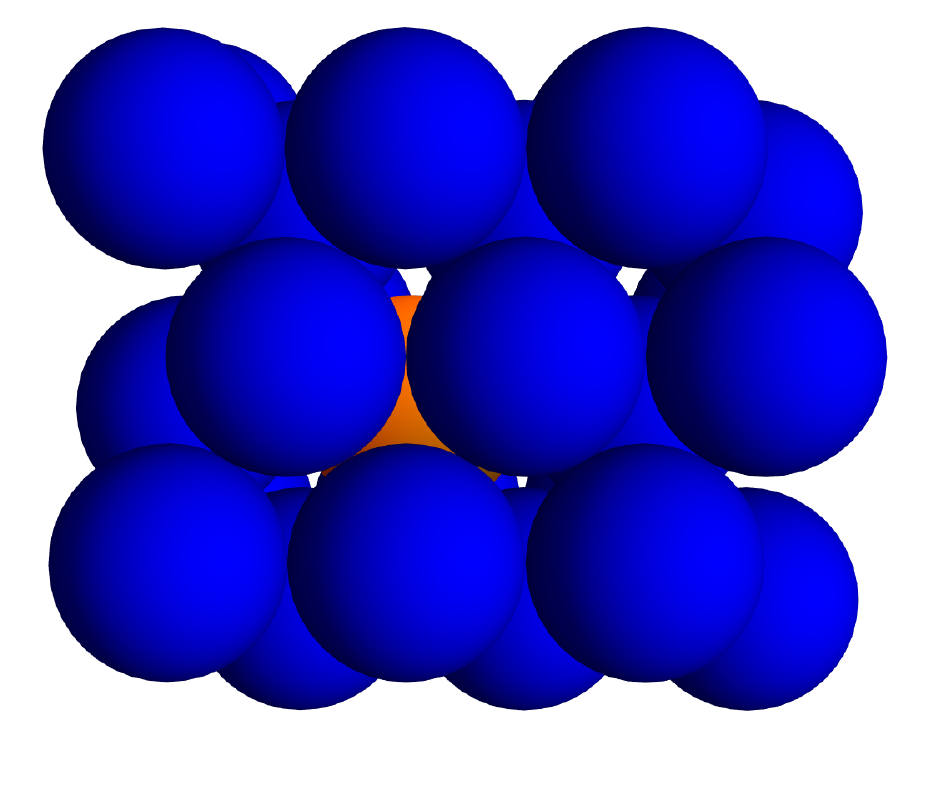}
                
     }%
     \subfigure[\label{fig:thick_net}]{
            \includegraphics[width=0.4\textwidth]{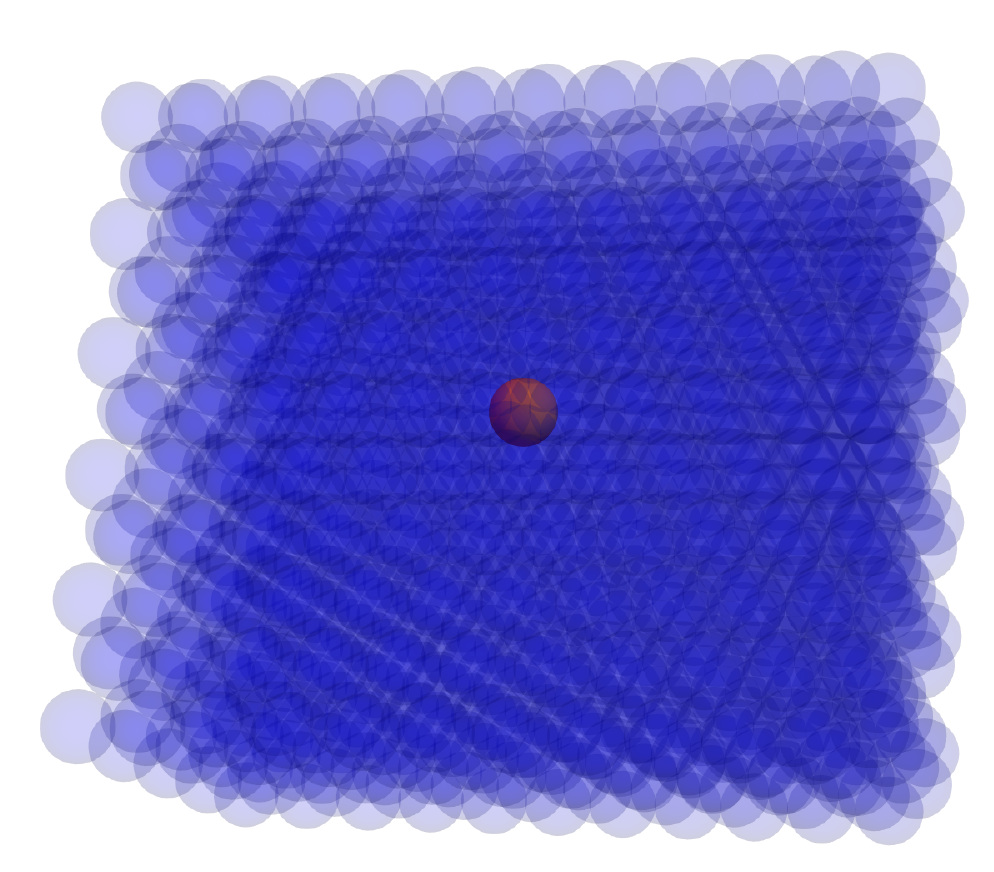}

    }\\
    
    \subfigure[ \label{fig:super_sandwich_config}]{
         \includegraphics[width=0.6\textwidth]{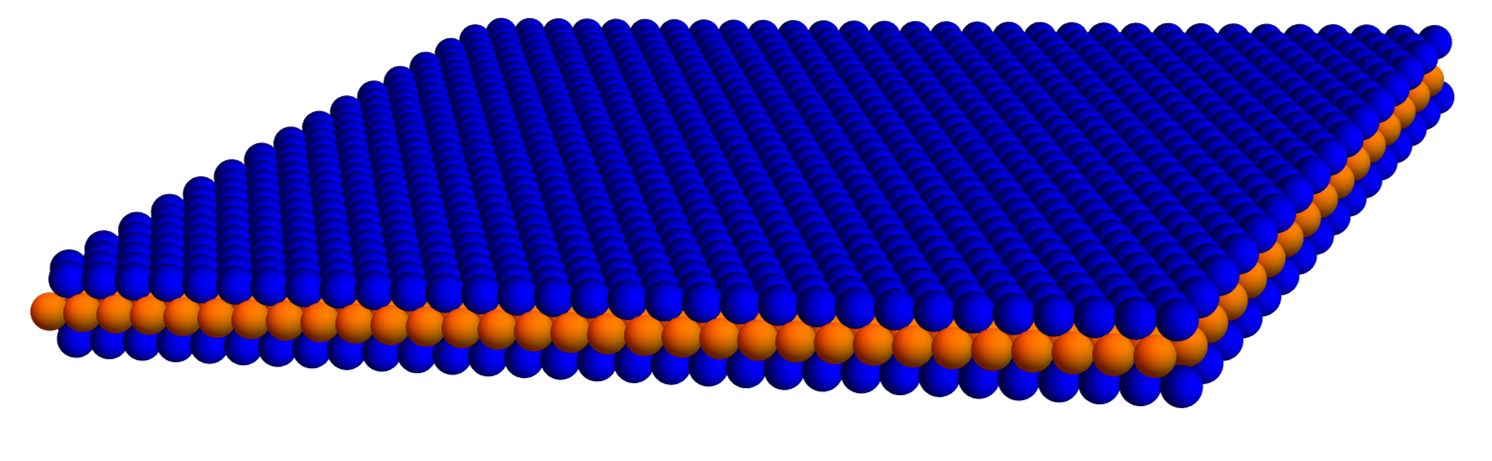}
 }
\caption{(a) Configuration of modes in $D=3$ in a hexagonal close-packed with a single mode in subsystem A and $N_B=18$ modes in subsystem B. (b) Example of a configuration we used to compute the LN in $D=3$. The blue spheres represent the regions of support of the $N_B = 1088$ modes, while the orange sphere is where the single mode in subsystem A is supported. (c) Configuration in $D=3$. The blue spheres represent the regions of support of the $N_B = 1922$ modes, while the orange spheres represent the regions of support of the $N_A=961$ modes in subsystem A. We find that LN vanishes in this configuration. }
\label{fig:sphere_configs_3D}
 \end{figure*}

 Next, we apply the same strategy to the case of $D=3$ spatial dimensions; namely, we  generalize the results of Sec.~\ref{sec:3} by adding new modes to subsystem $B$, supported in disjoint regions located close to subsystem $A$. In $D=3$, the densest regular arrangement of spheres can be achieved by placing them either in a face-centered cubic configuration, or in a hexagonal close-packed configuration. We choose the latter. We illustrate one such configuration in Fig.~\ref{fig:1p5HCP_Cell}.

 We have computed analytically the covariance matrix for this set up, including up to $N_B=1088$ modes in subsystem $B$, while keeping $N_A=1$  [see Fig.~\ref{fig:thick_net}]. From this covariance matrix, we have  computed the LN (with the assistance of software for  symbolic calculations, such as {\em Mathematica}) for a massless noninteracting scalar field and for the  family of smearing functions introduced in \eqref{eq:family_test_funcs1}. 
 
 Contrary to the situation in $D=2$, in $D=3$ we have found that the LN is {\em zero} for $N_B$ up to 1088. This is true even for smearing functions with $\delta$ equal or close to 1, for which we found entanglement in $D=2$ for $N_B\geq 5$.

 This result confirms the trend we found in the last subsections: entanglement is  ``weaker'' or ``sparser'' in higher  dimensions.

We have further extended our calculations by increasing also the number of modes in subsystem A. One such configuration is showed in Fig.~\ref{fig:super_sandwich_config}. We have increased the number of modes until $N_A=961$ and $N_B=1922$ and obtained LN equal to zero in all cases. 
Entanglement for $D=3$ is too weak or diluted to be captured using the finite number of modes we have used so far. In the next two sections, we extend the family of modes and the way they are distributed in space.

\section{Other smearing functions}\label{sec:5}

In this section, we extend our calculation in different directions, with the goal of checking whether the absence of entanglement between pairs of modes supported in disjoint regions in $D\geq 2$ (or finite sets of them for $D\geq 3$) is a peculiarity of the concrete family of modes we have used so far. We generalize our calculations in different directions, by (i) considering other families of smearing functions including some without a definite sign, and (ii) by mixing field and momentum operators. In none of these cases have we found entanglement between subsystems $A$ and $B$ in $D \geq 2$ for pairs of modes supported in disjoint spherical regions.

\subsection{Other positive semidefinite smearing functions}

In addition to the smearing functions introduced in Eq.~\eqref{eq:family_test_funcs1}, we have explored the following families of non-negative smearing functions, all spherically symmetric around a center $\vec x_i$ and where $r$ is the distance to the center in units of the radius $R$ of the region of support:

\begin{enumerate}
\item 

$   h^{(n)} \left(r\right) = A_n\, \Theta\left(1- r\right) \, \cos^n\left(\frac{\pi}{2}r\right)\,, \quad n>1\in \mathbb{N}\,, $ 

with $A_{n}$ a normalization constant. This family has been used before in \cite{Reznik:2002fz}. 

\item 
$   g\left(r\right) = A \, \Theta\left(1- r\right) \, \exp\left(- \frac{1}{1-r^2}\right)\,. $

\item $$ w^{(\delta)}\left(r\right) = A_{\delta} \begin{cases}
1 &  0<r\leq 1\\ 
-\frac{1}{\delta}(r-1) +1 & 1<r\leq 1+\delta\\ 
0 & r > 1+ \delta
\end{cases}\,. $$
This family has been used before in \cite{martin_real-space_2021}.

\item   $j^{(n)}(r) = A_{n}\, \Theta\left(1- r\right) \left(1 - r^n \right)\,,\quad n>1 \,\in\, \mathbb{N}\,.$  
\end{enumerate}

All these functions and their first derivatives are continuous ($g(r)$ is actually smooth). We argue in Appendix~\ref{app:smooth_functions} that this is sufficient for our purposes.

\subsection{Non-semi-positive definite smearing functions}
We have explored the following family of $L^2$-orthonormal functions of compact support,  
\begin{equation}
    k^{(n)}(r)=
    \frac{\sin{2\pi n r}}{2\pi n r}\,\Theta(1-r)\, .\label{SF3D.sinc}
\end{equation}
Again, these functions and their first derivatives are continuous. When these functions are used to define  modes supported in disjoint regions, we find no entanglement between them. In the next section, we will use these functions to define  modes supported {\em in the same region} (this is possible since these functions are orthogonal to each other for different values of $n$), in which case we do find entanglement (see Sec.~\ref{overlapping} below).

\subsection{Combinations of field and momentum}

The modes  of the field  used in the calculations presented in  previous sections were all defined from pairs of operators of the form of a pure-field and a pure-momentum operator, $(\hat \Phi_i, \hat \Pi_i)$, both constructed from the same smearing function [up to an exact factor $c$ in $\hat \Pi_i$, needed to ensure that both operators have the same units (action)].

We  extended  here these calculations by considering modes defined from (the algebra generated by) pairs of canonically conjugated operators of the following form
\begin{align} \label{modesmix}
   \hat{O}^{(1)}_A & = \frac{1}{\sqrt{2N}} \sum_{i=1}^{N}\left(\hat{\Phi}^{(2i-1)}_A - \hat{\Pi}^{(2i)}_A\right)\, , \\ 
   \hat{O}^{(2)}_A & =  \frac{1}{\sqrt{2N}}\sum_{i=1}^N \left(-\hat{\Phi}^{(2i-1)}_A + \hat{\Pi}^{(2i)}_A\right)\, ,
\end{align}
where $\hat{\Phi}^{(n)}_A$ and $\hat{\Phi}^{(n)}_A$ indicate field and momentum operators, respectively, smeared with the element $k^{(n)}(r)$ of the family of orthonormal  functions written in Eq.~\eqref{SF3D.sinc}. 

Therefore, each of these two operators $\hat{O}^{(i)}_A$ is made by combining pure-field and pure-momentum operators, each constructed from different smearing functions. The orthonormality of the smearing functions $k^{(n)}(r)$ guarantees that  $\hat{O}^{(1)}_A$ and $\hat{O}^{(2)}_A$ are canonically conjugate. 

Subsystem $B$ is defined in the same way, with support in a spherical region as close as possible to the support of subsystem $A$. We have explored different values of $N$, from 1 to 10 in  \eqref{modesmix}, and we have not found entanglement between the two subsystems in any case. \\

The analysis of this section reveals that the absence of entanglement in $D\geq 2$ between pairs of modes supported in two disjoint spherical regions is not unique to the smearing functions we have chosen, but a rather generic fact. 

\section{\label{sec:6}Pairs of entangled modes}

The analysis so far shows that, for $D\geq 2$, finding entanglement between  two field modes is not an easy task, or at least not as easy as one would have thought. In fact, in none of the modes explored so far  have we found pairwise entanglement for $D\geq 2$. We have presented results for a massless scalar field, but  introducing a mass only makes entanglement weaker.  Does this mean that it is actually impossible to find pairs of modes of the field which are entangled? Absolutely not. Entanglement is intrinsic to every quantum state of multimode systems, even in ordinary quantum mechanics, in the sense that, for every quantum state, one can find subsystems that are entangled \cite{Zanardi:2004zz,ABRM_2022}. 

As a simple illustrative example, consider two uncoupled harmonic oscillators prepared in the ground state (the product of ground states of each oscillator). This is a product state, and therefore, the subsystems defined by $(\hat x_A,\hat p_A)$ and $(\hat x_B,\hat p_B)$ are obviously not entangled. However, it is not difficult to find other partitions of the systems for which entanglement shows up in the ground state. A simple choice is made by the two subsystems $(\hat x_1,\hat p_1)$ and $(\hat x_2,\hat p_2)$, where 
\bea 
\hat x_1&=&\cosh z\, \hat x_A+\sinh z\,\hat x_B\, , \nonumber \\
\hat p_1&=&\cosh z\, \hat p_A-\sinh z\,\hat p_B\, ,\nonumber \\
\hat x_2&=&\sinh z\, \hat x_A+\cosh z\,\hat x_B\, ,\nonumber \\
\hat p_2&=&-\sinh z\,\hat p_A+\cosh z\,\hat p_B\, , \nonumber 
\eea
with $z\in \mathbb{R}$. Subsystems $(\hat x_1,\hat p_1)$ and $(\hat x_2,\hat p_2)$ are obtained by mixing the original oscillators, but each pair defines a licit mode of the system. It is straightforward to check that the LN between these two modes is $E_{\mathcal{N}}= \frac{2 \, z}{\ln 2}$; therefore, 
there is entanglement between these two modes for all $z\neq 0$, and it grows monotonically with $z$.

This example reminds us about the well-known fact that entanglement is not a property of a quantum state alone; it is an attribute of a state {\em and} a choice of subsystems. Furthermore, for every quantum state of a multimode system there are choices of subsystems for which entanglement is present~\cite{ABRM_2022}. When we simply say that a quantum state is not entangled, it is because we implicitly assume a natural or physically preferred set of modes of the systems.  

Coming back to field theory in Minkowski spacetime, it is straightforward to find modes which are entangled in the vacuum, by simply mimicking the example of the two harmonic oscillators. Consider any pairs of modes $(\hat \Phi_A, \hat \Pi_A)$ and $(\hat \Phi_B, \hat \Pi_B)$ considered in previous sections, for which we found that the reduced state is separable. From them, we can construct new  modes 
$(\hat \Phi_1, \hat \Pi_1)$ and $(\hat \Phi_2, \hat \Pi_2)$, where 
\bea 
\hat \Phi_1&=&\cosh z\,\hat \Phi_A+\sinh z\, \hat \Phi_B\, , \nonumber \\
\hat \Pi_1&=&\cosh z\,\hat  \Pi_A-\sinh z\,\hat \Pi_B\, ,\nonumber \\
\hat \Phi_2&=&\sinh z\, \hat \Phi_A+\cosh z\,\hat \Phi_B\, ,\nonumber \\
\hat \Pi_2&=&-\sinh z\, \hat \Pi_A+\cosh z\, \hat \Pi_B\, , \nonumber
\eea
with $z\in \mathbb{R}$. We have shown that there exists a minimum value of $|z|$ above which the new subsystems are entangled. (Contrary to the example of harmonic oscillators, $|z|$ must be above a nonzero threshold. The reason is that, in field theory, the reduced state describing the two modes together is always mixed (nonzero entropy). This mixedness acts as a source of noise for entanglement, requiring a minimum amount of ``squeezing'' intensity $|z|$ to entangle the modes. The threshold value of $|z|$ depends on the smearing functions chosen to define the initial pair of modes.)

Although we can define a new pair of modes, $(\hat \Phi_1, \hat \Pi_1)$ and $(\hat \Phi_2, \hat \Pi_2)$, that are entangled they are not supported in disjoint regions, since each mode is a combination of a part supported in region A and a part in B. In this sense, these modes are nonlocal and somewhat unnatural.

In the remaining of this section, we describe other---less obvious and therefore more informative---examples of pairs of entangled modes in field theory. 

\subsection{Independent modes with overlapping support}\label{overlapping}

Motivated from the intuition that entanglement falls off very rapidly with distance, we consider two modes defined in the same region of space. This can be done by defining each mode from pairs $(\hat \Phi_n,\hat \Pi_n)$ using the smearing functions in the family  $k^{(n)}(r)$ defined in  \eqref{SF3D.sinc}. These functions are orthogonal to each other for different values of $n$, hence they define independent (commuting) modes. We will use $n=n_A$ to define mode A, and $n=n_B\neq n_A$ for mode B. 

\begin{figure}[h]
\begin{center}
\includegraphics[width=0.5\textwidth]{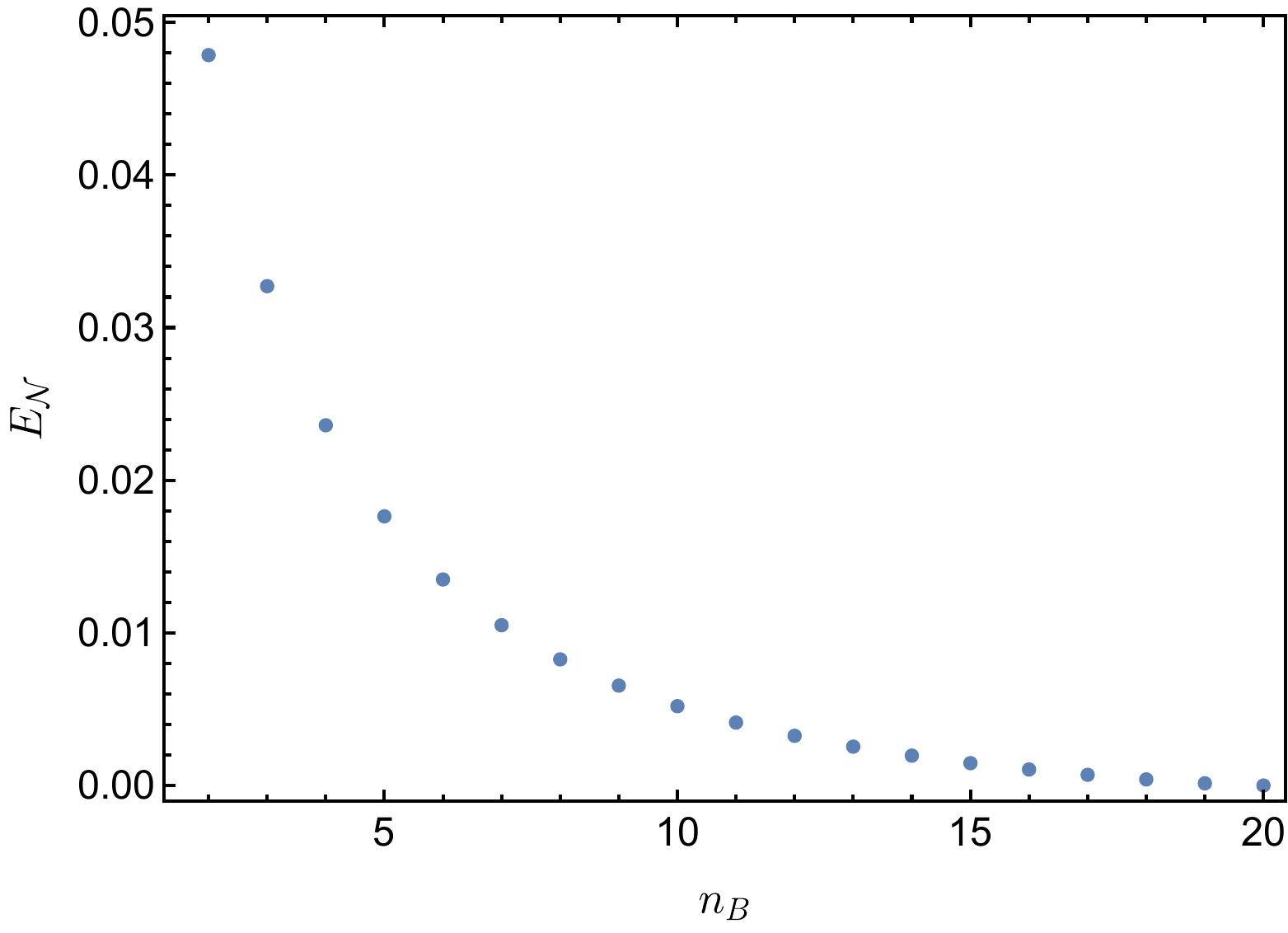}
\caption{LN between two modes of the form $(\hat \Phi_i,\hat \Pi_i)$, $i=A,B$, defined in the same region of space, from the family of orthonormal functions $k^{(n)}(r)$ introduced in  \eqref{SF3D.sinc}. System $A$ is defined using $k^{(n_A)}(r)$ and similarly for system $B$, with $n_A\neq n_B$. We plot the LN vs $n_B$, for $n_A=1$. Two messages can be extracted from this plot: (i) the two modes are entangled, and (ii) entanglement is larger the closer $n_A$ and $n_B$ are.  }\label{fig.LN.vs.n2}
\end{center}
\end{figure}

We have found that such two modes {\em are entangled} in $D=3$. The entanglement is largest when $n_A=1$ and $n_B=2$ and decreases with both $n_A$ and $n_B$. These results are illustrated for $D=3$ in Fig.~\ref{fig.LN.vs.n2}. 

\subsection{Rindler modes}
\begin{figure*}[t]
    \centering
    \subfigure[\label{fig:ball&shell}]{
    \begin{tikzpicture}
    \node at (0,0) {\includegraphics[width=0.4\textwidth]{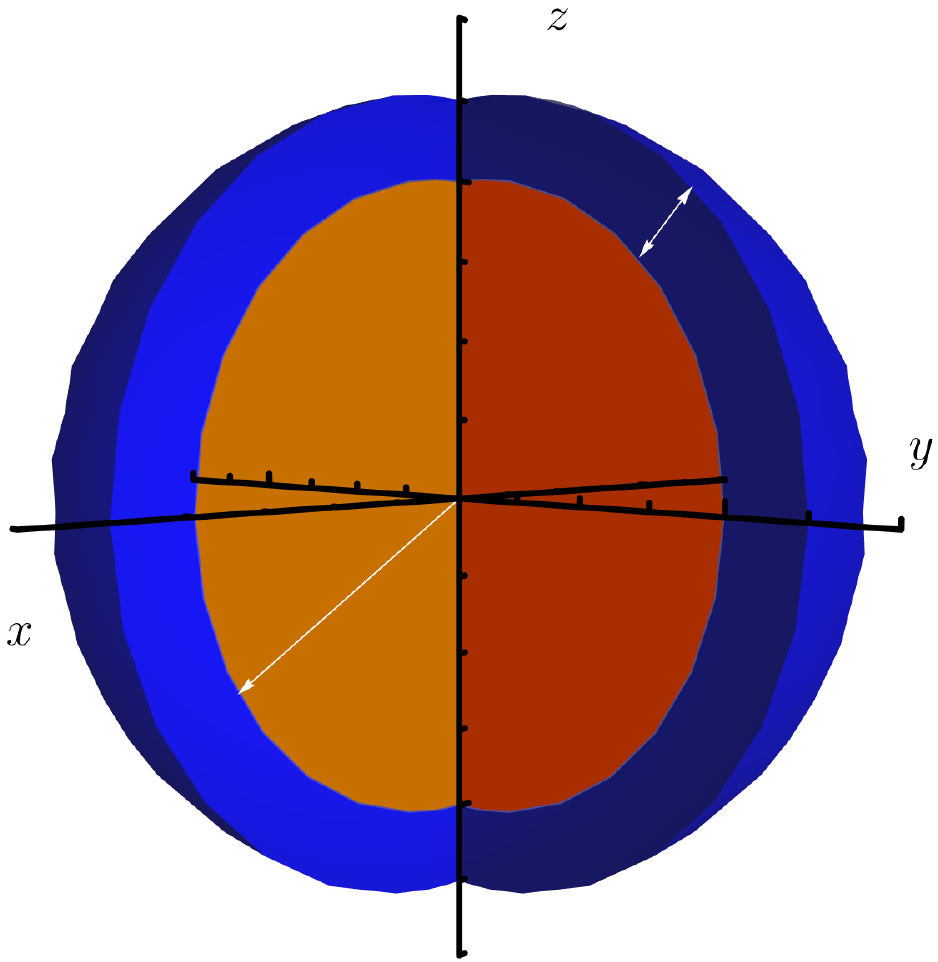}};
    \node[white,rotate=55] at (1.25,2.2) {$d_B$}; 
    \node[white, rotate=42] at (-1.25,-0.75) {$R_A=R_B$};
    \end{tikzpicture}
    }~\subfigure[]{
     \begin{tikzpicture}
    \node at (0,0) { \includegraphics[width=0.4\textwidth]{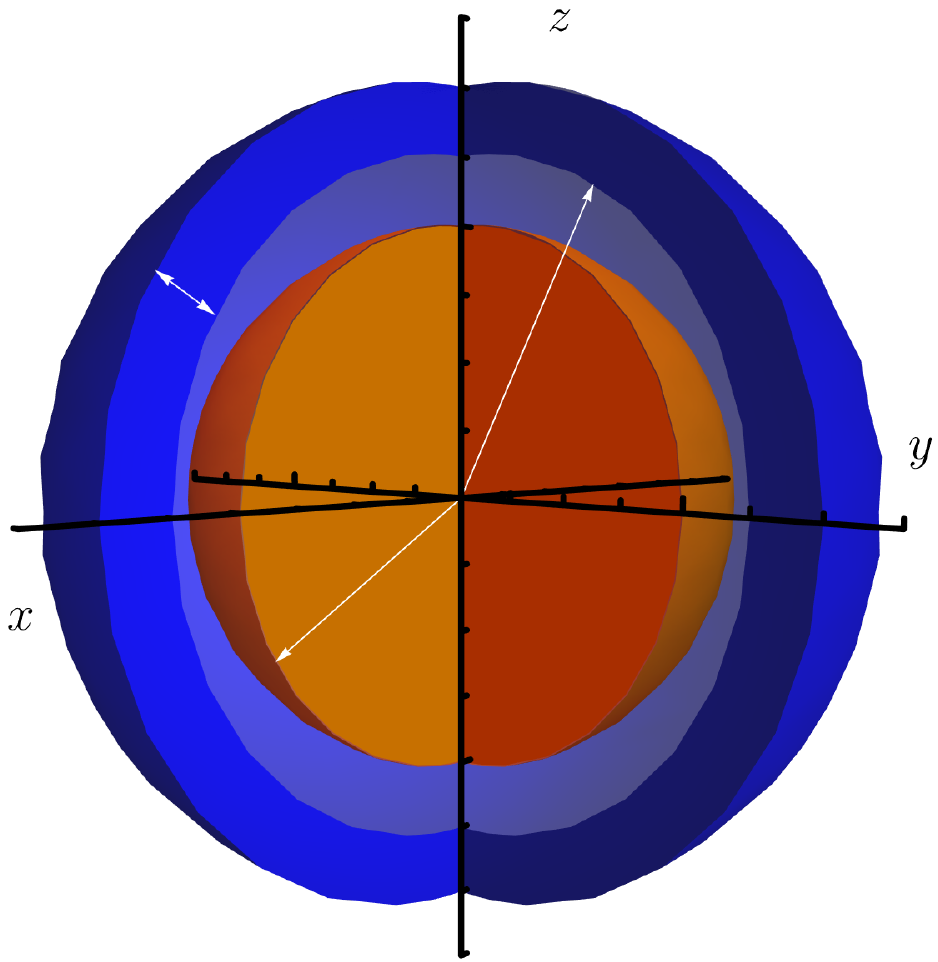}}; 
     \node[white,rotate=75] at (0.75,1) {$R_B$}; 
    \node[white, rotate=50] at (-1.2,-0.75) {$R_A$};
   \node[white, rotate=-30] at (-2,1.75) {$d_B$};
    \end{tikzpicture}
    }
    \caption{Regions of support for two field modes. The orange sphere is where the mode defining subsystem $A$ is supported, while the mode in $B$ is supported in the blue shell. The two panels illustrate the freedom we have in the distance between the sphere and the shell, as well as the thickness of the shell.  (a) The distance between the sphere and the shell is zero, such that, $R_A = R_B$. (b) The sphere and the shell are separated by  a nonvanishing distance, such that $R_B - R_A >0 $.}
\label{fig:ball&shell3D}
\end{figure*}
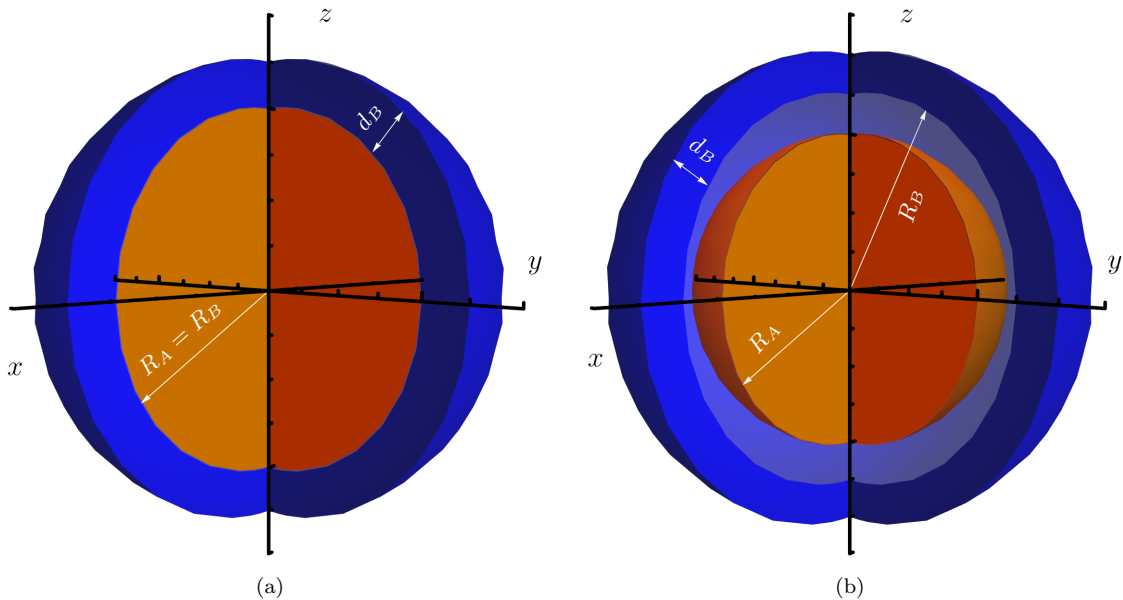
A well-known example of pairs of modes that are entangled in the vacuum is that of Rindler modes. For completeness, in this subsection we use our tools to check that the right-Rindler and left-Rindler modes are indeed entangled. This is an example of modes with support in disjoint regions which, nevertheless, are entangled.

Let $\hat{a}^{\mathrm{R}}_{\omega\vec{k}_{\perp}}$ and $\hat{a}^{\mathrm{L}}_{\omega\vec{k}_{\perp}}$ be standard right and left, respectively, Rindler annihilation operators (see, for instance, \cite{Crispino:2007eb}), where $\omega$ is the Rindler frequency, and $\vec{k}_{\perp} = (k_x,k_y)$ is the momentum in the directions perpendicular to the Rindler acceleration, which we assume to be in the $z$-direction. From these operators and their adjoints, we build two modes $(\hat{X}^{\mathrm{R}}_{\omega\vec{k}_{\perp}}, \hat{P}^{\mathrm{R}}_{\omega\vec{k}_{\perp}})$ and $(\hat{X}^{\mathrm{L}}_{\omega\vec{k}_{\perp}}, \hat{P}^{\mathrm{L}}_{\omega\vec{k}_{\perp}})$:
\begin{align}
    \hat{X}^{\mathrm{R}}_{\omega\vec{k}_{\perp}} &= \frac{1}{\sqrt{2}}\left(\hat{a}^{\mathrm{R}}_{\omega\vec{k}_{\perp}} + \hat{a}^{\mathrm{R}\,\dagger}_{\omega\vec{k}_{\perp}} \right)\,,
    \\
    \hat{P}^{\mathrm{R}}_{\omega\vec{k}_{\perp}} &= \frac{-i}{\sqrt{2}}\left(\hat{a}^{\mathrm{R}}_{\omega\vec{k}_{\perp}} - \hat{a}^{\mathrm{R}\,\dagger}_{\omega\vec{k}_{\perp}} \right)\,,
\\    
\hat{X}^{\mathrm{L}}_{\omega'\vec{k}'_{\perp}} &= \frac{1}{\sqrt{2}}\left(\hat{a}^{\mathrm{L}}_{\omega'\vec{k}'_{\perp}} + \hat{a}^{\mathrm{L}\,\dagger}_{\omega'\vec{k}'_{\perp}} \right)\,, 
\\    
\hat{P}^{\mathrm{L}}_{\omega'\vec{k}'_{\perp}} &= \frac{-i}{\sqrt{2}}\left(\hat{a}^{\mathrm{L}}_{\omega'\vec{k}'_{\perp}} - \hat{a}^{\mathrm{L}\,\dagger}_{\omega'\vec{k}'_{\perp}} \right)\, .
\end{align} 
 They are independent of each other (i.e., they commute) and satisfy canonical commutation relations,  $[\hat{X}^{\mathrm{R}(\mathrm{L})}_{\omega\vec{k}_{\perp}},\hat{P}^{\mathrm{R}(\mathrm{L})}_{\omega'\vec{k}'_{\perp}}]=i\, \delta(\omega-\omega')\delta^2(\vec{k}_{\perp}-\vec{k}_{\perp}) $. 
 
Following the  procedure introduced in  Sec.~\ref{sec:2}, we  compute the components of the covariance matrix $\sigma_{RL}(\omega,\vec{k}_{\perp},\omega',\vec{k}'_{\perp})$ of the reduced system for these two modes,  supported in different wedges, in the Minkowski vacuum. They are
\begin{align}
    \braket{ (\hat{X}^{\mathrm{R(L)}}_{\omega\vec{k}_{\perp}})^2} &=  \frac{1}{2}\tanh{\left(\pi\omega/a\right)}^{-1}\delta^3(0)\,,
\\    
\braket{ (\hat{P}^{\mathrm{R(L)}}_{\omega\vec{k}_{\perp}})^2} &=  \frac{1}{2}\tanh{\left(\pi\omega/a\right)}^{-1}\delta^3(0)\,,
\end{align} 
and 
\begin{equation}
    \braket{ \{\hat{X}^{\mathrm{R}}_{\omega\vec{k}_{\perp}},\hat{X}^{\mathrm{L}}_{\omega'\vec{k}'_{\perp}}\}} =\sinh{(\omega\pi/a)}^{-1}\delta(\omega - \omega')\delta^{2}(\vec{k}_{\perp} + \vec{k}'_{\perp})\,,
\end{equation} and \begin{equation}
     \braket{ \{\hat{P}^{\mathrm{R}}_{\omega\vec{k}_{\perp}},\hat{P}^{\mathrm{L}}_{\omega'\vec{k}'_{\perp}}\}} = -\sinh{(\omega\pi/a)}^{-1}\delta(\omega - \omega')\delta^{2}(\vec{k}_{\perp} + \vec{k}'_{\perp}) \; .
\end{equation}
The constant $a$ is the acceleration of the Rindler frame.  The presence of Dirac deltas  is a result of the normalization of the modes and, as usual, it can be removed by using wave packets instead of plane waves.

Note that $\sigma_{RL}$ is ``diagonal'' in the labels $(\omega,\vec{k}_{\perp})$ and $(\omega',\vec{k}'_{\perp})$, so right and left modes with different labels $(\omega,\vec{k}_{\perp})$ are uncorrelated and unentangled, as expected.  

From the previous expressions, we compute the partially transposed covariance matrix and its symplectic eigenvalues  for $\omega'= \omega$ and $\vec{k}_{\perp} =-\vec{k}'_{\perp}$.  The smallest of these  eigenvalues  is 
\begin{equation}\label{tildenuR}
    \tilde{\nu}_- = \delta^3(0) \, \tanh\left( \frac{\omega \pi }{2\,a}\right) \,.
\end{equation}
This eigenvalue is smaller than 1 for any $a\neq 0$. Consequently, right and left Rindler modes with the same labels $(\omega,\vec{k}_{\perp})$ are entangled in the Minkowski vacuum as long as $a\neq 0$, and their entanglement grows monotonically with $a$. 

This calculation refers to Rindler modes defined using plane waves for which, strictly speaking, all quantities above blow up. The calculations can be made finite by replacing plane waves by wave packets, and the finite part of \eqref{tildenuR} should be understood as the limiting result when the support of the wave packets tends to the entire Rindler wedge. It is an interesting question whether the use of wave packets with finite support  requires a minimum acceleration $a$ for right and left modes to be entangled.

\subsection{Entanglement between spherical shells} 

Since correlations decrease with the distance between the two subsystems, we have explored other configurations for two modes, where the ``contact'' between $A$ and $B$ is maximized. Such is the case when subsystem $A$ has support within a sphere, while subsystem $B$ is supported in a spherical shell surrounding it. The region of support of each mode is illustrated in Fig.~\ref{fig:ball&shell3D}. 

In this section, we construct modes from pairs of operators of the form $(\hat \Phi_i,\hat \Pi_i)$, $i=A,B$, similarly to Sec.~\ref{sec:3}, with the difference that, in this section, we use the following smearing functions: 
\begin{equation}
    f_A(r;R_A)=
    \cos^2\left(\frac{\pi}{2}\frac{r}{R_A}\right)\, \Theta(R_A-r)\, ,\end{equation}
for the sphere (subsystem $A$), and 
\begin{equation}\begin{split}
    &f_B(r;R_B,d_B)=\\
   & \begin{cases}
   \sin^2\left(\pi\frac{r-R_B}{d_B}\right) \qquad R_B\leq r\leq R_B+ d_B\\
    0,\qquad \qquad \qquad \qquad  \text{otherwise}
    \end{cases}\end{split}\label{SFf2}
\end{equation}
for the shell (subsystem $B$). 
These functions and their first derivatives are continuous.

While the smearing function in the sphere depends only on one parameter, namely the radius $R_A$, the mode in the shell is parametrized by the inner radius $R_B$ and the thickness of the shell $d_B$. We have checked that for a massless scalar field the quantities we evaluate in this section (the LN) do not change if we rescale these three parameters simultaneously. Therefore, we have only two independent parameters, which we choose to be $R_B$ and $d_B$ measured in units of $R_A$. Another freedom we play with is the dimension of space $D$. 

We compute the LN between subsystems $A$ and $B$ and the way it changes with $R_B$, $d_B$, and $D$. The results of this section are obtained numerically.

\begin{figure}
    \centering
    \includegraphics[width=0.5\textwidth]{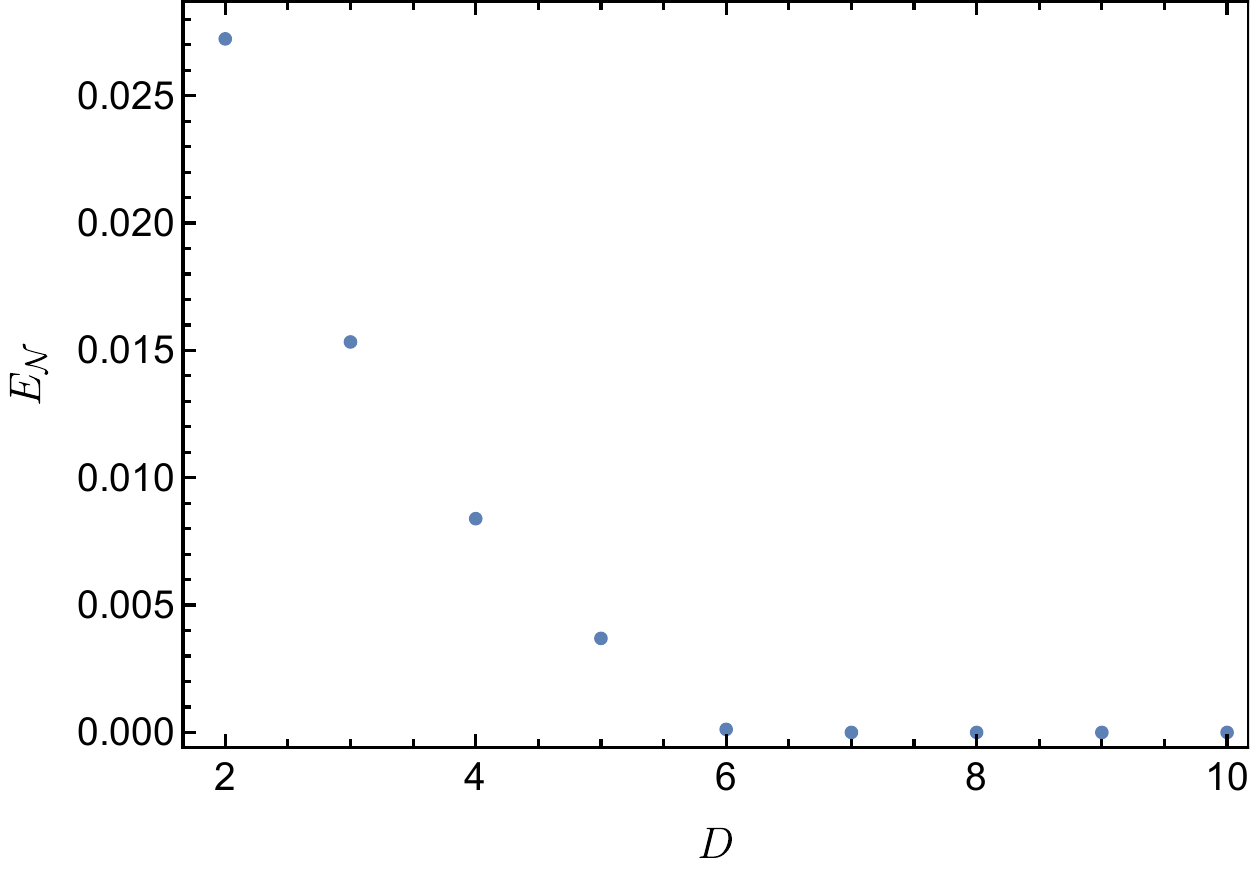}
    \caption{LN as a function of the spatial dimension, $D$, for the configuration of two modes depicted in Fig.~\ref{fig:ball&shell3D} (a), i.e., when there is no gap between the sphere and the  shell, $R_A=R_B =1$. We choose $d_B=0.5$ for this plot.}
    \label{fig:LN_vs_D_Rs1}
\end{figure}

First, Fig.~\ref{fig:LN_vs_D_Rs1} shows the value of the LN when there is no gap between the shell and the sphere $(R_A=R_B)$ (the optimal case) and for a fixed value of $d_B$. The LN is evaluated as a function of the dimensionality of space $D$. The main lessons from this plot are that (i) the LN is different from zero, and (ii) it decreases in higher dimensions, completely vanishing for $D>6$ for the values of $d_B$ we have chosen.

\begin{figure}
    \centering
    \includegraphics[width=0.5\textwidth]{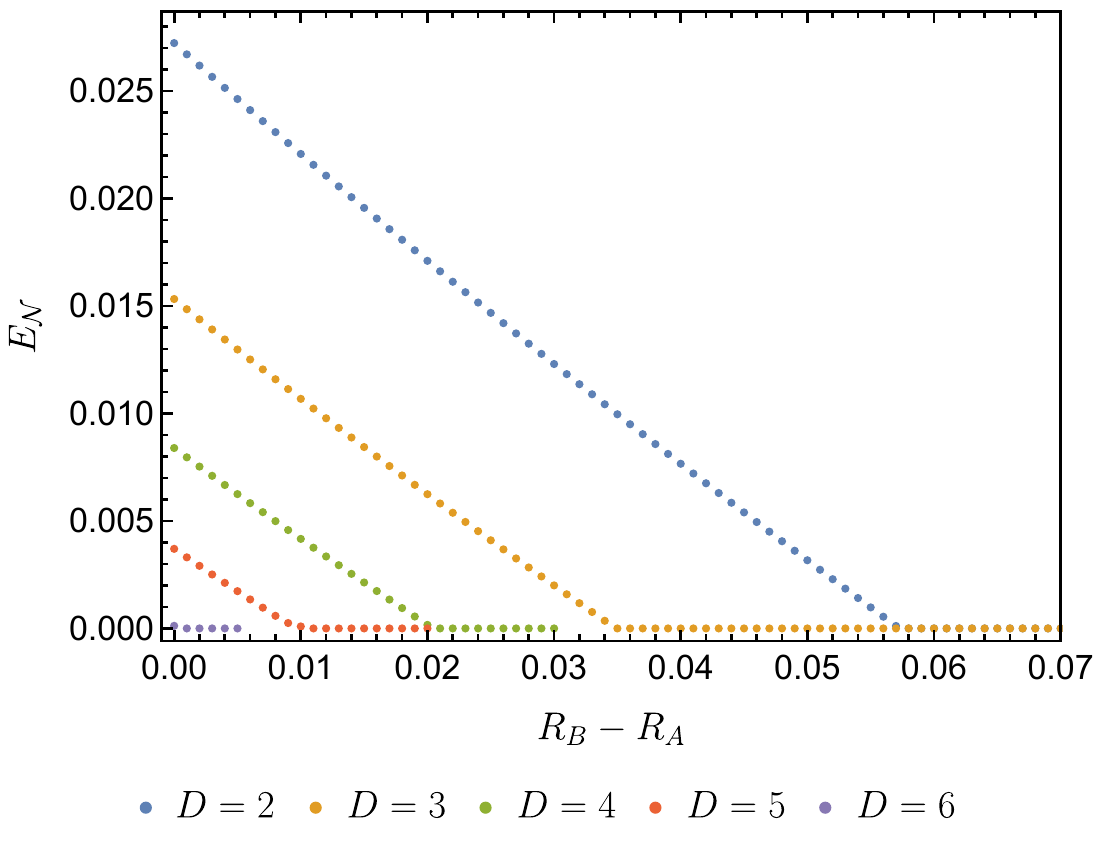}
    \caption{LN for the same configuration as in Fig.~\ref{fig:LN_vs_D_Rs1}, now plotted versus the separation between the sphere and the shell, for different spatial dimensions $D$. We used $d_B = 0.5$ for the thickness of the shell (all distanced measured in units of $R_A$).}
    \label{fig:LN_vs_RbmRa_D2to6_delta0p5}
\end{figure}

Next, we study how the LN depends on the distance between $A$ and $B$, as quantified by  $R_B-R_A$. This is shown in Fig.~\ref{fig:LN_vs_RbmRa_D2to6_delta0p5} for several values of $D$. As expected, the LN quickly falls off with the distance between both subsystems and disappears beyond some threshold distance.  Once more, we see that entanglement is weaker in higher dimensional theories.

\begin{figure}
    \centering
    \includegraphics[width=0.5\textwidth]{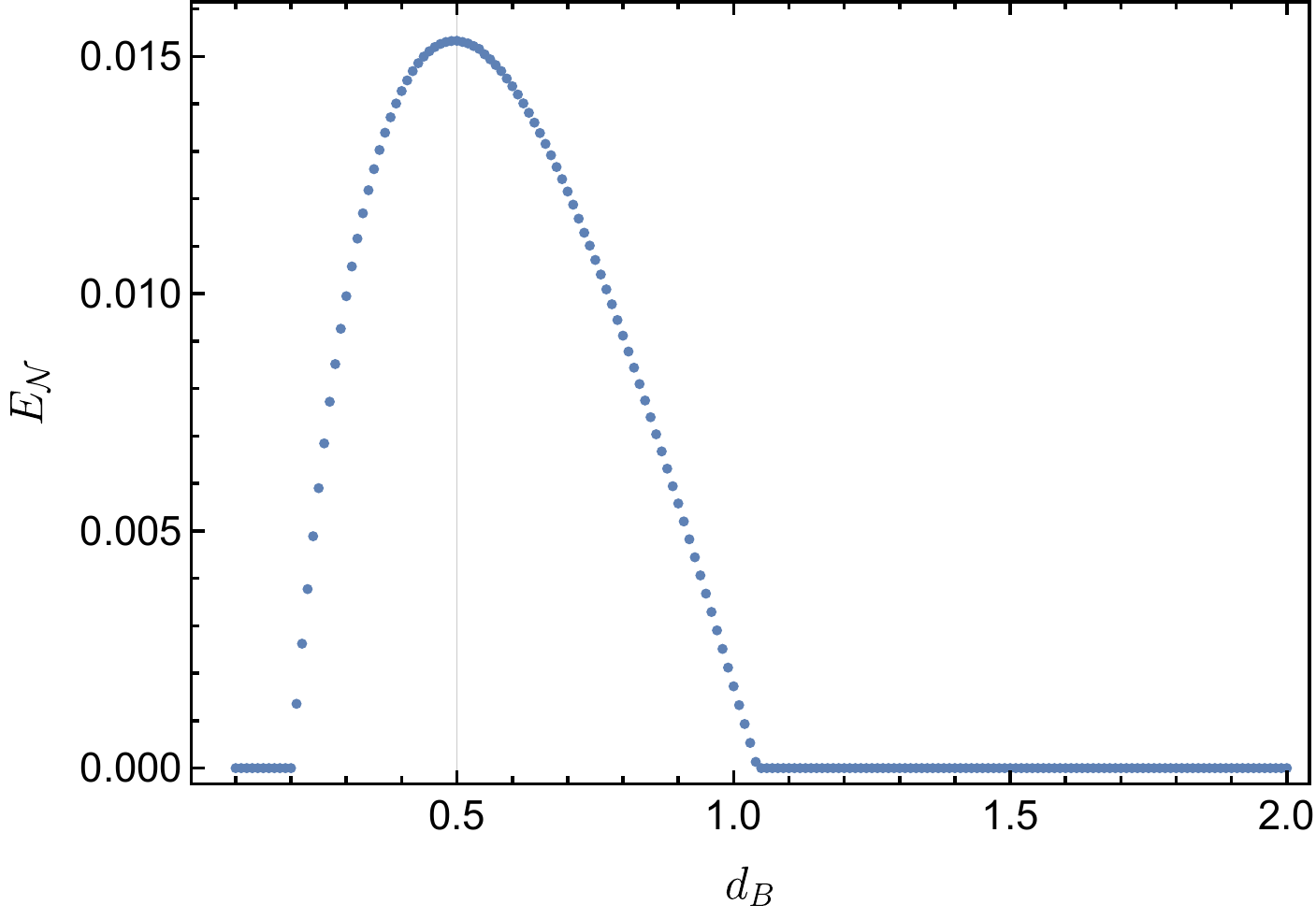}
    \caption{LN as a function of the width of the  shell, when there is no gap between the sphere and the shell. Interestingly, the LN is different from zero only in a finite interval of $d_B$.}
    \label{fig:LN_vs_dB_D3}
\end{figure}

Finally, we study how the LN depends on the thickness of the shell $d_B$. Figure~\ref{fig:LN_vs_dB_D3} shows a curious result: the LN is different from zero in a finite interval and vanishes when $d_B$ is either bigger or smaller than this interval. 

As a final curiosity, we plot in Fig.~\ref{fig:Several_shells} an ``onionlike'' configuration, in which we increase the  number of spherical shells and assign one mode per shell alternating between modes in subsystem $A$ and $B$.  Figure~\ref{fig:LN_Several_shells} shows what is expected: the LN grows monotonically with the number of layers. 

\savebox{\mybox}{\includegraphics[width=0.5\textwidth]{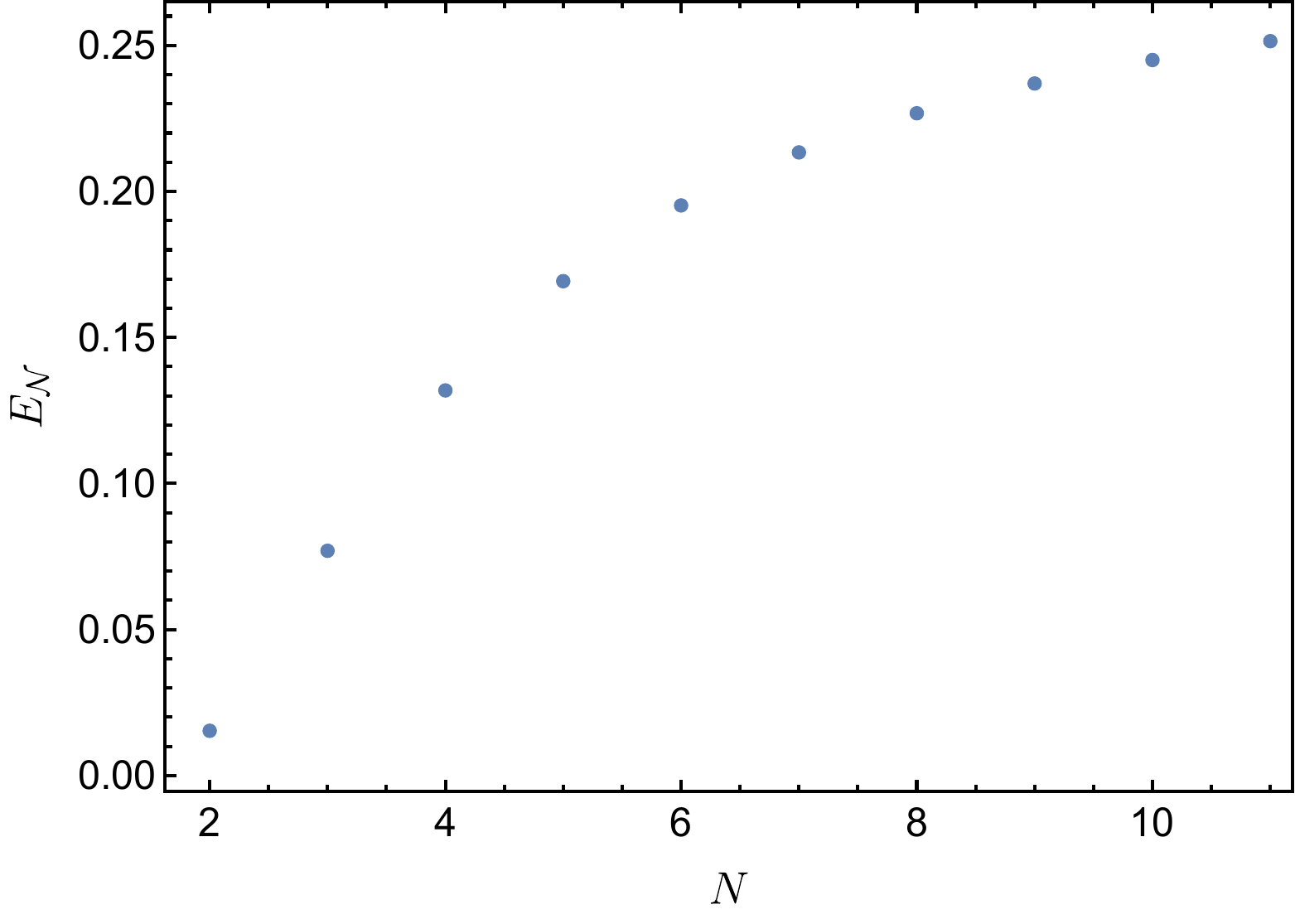}}
\begin{figure*}
    \centering
    \subfigure[\label{fig:Several_shells}]{
     \begin{minipage}{0.45\textwidth}
        \centering
        \vbox to \ht\mybox{%
            \vfill
           
         \includegraphics[width=0.8\textwidth]{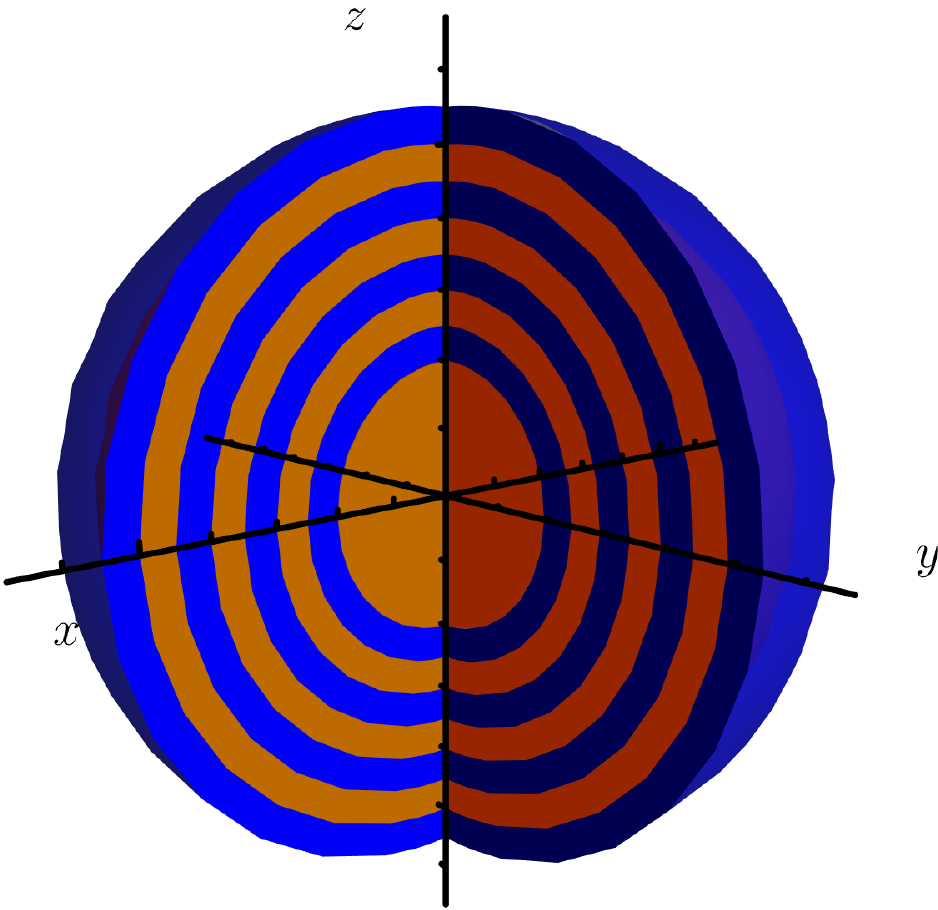}

            \vfill
        }
       
    \end{minipage}}~\subfigure[\label{fig:LN_Several_shells}]{\begin{minipage}{0.45\textwidth}
        \centering
        \usebox{\mybox}
    
    \end{minipage}}
    \caption{(a) Regions of support of modes in an onionlike configuration. The orange  sphere and shells belong to subsystem $A$, and the blue shells belong to subsystem $B$. (b) LN as a function of the total number of degrees of freedom $N=N_A+N_B$. As expected from our previous results, the LN grows with the number of shells.}
    \end{figure*}

In summary, in this subsection, we have (finally) found configurations of two modes compactly supported in disjoint regions that are entangled in $D\geq 2$ spatial dimensions. Entanglement is fragile, in the sense that it disappears when increasing the distance between the modes or the dimension of space. These results contain a valuable message: entanglement can be found in pairs of disjoint and compactly supported modes, but one needs to  carefully choose their spatial configuration. This  is in consonance  with recent results  in \cite{deSLTorres:2023aws}, which indicate that the entanglement between regions $R_A$ and $R_B$ is largely concentrated in modes sharply supported near the boundaries.

\section{Discussion}\label{sec:discussion}
Entanglement in quantum field theory has been discussed in detail  from diverse perspectives. The entanglement entropy associated with a region of space $R$ is perhaps the most studied quantity in this context \cite{Calabrese:2004eu}; it has played an important role in many developments in theoretical physics, ranging from black holes \cite{Solodukhin:2011gn,Srednicki:1993im,Bombelli:1986rw,Sorkin:1985bu} to the quantum nature of spacetime \cite{Jacobson:1995ab,Ryu:2006bv,Bianchi:2012ev}. This entropy also presents some inconveniences. In the first place, it is intrinsically divergent, requiring a regulator to extract a finite value from it, a procedure that introduces ambiguities. Moreover, the Hilbert space of the field theory is {\em not} of the form $\mathcal{H}_R\otimes \mathcal{H}_{\bar R}$, with $\mathcal{H}_R$ the Hilbert space of the field degrees of freedom within region $R$ and $\mathcal{H}_{\bar R}$ the analog for the region $\bar R$ complementary to $R$. This implies that one is outside the standard realm to define entanglement in quantum mechanics, making it unclear how to interpret this entropy in terms of entanglement between region $R$ and its complement \cite{Hollands:2017dov}. 

On the other hand, starting from the Reeh-Schlieder theorem \cite{reehschlider}, it is possible to show that the field degrees of freedom supported within two compact regions of space, $R_A$ and $R_B$, which are separated from each other, are entangled for any state satisfying the Reeh-Schlieder property. The separation guarantees that entanglement is finite \cite{Verch:2004vj,Hollands:2017dov}.  In Minkowski spacetime, the Reeh-Schlieder property holds for any state of finite energy, including the vacuum, implying that entanglement is ubiquitous in this theoretical paradigm. 

These results, although of great conceptual interest, involve subsystems with infinitely many degrees of freedom and do not tell us how much finite sets of field modes are entangled, or even if they are entangled at all. The primary goal of this paper is  to introduce a strategy to address these questions and answer them in some concrete examples. 

The calculations in this article are restricted to a free scalar theory in $D+1$-dimensional Minkowski spacetime. We extract individual degrees of freedom (modes) of the field, localized in a compact region of space, by smearing the field and its conjugate momentum against functions of compact support. Each mode defines an algebra isomorphic to the algebra of an ordinary harmonic oscillator. This strategy provides a way of extracting a finite set of modes out of the field in a local and covariant manner, to which the standard tools of quantum mechanics to evaluate entanglement can be applied. In particular, our strategy is free of divergences  plaguing other approaches. 

The focus on a finite number of modes is motivated by the finite capabilities of observers. The resulting system has all the benefits of a lattice field theory regarding conceptual  and computational simplicity, while keeping the richness and subtleties of the continuum theory. In contrast to lattice field theory, we do not truncate the degrees of freedom prior to quantization. 
The concrete family of modes under consideration depends on the choice of smearing functions. We have explored different families in this work and have focused on results that are common to all of them. For some of our smearing functions, we obtain results analytically in $D$ spatial dimensions. For other smearings, we proceed numerically. We have tested our numerical tools against the analytical results when they are available and checked that they agree to high precision. 

Using the Gaussianity of the Minkowski vacuum, we have computed the reduced state describing a finite number of modes and evaluated its entropy, mutual information, and entanglement. In particular, we have checked that the reduced state describing finite dimensional subsystems is always mixed, in agreement with general results \cite{Ruep:2021fjh}.  

The main lesson from our analysis is that it is difficult to find pairs of modes supported in disjoint regions of space separated by a nonzero distance that are entangled. The difficulty increases with the dimensionality of space. Namely, in $D=1$ it is relatively easy to find pairs of such modes which are entangled, but this task becomes increasingly challenging for $D\geq 2$. In fact, we find that the regions of support of two modes need to be carefully chosen to find any entanglement for $D\geq 2$. One example of  a configuration for which we find entanglement is when one mode is supported within a sphere while the other mode is supported on a spherical shell surrounding the sphere. This configuration is efficient in minimizing the distance between both modes and is able to capture pairwise entanglement. (Another configuration in which we have found pairwise entanglement for $D\geq 2$ is when two independent modes coexist in the same region of support.)

In the cases where we find entanglement between a pair of modes, or between two subsystems each made of a finite number of them, we have checked that the entanglement quickly disappears when the distance between the subsystems increases. 

Hence, we conclude that entanglement in field theory is not as prevalent as normally thought. It is ubiquitous when considering subsystems containing infinitely many modes, but not for finite dimensional systems. This is an important lesson which, to the best of our knowledge, has not been pointed out before. Furthermore, this result is compatible with the Reeh-Schlieder theorem, which guarantees that given a field mode supported in a region $R_A$, and a second region $R_B$, there exists at least one mode in $R_B$ that is entangled with the given mode in $R_A$ \cite{Hollands:2017dov}. However, the theorem does not tell how many modes in $R_B$ are entangled with the given mode in $R_A$, or how complicated such a mode is. Our results show that one needs to carefully select the mode in $R_B$ to find any entanglement. This is compatible and complementary to the results recently obtained in \cite{deSLTorres:2023aws}, which indicate that the entanglement between regions $R_A$ and $R_B$ is largely concentrated in modes sharply supported near the boundaries. 

 We finish by pointing out the relation  between our results and the protocol of entanglement harvesting in quantum field theory~\cite{Reznik:2002fz,Reznik:2003mnx,Pozas-Kerstjens:2015gta}. This protocol couples the field theory to two nonrelativistic systems, which play the role of detectors. These detectors are turned on only for a finite amount of time and are separated in space in such a way that they remain spatially separated during the interval they are on. The detectors are prepared in their respective ground states, so the initial state of the system made of the two detectors is a product state with no entanglement. After the interaction, one is interested in knowing whether the two detectors end up in an entangled state. If they do because the detectors do not interact with each other, nor do they interact via the field since they remain spatially separated, then the only possible origin of the entanglement  is entanglement in the field itself which has been swapped to the detectors by means of the interaction. In this sense, the detectors ``harvest'' entanglement from the field.




Our calculations raise the question of where the entanglement harvested in the detectors is coming from since generic pairs of field modes are not entangled. 
An answer to this question requires a detailed analysis of which modes  of the field---and how many of them---detectors actually couple to.  Such analysis goes beyond the scope of this paper and will be reported in \cite{harvesting}.

\acknowledgments

The content of this paper has benefited enormously from discussion with: A.~Ashtekar, E.~Bianchi, B.~Elizaga-Navascues, A.~Delhom, G.~Garcia-Moreno, S.~Hollands, E.~Martin-Martinez, J. Polo-Gomez, A. del Rio, K. Sanders, and V. Vennin. I.A.\ and D.K.\ are supported by the NSF Grant No. PHY-2110273 and by the Hearne Institute for Theoretical Physics. S. N. is supported by the Universidad de Valencia, within the Atracci\'o de Talent Ph.D Fellowship No. UV-INV- 506 PREDOC19F1-1005367.

\appendix

\section{\label{app:details}DETAILS}

In this appendix, we provide some of the details that have been omitted in Secs.~\ref{sec:3} and~\ref{sec:4}.  In particular, we compute the components of the covariance matrix for a system consisting of $N$ spacelike separated modes of a scalar field in $D+1$-dimensional Minkowski spacetime. We assume that the $N$-modes are supported in spacelike separated spherically symmetric regions of radius $R$. For simplicity, all the modes we consider are extracted using the same smearing function, which in addition we take to belong to the family presented in Eq.~\eqref{eq:family_test_funcs1}, that is 

\begin{equation}
    \hat{\Phi}_i = \int \mathrm{d}^Dx\,f_i^{(\delta)}(\vec{x})\, \hat{\phi}(\vec{x})\,,
\end{equation} and \begin{equation}
      \hat{\Pi}_i = c \int \mathrm{d}^Dx\,f_i^{(\delta)}(\vec{x})\, \hat{\pi}(\vec{x})\,,
\end{equation} with $i = 1,\dots ,N$ and such that $|\Delta \vec{x}_{ij}| > 2\,R$ for every $i \neq j$ (to ensure that the regions are truly disjoint). 

Imposing canonical commutation relations between the operator pairs $(\hat{\Phi}_i, \hat{\Pi}_i)_{i=1,\dots, N}$  (that is,  $[\hat{\Phi}_i, \hat{\Pi}_j] = i \delta_{ij}$ ) allows us to compute the normalization constant $A_{\delta}$ for any dimension, 
\begin{equation}\label{fFourier}
    A_{\delta} = c^{-1/2}\, R^{-D/2}\pi^{-D/4} \sqrt{\frac{\Gamma(1 + D/2 +2 \delta)}{\Gamma(1+2\delta)}}\,. 
\end{equation}

As we already mentioned, one of the advantages of the functions~\eqref{eq:family_test_funcs1} is that their $D$-dimensional Fourier transform has a simple expression in terms of Bessel functions. In particular, we find (see Theorem. 4.15 in Sec.~IV.4 of~\cite{stein1971introduction})
\begin{equation*}\begin{split}
    &\tilde{f}_i^{(\delta)} (\vec{k})=  \int \mathrm{d}^Dx\, \mathrm{e}^{i \vec{k}\cdot\vec{x}} f_i^{(\delta)} \left(\vec{x}\right)\\
    &=  \mathrm{e}^{i \vec{k}\cdot \vec{x}_i}   A_{\delta}\Gamma(\delta + 1) 2^{\delta}(2\pi)^{\frac{D}{2}} \left(k R\right)^{-(\frac{D}{2}+\delta)}\,J_{\frac{D}{2}+\delta} (kR)\,, 
\end{split}
\end{equation*} where $J_{\frac{D}{2}+\delta} (kR)$ is the Bessel function of the first kind of order $\frac{D}{2}+\delta$. 
The relevant two-point functions in the Minkowski vacuum can be computed straightforwardly, and one finds 
\begin{equation*}
    \braket{\{\hat{\Phi}_i, \hat{\Phi}_j\}} = R^{D+1}\int \frac{\mathrm{d}^Dq}{(2\pi)^D} \frac{|\tilde{f}_i^{(\delta)}(q)|^2\, e^{i \vec{q}\cdot \vec{\rho}_{ij}}}{\sqrt{q^2 + \mu^2}}\,,
\end{equation*}
\begin{equation*}
     \braket{\{\hat{\Pi}_i, \hat{\Pi}_j\}} =c^2 R^{D-1}\int \frac{\mathrm{d}^Dq}{(2\pi)^D} |\tilde{f}_i^{(\delta)}(q)|^2\, e^{i \vec{q}\cdot \vec{\rho}_{ij}}\sqrt{q^2 + \mu^2}\,,
\end{equation*} and 
\begin{equation*}
    \braket{\{\hat{\Phi}_i,\hat{\Pi}_j\}} = 0\,,
\end{equation*} where $\vec{q} = R \vec{k}$, $\vec{\rho}_{ij} = \frac{\Delta \vec{x}_{ij}}{R} = \frac{\vec{x}_i - \vec{x}_j}{R}$ is the dimensionless distance between the centers of the spherical regions centered at $\vec{x}_i$ and $\vec{x}_j$, and $\mu = m \,R$ is the dimensionless mass. 

Interestingly, in the massless limit ($\mu = 0$) the integrals in the correlation functions above can be solved analytically using the following integrals: 
\begin{widetext}
\begin{equation}\label{eq:tisp_2bessel}
    \int_0^{\infty} dt\,t^{-\beta} J_{\nu}(\alpha t) J_{\mu}(\alpha t) = \frac{\alpha^{\beta -1 } \Gamma(\beta) \Gamma\left(\frac{\nu+\mu-\beta+1}{2}\right)}{2^{\beta} \Gamma\left(\frac{-\nu + \mu + \beta +1}{2}\right) \Gamma \left( \frac{\nu + \mu + \beta+1}{2}\right) \Gamma \left( \frac{\nu - \mu + \beta +1}{2}\right)}\,,
\end{equation}

 if $\mathrm{Re}(\nu + \mu +1) > \mathrm{Re}\, \beta > 0 $, $\alpha>0$ (Eq. 6.574.2 in~\cite{gradshteyn_table_2000}) and 
 \begin{equation}\label{eq:tisp_3bessel}
\begin{split}
    \int_0^{\infty} dt\, t^{\beta - 1} J_{\alpha} (at) J_{\mu}(bt) J_{\nu}(ct) =& \frac{2^{\beta-1}a^{\alpha}b^{\mu} c^{-\alpha - \mu - \beta } \Gamma \left( \frac{\alpha + \mu + \nu + \beta}{2}\right) }{\Gamma (\alpha +1) \Gamma(\mu +1) \Gamma \left(1-\frac{\alpha + \mu - \nu + \beta}{2} \right) }\\
   & \times F_4 \left(\frac{\alpha + \mu - \nu + \beta}{2}, \frac{\alpha + \mu + \nu + \beta}{2}; \alpha +1 , \mu +1; \frac{a^2}{c^2}, \frac{b^2}{c^2} \right) \,, 
\end{split}
\end{equation}
 \end{widetext}
which holds if $\mathrm{Re}(\alpha + \mu + \nu + \beta) > 0$, $\mathrm{Re} \beta < \frac{5}{2}$, $a > 0$, $b>0$, $c>0$ and $c>a+b$ (Eq. 6.578.1 in~\cite{gradshteyn_table_2000}) where 
\begin{equation}
    F_4 (\alpha, \beta, \gamma, \gamma'; x,y) = \sum_{m=0}^{\infty} \sum_{n=0}^{\infty} \frac{(\alpha)_{m+n} (\beta)_{m+n}}{(\gamma)_{m}(\gamma')_n m! n! } x^m y^n \,,  
\end{equation} with $|\sqrt{x}| + |\sqrt{y}| <1\,,$ is the Appell hypergeometric function of the fourth kind~\footnote{Notice that when $x=y$ and $\gamma_1=\gamma_2$, this function reduces to a generalized hypergeometric function, that is $F_4(\alpha, \beta, \gamma, \gamma, x,x) = \, _3F_2\left(\alpha ,\beta ,\gamma -\frac{1}{2};\gamma ,2 \gamma -1;4 x\right)$}. Using Eqs.~\eqref{eq:tisp_2bessel} and~\eqref{eq:tisp_3bessel}, the two point correlations in the massless limit ($\mu=0$) can be written as 
\begin{equation}
     \braket{\{\hat{\Phi}_i, \hat{\Phi}_j\}} = 2N_{\delta}^2\,\frac{R}{c}\, \left\{ \begin{matrix}
     \mathcal{J}^{D}(-1,\delta) & i=j\\
     \mathcal{L}^{D}(-1,\delta, \rho_{ij})& i\neq j
     \end{matrix}\right.\,,
\end{equation}
\begin{equation}
     \braket{\{\hat{\Pi}_i, \hat{\Pi}_j\}} = 2N_{\delta}^2\,\frac{c}{R}\, \left\{ \begin{matrix}
     \mathcal{J}^{D}(1,\delta) & i=j\\
     \mathcal{L}^{D}(1,\delta, \rho_{ij})& i\neq j
     \end{matrix}\right.\,,
\end{equation} where
\begin{equation}
      \mathcal{J}^{D}(\lambda, \delta)= 2^{-1-2\delta +\lambda}\frac{\Gamma\left(\frac{D+\lambda}{2}\right)\Gamma\left(1+2\delta - \lambda\right)}{\Gamma\left(1+\delta-\frac{\lambda}{2} \right)^2 \Gamma\left(\frac{D-\lambda}{2}+2\delta+1 \right)}\,,
\end{equation}
\begin{equation}
\begin{split}
    \mathcal{L}^{D}(\lambda, \delta, \rho) =&\rho^{-(D+\lambda)} \frac{ \Gamma\left(\frac{D+\lambda}{2}\right)\Gamma\left(D/2\right)}{2^{1+2\delta-\lambda}\Gamma\left(\frac{D}{2}+1 + \delta\right)^2\Gamma\left(-\frac{\lambda}{2}\right)}\times \\
    &{}_3F_2\left[\begin{matrix}
    1+\frac{\lambda}{2},\frac{D+\lambda}{2},\frac{D+1}{2}+\delta\\
    \frac{D}{2} + 1+\delta, D+1+2\delta
    \end{matrix}\,; \frac{4}{\rho^2}\right]\,,
\end{split}
\end{equation} and 
\begin{equation}
    N_{\delta}^2 = \frac{2^{2\delta} \Gamma\left(1 + \frac{D}{2} + 2\delta\right)\Gamma\left(1 + \delta\right)^2}{\Gamma\left(1+2\delta\right)\Gamma\left(D/2\right)}\,.
\end{equation}

\section{\label{app:smooth_functions}\label{Ap:B} FROM FINITELY DIFFERENTIABLE TO SMOOTH SMEARING FUNCTIONS}

 The set of smooth (i.e., infinitely differentiable) functions of compact support, $C^{\infty}_0(\mathbb{R}^D)$, provides a mathematically convenient habitat for the  smearing functions used to define smeared operators supported on a compact region of space. In some parts of this paper, however, we have used smearing functions which are only finitely differentiable. This is true, in particular, for the smearing functions $f^{(\delta)}$ introduced in Sec.~\ref{sec:3}. For them, we were able to derive results analytically. In this appendix, we address the question of whether the finite differentiability of these functions limits the range of validity of the results obtained from them. We show that this is not the case.\footnote{P.R.M. thanks K. Sanders for pointing out the argument spelled out in this Appendix.}  (We restrict in this appendix to massless fields in $D\geq 2$, but the arguments can be extended to the massive case and $D=1$.)

Our argument goes as follows. In the first place, we identify a  space of (not necessarily smooth) functions for which the smeared operators constructed from them are all well defined  in the standard Fock space in Minkowski spacetime. Secondly, we argue that, for each smearing function in this family, there always exists another function in $C^{\infty}_0(\mathbb{R}^D)$ producing the same physical predictions with arbitrarily high accuracy. This implies that restriction to this family of functions does not limit the validity of our results. Finally, we show that the  finitely-differentiable  functions we have used in some portions of this paper belong to this family.\\

\begin{defn} (\textit{Definition 1.31 in~\cite{bahouri_fourier_2011}}\label{def:hom_sobolev_spaces}) 
Let $s\in \mathbb{R}$. The homogeneous Sobolev space $\dot{H}^s(\mathbb{R}^D)$ is the space of tempered distributions $f$ over $\mathbb{R}^D$, with locally integrable Fourier transform and satisfying 
\begin{equation}\label{eq:sobolev_norm}
    ||f||^2_{s} := \int_{\mathbb{R}^D} \frac{d^Dk}{(2\pi)^D}\, |\vec k|^{2s} \, |\tilde{f}(\vec k)|^2  < \infty
\end{equation}
where $\tilde{f}$ denotes the Fourier transform of $f$. 
\end{defn}

(\textit{Proposition 1.34 in~\cite{bahouri_fourier_2011}}\label{prop:hilbert_space})
$\dot{H}^s(\mathbb{R}^D)$ is a Hilbert space if and only if $s< D/2$, with inner product

\begin{equation}
    (f|g)_{s} = \int_{\mathbb{R}^D} \frac{d^Dk}{(2\pi)^D} \, |\vec k|^{2s} \tilde{f}(\vec k) \, \bar{\tilde{g}}(\vec k)\, ,
\end{equation}
(bar denotes complex conjugation).

The connection between homogeneous Sobolev spaces and the content of this paper becomes clear by noticing that the vacuum expectation value of smeared operators can be written in terms of this inner product as

\begin{align}
\label{eq1}
&\braket{\hat{\Phi}^2[f]} =  \frac{1}{2 }\, (f,f)_{-1/2} \, , \\
\label{eq2} 
&\braket{\hat{\Pi}^2[f]} =  \frac{1}{2 }\, (f,f)_{1/2} \, , \\
\label{eq3} 
&\braket{\{\hat{\Phi}[f],\hat{\Phi}[g]\}} = \Re \, (f,g)_{-1/2} \, , \\
\label{eq4} 
&\braket{\{\hat{\Pi}[f],\hat{\Pi}[g]\}}  = \Re \, (f,g)_{1/2} \, , \\
&\braket{\{\hat{\Phi}[f],\hat{\Pi}[g]\}} = 0  \, ,
\end{align} 
with $\Re$ denoting the real part.
For completeness, we prove here one of these equalities; the rest are proven similarly as follows:
\begin{widetext}
\begin{equation}
\begin{split}
    \braket{\{\hat{\Phi}[f], \hat{\Phi}[g]\}} &=    \int d^Dx f(\vec x)\int d^Dx' g(\vec x') \int \frac{d^Dk}{(2\pi)^D} e^{i \vec{k}\cdot \vec{x}} \int \frac{d^Dk'}{(2\pi)^D}   e^{i \vec{k}' \cdot \vec{x}'} \braket{\{\hat{\phi}_{\vec{k}},\hat{\phi}_{\vec{k}'}\}} \\
    &=   \int \frac{d^Dk}{(2\pi)^D} \frac{1}{2|\vec k|}  \, \Big(\tilde{f}\, \bar{ \tilde{g}}+\tilde{g}\, \bar{ \tilde{f}}\Big) = \Re (f|g)_{-1/2}
\end{split}
\end{equation}
\end{widetext}
where we have used $\hat{\Phi}(x)=\int \frac{d^Dk}{(2\pi)^D}\, \hat{\phi}_{\vec{k}} \, e^{i \vec{k}\cdot \vec{x}}$, and $\braket{\hat{\phi}_{\vec{k}}\hat{\phi}_{\vec{k}'}} =\frac{1}{2|k|}\delta(\vec k+\vec k')$.

This automatically implies that if we smeared the field $\hat \Phi$ with functions in $ \dot{H}^{-1/2} (\mathbb{R}^D)$, and the momentum $\hat \Pi$ with functions in $\dot{H}^{1/2} (\mathbb{R}^D)$, then the second moments \eqref{eq1}--\eqref{eq4} are all finite  for $D\geq 2$. This in turn implies that the action of these operators in the Minkowski vacuum produces a state with finite norm. We will not make such a distinction and restrict all smearing functions to belong to the intersection  $ \dot{H}^{-1/2} (\mathbb{R}^D) \cap \dot{H}^{1/2} (\mathbb{R}^D)$, so they can be used to smear both field and momentum operators. This restriction will bring additional conveniences, as we discuss below.

Furthermore, if $\hat a^{\dagger}_w$ is the creation operator associated with a function $w(\vec x)\in C^{\infty}_0(\mathbb{R}^D)$, then the commutators $\big[\hat{\Phi}[f],\hat a^{\dagger}_w\big]=\frac{1}{2}(f,w)_{-1/2}$, and $\big[\hat{\Pi}[f],\hat a^{\dagger}_w\big]=(f,w)_{1/2}$ are finite. This implies that the operators $\hat{\Phi}[f]$ and $\hat{\Pi}[f]$ produce states of finite norm  when acting on a dense subspace of the entire Fock space. Hence, as long as the smearing functions belong to $ \dot{H}^{-1/2} \cap \dot{H}^{1/2}$, the associated operators are well defined. 

Secondly, the intersection $ \dot{H}^{-1/2} (\mathbb{R}^D) \cap \dot{H}^{1/2} (\mathbb{R}^D)$ is a subspace of $L^2(\mathbb{R}^D)$ (\textit{Proposition 1.32 in~\cite{bahouri_fourier_2011}}\label{prop:inclusions}). Since $C^{\infty}_0(\mathbb{R}^D)$ is dense in $L^2(\mathbb{R}^D)$, it is also dense in $ \dot{H}^{-1/2} (\mathbb{R}^D) \cap \dot{H}^{1/2} (\mathbb{R}^D)$. This automatically implies that, given any function $f\in \dot{H}^{-1/2} (\mathbb{R}^D) \cap \dot{H}^{1/2} (\mathbb{R}^D)$, and a small real number  $\epsilon> 0$, one can always find a function $f_{\epsilon} \in C^{\infty}_0(\mathbb{R}^D)$ such that $||f-f_{\epsilon}||_{\pm 1/2}<\epsilon$, and consequently, the second moments  \eqref{eq1}--\eqref{eq4} defined from $f$ and $f_{\epsilon}$ are equally close. 

Finally, it remains to check that all the functions we use in this paper actually belong to $\dot{H}^{-1/2} (\mathbb{R}^D) \cap \dot{H}^{1/2} (\mathbb{R}^D)$. For the family of functions $f^{(\delta)}(x)$ introduced in Sec.~\ref{sec:3}, this has been shown in Appendix A, whenever $\delta\geq 1$ and $D>1$. To see this, recall that the Fourier transform of the functions in this family consists of products of  monomials and Bessel functions [see Eq.~\eqref{fFourier}], and one can easily check that they are all locally integrable for $\delta\geq 1$ and $D\geq 1$. In addition, in Appendix A we showed that $k^{\pm 1/2} \tilde{f}^{(\delta)}(k) \in L^{2} (\mathbb{R}^D)$ (by showing explicitly that the integral can be solved analytically) for $\delta \geq 1$ and $D\geq 2$. Notice that this implies $||f||_{\pm 1/2} = \int d^Dk \, |k|^{\pm 1}|\tilde{f}^{(\delta)}|^2 <\infty $. Hence, we conclude that $f^{(\delta)}(x) \in \dot{H}^{-1/2} (\mathbb{R}^D) \cap \dot{H}^{1/2} (\mathbb{R}^D)$ for $\delta \geq 1$ and $D\geq 2$.  

The case $\delta=0$ (Heaviside step function) is special since $f^{(0)}(x)\in H^{-1/2} (\mathbb{R}^D)$ but $f^{(0)}(x)\notin H^{1/2} (\mathbb{R}^D)$ (which leads to the presence of the UV divergences in the momentum self-correlations, as first noticed in~\cite{Martin:2015qta}). 

We have also checked explicitly that all the smearing functions introduced in Sec.~\ref{sec:5} belong to $ \dot{H}^{-1/2} (\mathbb{R}^D) \cap \dot{H}^{1/2} (\mathbb{R}^D)$.

\bibliographystyle{apsrev4-1}
\bibliography{MinkBib.bib}
\end{document}